\documentclass[useAMS,usenatbib]{mn2e}
\usepackage{graphicx}
\title[Symbiotic Stars towards the Galactic Bulge]{Symbiotic stars and other H$\alpha$ emission line stars towards the Galactic Bulge\thanks{Based on observations made with the Anglo-Australian Telescope at Siding Spring Observatory, the 1.3-m Warsaw Telescope at Las Campanas Observatory of the Carnegie Institution for Science, Chile, the Very Large Telescope at Paranal Observatory under programme 079.D-0764(A), the Southern African Large Telescope (SALT) under programme 2012-1-RSA-009, the New Technology Telescope at La Silla Observatory under programmes 071.D-0448(A) and 079.D-0764(B), and the South African Astronomical Observatory (SAAO) 1.9-m and 1.0-m telescopes.}}
\author[Miszalski, Miko{\l}ajewska and Udalski]{Brent Miszalski$^{1,2}$\thanks{E-mail: brent@saao.ac.za}, Joanna Miko{\l}ajewska$^{3}$ and Andrzej Udalski$^{4}$\\
$^{1}$South African Astronomical Observatory, PO Box 9, Observatory, 7935, South Africa\\
$^{2}$Southern African Large Telescope Foundation, PO Box 9, Observatory, 7935, South Africa\\
$^{3}$Nicolaus Copernicus Astronomical Centre, Bartycka 18, 00716 Warsaw, Poland\\
$^{4}$Warsaw University Observatory, Al. Ujazdowskie 4, PL-00-478, Warsaw, Poland\\
}
\begin{document}

\date{Accepted . Received ; in original form }

\maketitle
\begin{abstract}
Symbiotic stars are interacting binaries with the longest orbital periods and their multi-component structure makes them rich astrophysical laboratories. The accretion of a high mass loss rate red giant wind on to a white dwarf (WD) makes them promising Type Ia supernovae (SNe Ia) progenitors. Systematic surveys for new Galactic symbiotic stars are critical to identify new promising SNe Ia progenitors (e.g. RS Oph) and to better estimate the total population size to compare against SNe Ia rates. Central to the latter objective is building a complete census of symbiotic stars towards the Galactic Bulge. Here we report on the results of a systematic survey of H$\alpha$ emission line stars covering 35 deg$^2$. It is distinguished by the combination of deep optical spectroscopy and long-term lightcurves that improve the certainty of our classifications. A total of 20 bona-fide symbiotic stars are found (13 S-types, 6 D-types and 1 D$'$-type), 35\% of which show the symbiotic specific Raman-scattered OVI emission bands, as well as 15 possible symbiotic stars that require further study (6 S-types and 9 D-types). Lightcurves show a diverse range of variability including stellar pulsations (semi-regular and Mira), orbital variations and slow changes due to dust. Orbital periods are determined for 5 S-types and Mira pulsation periods for 3 D-types. The most significant D-type found is H1-45 and its carbon Mira with a pulsation period of 408.6 days, corresponding to an estimated period-luminosity relation distance of $\sim$$6.2\pm1.4$ kpc and $M_K=-8.06\pm0.12$ mag. If H1-45 belongs to the Galactic Bulge, then it would be the first bona-fide luminous carbon star to be identified in the Galactic Bulge population. The lack of luminous carbon stars in the Bulge is a longstanding unsolved problem. A possible explanation for H1-45 may be that the carbon enhancement was accreted from the progenitor of the WD companion. A wide variety of unusual emission line stars were also identified. These include central stars of PNe (one [WC10-11] Wolf-Rayet and 5 with high density cores), 2 novae, 2 WN6 Wolf-Rayet stars, 2 possible Be stars, a B[e] star with a bipolar outflow, an ultracompact HII region and a dMe flare star. Dust obscuration events were found in two central stars of PNe, increasing the known cases to 5, as well as one WN6 star. There is considerable scope to uncover several more symbiotic stars towards the Bulge, many of which are currently misclassified as PNe, provided that deep spectroscopy is combined with optical and near-infrared lightcurves. 

\end{abstract}


\section{Introduction}
Symbiotic stars are the longest orbital period interacting binaries composed of an evolved cool giant and an accreting hot, luminous companion (usually a white dwarf, WD) surrounded by a dense ionized nebula. Depending on the nature of the cool giant, two main classes of symbiotic stars have been defined: S-types (stellar) with normal M giants, and orbital periods of the order of a few years, and D-types (dusty) with Mira primaries surrounded by a warm dust (Whitelock 1987). They host a rich circumstellar environment produced by the ionisation of the giant's high mass loss rate wind by the WD. Several phenomena are observed including ionized and neutral regions, dust forming regions, accretion/excretion disks, interacting winds, and bipolar outflows and jets (see e.g. Miko{\l}ajewska 2007, 2012 for recent reviews).

Symbiotic stars have obvious implications towards understanding all interacting binaries that include either evolved giants or accreting white dwarfs during any phase of their evolution. In particular, understanding symbiotic stars may also help to solve one of the timeliest problems in modern astrophysics - the missing progenitors of Type Ia supernovae (SNe Ia). Although there is general consensus that they result from thermonuclear disruption of a CO white dwarf reaching the Chandrasekhar mass, either due to mass accretion from a non-degenerate companion (single-degenerate or SD model) or mass transfer between and/or merger of two white dwarfs (double-degenerate or DD model), the progenitors of SNe Ia have never been observed before the explosion, and each of the proposed scenarios have their strengths and weaknesses (Di Stefano, Orio \& Moe 2013, and references therein). In regard to this problem, symbiotic stars are important not only to the SD scenario, where the progenitor is most likely a symbiotic nova similar to RS Oph (e.g. Dilday et al. 2012), but also to the DD scenario since most DD binaries are descended from symbiotic stars (e.g. Di Stefano 2010).

Driving the discovery for more symbiotic stars are two main factors. First, the potential to find other case studies like RS Oph, and in particular, ensuring that the most interesting objects are routinely monitored for nova-like activity (e.g. those with evidence for tidally distorted donors or orbital periods $\la$1000 days; Miko{\l}ajewska 2013). Second, to derive more realistic estimates of the total Galactic symbiotic population and, in turn, to compare symbiotic star birth rates against SNe Ia rates. Corradi et al. (2010a) and Corradi (2012) identified the Galactic Bulge symbiotic stars as a fundamental population to better characterise before estimates of the total number of Galactic symbiotics can be refined. Current estimates vary considerably from 3$\times10^3$ (Allen 1984), 3$\times10^4$ (Kenyon et al. 1993) to 3--4$\times10^5$ (Munari \& Renzini 1992; Magrini, Corradi \& Munari 2003). None of these, however, are in agreement with the very small number of observed symbiotic stars. Several major catalogues have been compiled including Allen (1984), Kenyon (1986), Miko{\l}ajewska, Acker \& Stenholm (1997) and Belczy\'nski et al. (2000). These efforts have amounted to no more than $\sim$300 symbiotics currently known. More recently, $\sim$1200 symbiotic star candidates were identified from the INT Photometric H$\alpha$ Survey of the Northern Galactic Plane (IPHAS, Drew et al. 2005). Spectroscopic follow-up of the sample has so far identified 14 new symbiotic stars (Corradi et al. 2008, 2010a, 2010b, 2011; Corradi \& Giammanco 2010; Corradi 2012).

A comparable resource to IPHAS in the Southern Galactic Plane is the AAO/UKST SuperCOSMOS H$\alpha$ Survey (SHS, Parker et al. 2005). The quantitative colour-colour candidate selection method developed by Corradi et al. (2008) is however problematic to adapt to the photographic SHS data whose colour-colour plane possesses a greater intrinsic dispersion than IPHAS. The Macquarie/AAO/Strasbourg H$\alpha$ (MASH) planetary nebulae (PNe) surveys effectively used difference imaging (Parker et al. 2006) and semi-automated techniques (Miszalski et al. 2008) to identify unresolved SHS H$\alpha$ emitters (e.g. compact PNe, symbiotic stars and other emission-line stars). As the MASH surveys were primarily concerned with PNe, mostly those H$\alpha$ emitters with the highest chance of being a PN were observed spectroscopically, leaving considerable scope for new symbiotic star discoveries.

Miszalski et al. (2009a) carefully selected a large sample of SHS H$\alpha$ emitters across $\sim$35 square degrees towards the Galactic Bulge and observed them using the AAOmega spectrograph (Sharp et al. 2006) with the Two-degree Field (2dF) facility on the 3.9-m Anglo-Australian Telescope (Lewis et al. 2002). 
A preliminary list of candidate symbiotic stars were published by Miszalski et al. (2009a), however only basic supporting information was given per object as the focus was only to flag non-PNe. A similar approach to that of Miszalski et al. (2009a) was taken in the Large Magellanic Cloud (LMC) by Miszalski et al. (2011b) where some promising symbiotic star candidates were identified. 

This paper contains a comprehensive reanalysis of the H$\alpha$ emitter survey that incorporates deep spectroscopy and long term $I$-band lightcurves to identify new symbiotic stars and other unusual H$\alpha$ emission line stars. It is structured as follows. Section \ref{sec:obs} describes the selection of H$\alpha$ emission line candidates and the spectroscopic and photometric observations of the sample. The overall properties and analysis of the sample are discussed in Sect. \ref{sec:sample}. A subset of the sample are discussed individually in Sections \ref{sec:stypes} (new S-type symbiotics), \ref{sec:dtypes} (new D-type symbiotics), \ref{sec:ddtypes} (a new D$'$-type symbiotic), \ref{sec:pstypes} (possible S-type symbiotics) and \ref{sec:pdtypes} (possible D-type symbiotics). Objects that were found not to be symbiotic stars are also discussed individually in Sections \ref{sec:egb6} (high density core PNe), \ref{sec:other} (miscellaneous objects) and Appendix \ref{sec:superpos} (superpositions). Section \ref{sec:discussion} discusses the completeness and depth of our survey, possible X-ray detections, the unique carbon Mira of H1-45 and its significance in carbon star formation, as well as the common tendencies as a function of symbiotic star type. We conclude in Sect. \ref{sec:conclusion} and several appendices contain finder charts and other observations of the sample.

\section{Observations}
\label{sec:obs}
\subsection{Multi-object spectroscopy}
We employ the same multi-object spectroscopy observations previously described by Miszalski et al. (2009a). The main observations were conducted in service-mode (PI: Miszalski) using the 4-m Anglo-Australian Telescope (AAT) with the 2dF/AAOmega facility (Lewis et al. 2002; Sharp et al. 2006) that can allocate up to 392 optical fibres of 2.1\arcsec\ diameter to science targets over a 2.1$^\circ$ diameter field of view. Additional observations were made using the 8.2-m Very Large Telescope (VLT) with the FLAMES facility (Pasquini et al. 2002) in visitor mode under programme ID 079.D-0764(A). FLAMES was used in the miniature integral field unit (mini-IFU) mode that can allocate up to 15 mini-IFUs for both science and sky targets over a 12.5\arcmin\ field of view. Each mini-IFU consists of a lenslet array that measures $2\times3$ arcsec$^2$ which feeds 20 fibres with a spatial sampling of 0.5\arcsec\ per lenslet.

Tables \ref{tab:aaomega} and \ref{tab:flames} give the field centres of 2dF/AAOmega and VLT FLAMES observations where most fibres were allocated to PNe to determine whether such observations were appropriate for measuring their nebular chemical abundances (Miszalski 2009) and the remainder to H$\alpha$ emission line candidates. Field centres were chosen to maximise the number of catalogued PNe in each field. Most AAOmega fields were observed with a large number of sky fibres (last column of Tab. \ref{tab:aaomega}) to assist in subtracting the highly variable sky background. Targets for spare fibres were chosen by inspection of full-resolution imaging data from the SHS and SuperCOSMOS Sky Survey (SSS, Hambly et al. 2001). Fields were inspected for H$\alpha$ emitters by blinking a colour-composite image made of SHS H$\alpha$ (red), SHS Short-Red (green) and SSS $B_J$ (blue) with the continuum-divided or `quotient' image (H$\alpha$ divided by Short-Red). This approach was first developed during the MASH-II survey (Miszalski et al. 2008). 

\begin{table*}
   \centering
   \caption{Log of AAOmega field centres observed.}
   \label{tab:aaomega}
   \begin{tabular}{lrrllllr}
      \hline
      Field & $\ell$ & $b$ & Coords & Date & Exposures & JD-2450000 & N(sky)\\
            & (deg) &  (deg) & (J2000) & (dd/mm/yy) & (s) & & \\
            \hline
N2	  & 357.86	&$+$1.86	 &  17 33 09.8 $-$29 45 12	& 26/03/07 &	3$\times$300, 3$\times$1200 & 4186 & 73  \\
S3	  & 355.89	&$-$3.26 &  17 48 44.4 $-$34 08 30	& 27/03/07 &	60, 2$\times$300, 2$\times$1200 & 4187 & 60\\
S18  & 1.02	   &$-$1.59 &  17 54 15.6 $-$28 52 25	& 29/05/08 &	60, 240, 1800 & 4616           &183      \\
S20  & 2.28	   &$-$3.26 &  18 03 42.8 $-$28 36 39	& 29/05/08 &	60, 240, 1800 & 4616           &100\\
S21  & 3.57	   &$-$2.06 &  18 01 51.1 $-$26 54 25	& 29/05/08 &	60, 240, 1800 & 4616           &100\\
S22  & 0.46	   &$-$3.72 &  18 01 26.4 $-$30 25 37	& 30/05/08 &	60, 240, 1800 & 4617           &100\\
N3   & 359.72	&$+$1.79	 &  17 38 01.8 $-$28 14 06	& 30/05/08 &	60, 240, 1800 & 4617           &100       \\
N4	  & 1.87	   &$+$1.76	 &  17 43 15.0 $-$26 25 08	& 30/05/08 &	60, 240, 1800 & 4617           &100       \\
N5	  & 358.93	&$+$3.80	 &  17 28 26.5 $-$27 47 56	& 30/05/08 &	60, 240, 1800 & 4617           &100       \\
N7	  & 355.32	&$+$2.81	 &  17 22 57.1 $-$31 20 39	& 30/05/08 &	60, 240, 1800  & 4617          &100       \\
S21a & 3.47	   &$-$2.26 &  18 02 24.2 $-$27 05 05	& 18/04/09 &	900, 2$\times$1200 &4940       &25     \\
S3a  & 355.89	&$-$3.26 &  17 48 44.4 $-$34 08 30	& 19/04/09 &	300, 5$\times$900  &4941 &25\\
S20a & 2.18  	&$-$2.95 &  18 02 15.9 $-$28 32 58	& 23/04/09 &	60, 300, 900, 1800  & 4945  &196   \\
            \hline
   \end{tabular}
\end{table*}

\begin{table*}
   \centering
   \caption{Log of VLT FLAMES field centres observed that appear in this work. Exposure times are given per grating setting.}
   \label{tab:flames}
   \begin{tabular}{lrrllrrrrrrl}
      \hline
      Field & $\ell$ & $b$ & Coords & Date & LR1 &	LR2 &	LR3 &	LR5 &	LR6 & LR7 & JD-2450000\\
            & (deg) &  (deg) & (J2000) & (dd/mm/yy) & (s) & (s) &(s) &(s) &(s) &(s) & \\
            \hline
            F54	&357.04	& $+$4.48 &	17 21 06.0 $-$28 58 53 & 	10/06/07 &	1200	&1200&	300, 1200&	1200&	600&	300 & 4261\\
            F157	&4.07	& $-$3.17	&18 07 17.8 $-$27 00 25 & 	11/06/07 &	1800	&600, 3600&	120, 1800&	1800&	120, 900&	600 & 4262\\
            \hline
   \end{tabular}
\end{table*}

The prepared target lists for AAOmega fields were allocated fibres with the \textsc{configure} program (Shortridge, Ramage \& Farrell 2006) that uses a simulated annealing algorithm to achieve an optimal configuration (Miszalski et al. 2006). FLAMES fields were configured using \textsc{fposs}, an ESO adapted version of \textsc{configure}, that we customised to show angular deviation limits for mini-IFU fibres to assist in field centre selection. The 580V and 385R volume-phase holographic AAOmega gratings provided a wavelength coverage of 3700--8850\AA\ at resolutions of 3.5\AA\ and 5.3\AA\ (FWHM), respectively. Miszalski et al. (2011a) gave details of the six grating settings used for the FLAMES observations, made at an average resolving power of $R\sim12000$, in addition to an account of the full data reduction procedure which includes relative flux calibration that we also applied to field F54. AAOmega data were reduced using the \textsc{2dfdr} program (see e.g. Croom et al. 2005). A typical spectrophotometric response function included with \textsc{2dfdr} for each spectrograph arm was divided through all data to remove the large-scale instrumental signature. This is by no means an accurate flux calibration since the inherent response of each fibre is different to the next and highly variable. The AAOmega spectra are therefore only useful for classification purposes and line ratios over small wavelength ranges. Some targets required manual tweaking of their spectra to resubtract background sky spectra minimising the sky residuals, resplice blue and red arm spectra, interpolate over cosmetic defects (particularly in the blue arm) and to stack multiple exposures. If more than one spectrum is available, then we have chosen to present only a representative spectrum (typically the average of multiple exposures), unless specified otherwise. 

\subsection{Additional spectroscopy}
Parker et al. (2006) published online spectroscopy of PPA1752-3542 and PPA1807-3158. An additional two objects were observed separate to the multi-object spectroscopy campaigns. We selected K2-17 (PN G336.8$-$07.2, Kohoutek 1977) for study because we identified an H$\alpha$ excess from the central star in SHS data, while Hen2-375 (PN G337.3$-$18.3, Henize 1967) appeared to be a good symbiotic star candidate with a red excess in the Two Microm All Sky Survey or 2MASS ($J-K_s=2.5$ mag, Skrutskie et al. 2006). We observed K2-17 with the Southern African Large Telescope (SALT; Buckley, Swart \& Meiring 2006; O'Donoghue et al. 2006) under programme 2012-1-RSA-009 and Hen2-375 with the SAAO 1.9-m telescope. Table \ref{tab:obs} summarises the observations taken with the Robert Stobie Spectrograph (RSS; Burgh et al. 2003; Kobulnicky et al. 2003) on SALT and the grating spectrograph (SpCCD) on the 1.9-m telescope. Basic pipeline calibrations were applied to the RSS data with the PySALT\footnote{http://pysalt.salt.ac.za} package (Crawford et al. 2010) and a relative flux calibration was made with the spectrophotometric standard star LTT7987 (Hamuy et al. 1994). The separate red and blue spectra of the central star of K2-17 were extracted with the \textsc{apall} \textsc{iraf} task and joined by matching the stellar continuum. In addition, the brightest part of the K2-17 nebula on the slit was extracted using a 20\arcsec\ wide aperture, equal in both red and blue spectra, with both spectra joined using the same scale factor for the central star spectrum. The Hen2-375 spectra were reduced in the usual fashion with \textsc{IRAF} and flux calibrated using the central star of NGC7293 (Oke 1990). 

\begin{table*}
   \centering
   \caption{Log of SALT RSS and SAAO 1.9-m SpCCD observations.}
   \label{tab:obs}
   \begin{tabular}{llllrllllr}
      \hline
      Object & Spectrograph &  Date       & Grating & Exptime & Slit Width & $\lambda$ & $\Delta\lambda$ & Dispersion & PA\\
      &    &  (DD/MM/YY)         &  &     (s)    & ('') & (\AA)   & (FWHM, \AA)     &   (\AA\ pix$^{-1}$) & ($^\circ$)\\
      \hline
      K2-17 & RSS &  25/05/12 & PG1300 & 2100 & 1.5 & 4275--6365 & 3.9 & 0.67 &0\\ 
      K2-17 & RSS & 20/06/12 & PG900 & 1350 & 1.5 & 5880--8880 & 5.5 & 0.96&0\\ 
      Hen2-375 & SpCCD & 28/06/12 & 6 & 3$\times$90, 1200 & 1.5 & 4055--5950 & 2.8 & 1.10 & 90 \\ 
      Hen2-375 & SpCCD & 28/06/12 & 6 & 60, 1800 & 1.5 & 5710--7560 & 2.5 & 1.05 &90\\ 
      \hline
   \end{tabular}
\end{table*}

\subsection{OGLE and MACHO lightcurves}
Several fields towards the Galactic Bulge have been extensively monitored by the OGLE\footnote{http://ogle.astrouw.edu.pl} (see e.g. Udalski 2009) and MACHO (Alcock et al. 1997) projects at optical wavelengths. Such lightcurves are ideally suited to study symbiotic stars that demonstrate a rich variety of variability over timescales of several hundred days including pulsations from the red giant component, ellipsoidal variability or eclipses due to orbital motion and dust obscuration events (Miko{\l}ajewska 2001; Gromadzki et al. 2009). This rich legacy dataset has only recently been used to study symbiotic stars towards the Galactic Bulge (Miszalski et al. 2009a, the precursor to this work; Gromadzki et al. 2009; Lutz et al. 2010) and the Magellanic Clouds (Miszalski et al. 2011; Kato, Hachisu \& Miko{\l}ajewska 2013). In particular, they offer significant advantages in distinguishing symbiotic stars apart from PNe (Miszalski et al. 2009a, 2011). Archival photographic plates also offer substantial opportunities to obtain long baseline lightcurves (Munari, Jurdana-{\v S}epi{\'c} \& Moro 2001; Munari \& Jurdana-{\v S}epi{\'c} 2002; Jurdana-{\v S}epi{\'c} \& Munari 2010).

We retrieved MACHO lightcurves from the recently updated virtual observatory compatible website.\footnote{http://macho.anu.edu.au} The MACHO catalogues were overlaid on top of available imaging to select the lightcurve corresponding to each source (referenced by their MACHO FTS number). Instrumental MACHO magnitudes were transformed into $V$ and $R$ magnitudes using the transformations given in Lutz et al. (2010) after Bessell \& Germany (1999). 

The procedure was more involved to gather the OGLE photometry which can originate from one or more of the four separate phases of the project. Compared to later OGLE phases, the OGLE-I phase had a more limited survey footprint and sparser sampling (Udalski et al. 1992). For simplicity we include only the OGLE-I lightcurve of PHR1806-2652 (Pigulski et al. 2003), relying mostly on the much better sampled OGLE-II (Udalski, Kubiak \& Szyma{\'n}ski 1997; Szyma{\'n}ski 2005), OGLE-III (Udalski et al. 2002a, 2002b, 2008; Szyma{\'n}ski et al. 2011) and OGLE-IV (e.g. Poleski et al. 2011; Udalski et al. 2012) photometry. Photometry from the OGLE-IV phase included in this work is somewhat preliminary, coming from the first 2--3 years of operations, and comes without a finalised photometric scale tied to OGLE-II and OGLE-III photometry. Where possible we attempted to put the OGLE-IV photometry on the OGLE-II/III system based on bright, non-variable stars overlapping between OGLE-III and OGLE-IV fields. This did not always work successfully, most likely because of uncharacterized systematic effects from the OGLE-IV photometry. Colour-related effects may also have played a role where a large spread in stellar $V-I$ colours appears in observed fields, not to mention many objects in our sample are highly variable and very red ($V-I>2.0$ mag). In most cases a mean offset from the OGLE-III lightcurve was estimated and applied to the OGLE-IV lightcurve. This was more difficult to do in the case of lightcurves with large amplitude semi-regular (S-types) or Mira (D-types) pulsations where the mean level is difficult to judge.

\section{The Sample}
\label{sec:sample}
\subsection{Classification}
Table \ref{tab:new} gives the names and coordinates of 20 new and 15 possible symbiotic stars that were classified as S-type or D-type (column 3) following the
criteria in Belczy\'nski et al. (2000). These criteria include:
\begin{itemize}
   \item (a) Detection of the late-type giant (e.g. TiO, CaII, etc. absorption lines).
   \item (b) Detection of strong emission lines of HI and HeI, together with emission lines of higher ionisation potential than 35 eV (e.g. [OIII]). 
   \item (c) Detection of the Raman-scattered OVI emission complex at $\lambda\lambda$6825, 7082 (Schmid 1989).
      
\end{itemize}
Note that (c) is a sufficient but not necessary condition as only symbiotics with the hottest WDs will be capable of producing this feature. Furthermore, the OVI bands may be redshifted by a few Angstroms (Schmid 1989). Generally, the combination of (a) and (b) is sufficient to classify an object as a symbiotic star. A smaller class of D$'$-type symbiotics, characterised by G-type giants with clear G-bands, absent TiO bands and often carbon enrichment (M\"urset \& Schmid 1999; Schmid \& Nussbaumer 1993). Only one D$'$ symbiotic was found in our sample (Sect. \ref{sec:shwi5}).

\begin{table*}
\centering
\caption{New and possible symbiotic stars.}
\label{tab:new}
\begin{tabular}{lllllrrrl}
\hline
Name & Coords (J2000) & Type & Field                             & Fig.                       & H$\alpha$/H$\beta$ & $E(B-V)$ & $E(B-V)$ & Nebula                  \\ 
     &                &      &                                   &                            &                    & (CaseB)  & (NIR)   &  present      \\
\hline                                                                                                                            
000.49$-$01.45 & 17 52 31.2 $-$29 15 34 & S & S18                & \ref{fig:s1}                & 268.5             & 4.2    & 1.6--2.0 & -                       \\          
001.70$-$03.67 & 18 04 04.9 $-$29 18 46 & S & S20 S20a           & \ref{fig:s1}                & 6.2               & 0.7    & 0.6 & -                       \\  
002.86$-$01.88 & 17 59 34.8 $-$27 25 43 & S & S21 S21a           & \ref{fig:s1}                & -                 & -      & 1.5 & -                       \\         
003.46$-$01.92 & 18 01 06.3 $-$26 55 59 & S & S21 S21a           & \ref{fig:s2}                & 59.3              & 2.8    & 1.3 & -                       \\         
354.98$-$02.87 & 17 44 53.0 $-$34 42 40 & D & S3 S3a             & \ref{fig:d1}                & 11.2              & 1.3    & -   & -                       \\         
355.28$-$03.15 & 17 46 48.2 $-$34 36 03 & S & S3 S3a             & \ref{fig:s2}                & 37.3              & 2.4    & 0.9 & -                       \\         
355.39$-$02.63 & 17 44 55.7 $-$34 14 18 & S & S3 S3a             & \ref{fig:s2}                & 27.2              & 2.1    & 1.1 & -                       \\         
356.04$+$03.20 & 17 23 21.2 $-$30 31 35 & S & N7                 & \ref{fig:s3}                & $<$87.8           & $<$3.2 & 1.2 & -                       \\ 
357.32$+$01.97 & 17 31 23.2 $-$30 08 44 & S & N2                 & \ref{fig:s3}                & 51.0              & 2.7    & 1.8 & -                       \\         
357.98$+$01.57 & 17 34 35.5 $-$29 48 22 & S & N2                 & \ref{fig:s3}                & $<$98.9           & $<$3.3 & 1.2 & -                       \\ 
358.46$+$03.54 & 17 28 13.1 $-$28 19 38 & S & N5                 & \ref{fig:s4}                & 28.5              & 2.1    & 0.3 & -                       \\ 
359.76$+$01.15 & 17 40 34.6 $-$28 31 41 & S & N3                 & \ref{fig:s4}                & 183.0             & 3.8    & 2.5 & -                       \\         
H1-45 & 17 58 21.8 $-$28 14 52 & D & S20a                        & \ref{fig:d1}                & 6.1               & 0.7    & 2.2 & N                       \\ 
Hen2-375 & 18 18 09.1 $-$57 11 14 & D & -                        & \ref{fig:d1}                & 4.1               & 0.3    & -   & Y                       \\ 
JaSt2-6 & 17 50 01.9 $-$29 33 25 & D & -                         & -                           &  -                & -      & 6.0 & Y                       \\         
JaSt79 & 17 51 53.5 $-$29 30 53 & D & S18                        & \ref{fig:d2}                & 43.6              & 2.5    & 3.0 & -                       \\         
K5-33 & 17 44 29.9 $-$27 20 40 & D & S3 S3a                      & \ref{fig:d2}                & 42.7              & 2.5    & -   & -                       \\         
NSV22840 & 17 35 58.5 $-$28 49 54 & S & N3                       & \ref{fig:s4}                & 45.6              & 2.6    & 2.0 & -                       \\         
PHR1757$-$2718 & 17 57 32.5 $-$27 18 25 & S & S21                & \ref{fig:s5}                & 55.5              & 2.7    & 1.8 & -                       \\         
ShWi5 & 18 03 53.7 $-$29 51 22 & D$'$& S22                         & \ref{fig:dd1}                & 10.6              & 1.2    & -   & -                       \\ 
\hline                                                                                                                                                     
001.33$+$01.07 & 17 44 37.8 $-$27 14 17 & S?& N4                 & \ref{fig:ps1}               & 216.3             & 4.0    & 2.2 & -                       \\          
001.37$-$01.15 & 17 53 21.2 $-$28 20 47 & D? & S18               & \ref{fig:pd1}               & 362.1             & 4.5    & -   & -                       \\          
001.71$+$01.14 & 17 45 13.6 $-$26 52 42 & S?& N4                 & \ref{fig:ps1}               & -                 & -      & 1.2 & -                       \\          
001.97$+$02.41 & 17 41 02.1 $-$25 59 34 & S?& N4                 & \ref{fig:ps1}               & 77.6              & 3.0    & 2.1 & -                       \\         
355.12$+$03.82 & 17 18 32.2 $-$30 55 44 & S? & N7                & \ref{fig:ps2}               & 14.86             & 1.5    & -   & -                       \\         
357.12$+$01.66 & 17 32 04.8 $-$30 28 55 & D? & N2                & \ref{fig:pd1}               & 125.3             & 3.5    & -   & -                     \\
Al2-B & 17 27 47.0 $-$28 11 03 & D? & N5                         & \ref{fig:pd1}               & 13.8:             & 1.5    & -   & -                       \\ 
Al2-G & 17 32 22.6 $-$28 14 29 & D? & N5                         & \ref{fig:pd2}               & 22.6              & 1.9    & -   & -                       \\         
M2-24 & 18 02 02.9 $-$34 27 47 & D? & -                          &    $(a)$                    & 6.2               & 0.7    & -   & Y                       \\ 
PHR1751$-$3349 & 17 51 15.0 $-$33 49 11 & D? & S3 S3a            & \ref{fig:pd2}               & 28.7              & 2.1    & -   & Y                      \\
PHR1803$-$2746 & 18 03 05.0 $-$27 46 44 & D? & S20 S20a S21 S21a & \ref{fig:pd2}               & 32.4              & 2.2    & -   & -                       \\         
PPA1746$-$3454 & 17 46 51.4 $-$34 54 05 & D? & S3 S3a            & \ref{fig:pd3}               & 25.4              & 2.0    & -   & -                       \\         
PPA1752$-$3542 & 17 52 05.9 $-$35 42 06 & S? & -                 & $(b)$                       & -                 & -      & -   & -                       \\         
PPA1807$-$3158 & 18 07 19.1 $-$31 58 08 & D? & -                 &  $(b)$                      & -                 & -      & -   & -                       \\         
Th3-9 & 17 23 59.2 $-$31 01 51 & S? & N7                         & \ref{fig:ps2}               & -                 & -      & -   & -                       \\ 
\hline
\end{tabular}
\begin{flushleft}
$(a)$ Zhang \& Liu (2003); $(b)$ Parker et al. (2006)
\end{flushleft}
\end{table*}

Columns 4 and 5 of Table \ref{tab:new} are the AAOmega (Tab. \ref{tab:aaomega}) or FLAMES (Tab. \ref{tab:flames}) field names used to observe the source, and the Figure numbers in which the spectra appear in this paper. In column 6 the observed H$\alpha$/H$\beta$ ratios are given. In cases where an upper limit is present the H$\beta$ emission line was coincident with a bad CCD column which reduced the measured H$\beta$ flux. The $E(B-V)$ reddening derived from these ratios under Case B conditions (Brocklehurst 1971) is given in column 7, but these values come with two strong caveats: (a) The accuracy is limited to 0.3-0.5 dex at best by the very poor AAOmega spectrophotometry, and (b) Case B conditions are not necessarily suitable for the very high electron densities encountered in symbiotic stars (e.g. Proga, Miko{\l}ajewska \& Kenyon 1994; Gutierrez-Moreno et al. 1995; Luna \& Costa 2005). Two exceptions are K2-17 and Hen2-375 with good quality longslit spectrophotometry. Column 8 gives an estimate of the $E(B-V)$ reddening based on the intrinsic NIR colours of the red giant, as determined from the observed spectral type (see Sect. \ref{sec:sptype}).

Finally, the last column of Tab. \ref{tab:new} specifies whether the object is surrounded by a resolved nebula. Nebulae are rather common amongst symbiotic stars and are thought to originate from an interaction between the red giant wind and the photoionising wind of the WD (see e.g. Corradi 1995; Corradi et al. 1999; Corradi 2003). Many of these resolved nebulae were identified from Figures \ref{fig:c1}, \ref{fig:c2}, \ref{fig:c3}, \ref{fig:c4} and \ref{fig:c5} that display colour-composite finder charts of each target in our sample made from SHS/SSS and 2MASS imaging. In some cases nebulae were also visible on their OGLE finder charts (Figures \ref{fig:ogleim1} and \ref{fig:ogleim2}). Additional imaging sources are specified in the detailed sections on the relevant objects.

Table \ref{tab:other} contains the same information as Tab. \ref{tab:new} for the non-symbiotic sample except for the reddening derived from the spectral type which is no longer defined. Two objects, M2-29 and PHR1805-2659, were largely dealt with in Miszalski et al. (2009a) and Miszalski et al. (2011a), but Section \ref{sec:egb6} also discusses M2-29 in relation to similar objects. Four objects in Tab. \ref{tab:other} are superpositions and are discussed separately in Appendix \ref{sec:superpos}. Section \ref{sec:other} discusses the remainder of the objects in Tab. \ref{tab:other}.

\begin{table*}
\centering
\caption{Other objects.}
\label{tab:other}
\begin{tabular}{lllllrrl}
\hline
Name & Coords (J2000) & Type & Field                                 &   Fig.            & H$\alpha$/H$\beta$   & $E(B-V)$     & Nebula \\
     &                &      &                                       &                   &                      & (CaseB)      & present\\
\hline                                                                                   
003.16$-$02.31 & 18 01 56.2 $-$27 22 55 & Nova & S21 S21a         &  \ref{fig:other1}            & -                    & -              & -                     \\
359.88$-$03.58 & 17 59 35.6 $-$30 51 32 & WN6 & S22                  &  \ref{fig:other1}            & -                    & -              & -                     \\
H2-32 & 17 56 24.2 $-$29 38 07 & Be? & S18                           &  \ref{fig:other1}            & 19.9                 & 1.8            & -                     \\
K2-17 & 17 09 35.8 $-$52 13 02 & HDC PN & -                  &  \ref{fig:egb61}            &  4.1                 & 0.3            & Y \\ 
M1-44 & 18 16 17.3 $-$27 04 33 & PN (superposition) & -              &  -                & -                    & -              & Y                     \\
M2-11 & 17 20 33.4 $-$29 00 38 & HDC PN & F54                &  \ref{fig:egb61}             & -                    & -              & Y           \\
M2-29 & 18 06 40.8 $-$26 54 56 & HDC PN & F157 S21a      &  $(c)$        & 5.8                  & 0.7            & Y \\ 
M3-8 & 17 24 52.1 $-$28 05 55 & PN (superposition) & N5              &  $(b)$       & -                    & -              &            -          \\
M3-38 & 17 21 04.6 $-$29 02 59 & HDC PN & F54               &  \ref{fig:egb62}             & -                    & -              & Y           \\
M4-4 & 17 28 50.3 $-$30 07 45 & PN (superposition) & N2              &  \ref{fig:superspec}        & -                    & -              & Y  \\
MPA1746-3412 & 17 46 18.5 $-$34 12 37 & [WC10-11] & S3 S3a        &  \ref{fig:other2}            & 12.9                 & 1.4              & -                       \\         
PHR1803$-$2748 & 18 03 31.2 $-$27 48 27 & B[e] & S20 S20a S21 S21a   &  \ref{fig:other2}            & 30.0                 & 2.2              & Y         \\
PHR1805$-$2659 & 18 05 43.5 $-$26 59 46 & dMe & S21 S21a         &  $(a)$                  & 4.4       & 0.4       &             -         \\
PHR1806$-$2652 & 18 06 56.0 $-$26 52 54 & HDC PN & F157 S21a &  \ref{fig:egb62}             & 8.7       & 1.5     & Y        \\ 
PPA1758$-$2628 & 17 58 37.0 $-$26 28 47 & UCHII & S21 S21a           &  \ref{fig:other2}            & 334.0                & 4.4            &   -                   \\
PPA1808$-$2700 & 18 08 01.4 $-$27 00 16 & PN (superposition) & F157  &  \ref{fig:superspec}         & -                    & -              & Y                 \\
Sa3-104 & 17 58 25.9 $-$29 20 48 & Be? & S18                         &  \ref{fig:other3}            & 16.9                 & 1.6            &    -                  \\
Th3-28 & 17 30 56.8 $-$26 59 10 & WN6 & N5                           &  \ref{fig:other3}            & -                    & -              &     -                 \\
V4579 Sgr & 18 03 37.9 $-$28 00 07 & Nova & S20 S20a S21a         &  \ref{fig:other3}            & -                    & -              &      -                \\
\hline                                                                
\end{tabular}
\begin{flushleft}
   $(a)$ Miszalski et al. (2009a). $(b)$ Miszalski et al. (2009b). $(c)$ Miszalski et al. (2011a).\\
\end{flushleft}
\end{table*}

The NIR colour-colour plane has been extensively applied in an attempt to separate symbiotic stars from other samples, especially PNe (e.g. Schmeja \& Kimeswenger 2001;  Miko{\l}ajewska 2004; Corradi et al. 2008). As our sample was H$\alpha$ selected, the NIR properties were only considered at a later stage and do not bias our survey. The classification of our sample was based mainly on the assessment of spectroscopic features and $I$-band variability (Sect. \ref{sec:lctext}), with NIR properties only later considered to characterise the symbiotic type. Tables \ref{tab:newmags} and \ref{tab:othermags} contain near- and mid-infrared magnitudes of our sample collated from the Two Micron All Sky Survey (2MASS, Skrutskie et al. 2006) and the Galactic Legacy Infrared Mid-Plane Survey Extraordinaire (GLIMPSE; see Benjamin et al. 2003, Churchwell et al. 2009) VizieR catalogue (II/293/glimpse), respectively. Inspection of the GLIMPSE images themselves showed the sample to consist entirely of point sources meaning no recourse to aperture photometry was required to determine accurate magnitudes.

\subsection{Lightcurves}
\label{sec:lctext}
Tables \ref{tab:lcnew} and \ref{tab:lcother} summarise the available lightcurves for the symbiotic (Tab. \ref{tab:lcnew}) and non-symbiotic (Tab. \ref{tab:lcother}) samples. Column 3 gives the Figure numbers in which the lightcurves appear in this paper.\footnote{Some of these lightcurves also appear in the long period variable catalogue of Soszy{\'n}ski et al. (2013).} The MACHO lightcurves are grouped together, whereas the OGLE lightcurves are mostly displayed according to what OGLE phases were available as specified in column 4. Column 5 uniquely identifies each MACHO lightcurve in the online archive\footnote{http://macho.anu.edu.au}. Finder charts for OGLE sources may be found in Figures \ref{fig:ogleim1} and \ref{fig:ogleim2}. These may be compared against the images in Figures \ref{fig:c1}, \ref{fig:c2}, \ref{fig:c3}, \ref{fig:c4} and \ref{fig:c5}, but note the very different resolutions (0.26\arcsec\ pixel$^{-1}$ of OGLE versus $\sim$1\arcsec\ pixel$^{-1}$ of SHS/SSS). Periodicities were searched for using the phase dispersion minimisation (PDM, Stellingwerf 1978) \textsc{iraf} task and the \textsc{period04}\footnote{http://www.univie.ac.at/tops/Period04} fourier analysis package (e.g. Lenz \& Breger 2004). In some S-type systems the periodic signature was found by inspection. In periodic systems we determined ephemerides by fitting several measured times of minima with a linear ephemeris (see columns 6 and 7 of Tab. \ref{tab:lcnew}). No strictly periodic sources were found amongst non-symbiotic sources. The average time span varies from 1.5--2.4 years (OGLE-IV only), 7.6--10.4 years (OGLE-III and OGLE-IV), 14.3--14.6 yrs (OGLE-II to OGLE-IV) and 18.6 years (OGLE-I to OGLE-IV). The last column in Tables \ref{tab:lcnew} and \ref{tab:lcother} describes the variability seen in each lightcurve.

\begin{table*}
\centering
\caption{Lightcurve information for new and possible symbiotic stars.} 
\label{tab:lcnew}
\begin{tabular}{llllllll}
\hline
Name & Type & Fig. & OGLE & MACHO FTS & $I_\mathrm{min}$ & Period & Description\\
     &      &      &      &           &  (JD-2450000) & (days) & \\
\hline
000.49$-$01.45 & S   & \ref{fig:oglethreefour1}                        & III IV & -                & 3736.4 & 1014.5  & semi-regular pulsations $+$ orbital\\
001.70$-$03.67 & S   & \ref{fig:macho}              & - & 114.20492.26          & - & -  & flat?\\ 
002.86$-$01.88 & S   & \ref{fig:oglethreefour1}                        & III IV & -                & 4442.6 & 1068.4 & semi-regular pulsations $+$ orbital\\
003.46$-$01.92 & S   & \ref{fig:oglethreefour1}                       & III IV & -                & 4560.2 & 801.0  & semi-regular pulsations $+$ orbital\\
354.98$-$02.87 & D   & \ref{fig:oglefour1}                       & IV & -                    & - & -  & relatively flat\\
355.28$-$03.15 & S   & \ref{fig:ogletwothreefour}        & II III IV & -             &- & -  & semi-regular pulsations\\
355.39$-$02.63 & S   & \ref{fig:oglethreefour1}                       & III IV & -                & 4431.4 & 854.5  & semi-regular pulsations $+$ orbital\\
357.98$+$01.57 & S   & \ref{fig:oglefour1}                       & IV & -                    & - & -  & semi-regular pulsations\\
358.46$+$03.54 & S   & \ref{fig:oglefour1}                       & IV & -                    & - & -  & semi-regular pulsations\\
H1-45          & D   & \ref{fig:oglethreefour2}, \ref{fig:macho}                        & III & 401.48290.73  & 3551.0    & 408.6  & carbon Mira pulsations\\
JaSt2-6        & D   & \ref{fig:ogletwothreefour}                               & II III IV& -              & 2760.9 & 605.4  & Mira pulsations\\
JaSt79         & D   & \ref{fig:oglethreefour2}                        & III IV & -                & 3835.4 & 424.8  & Mira pulsations\\
K5-33          & D   & \ref{fig:oglefour2}                       & IV & -                    & - & -  &  flat\\
NSV22840       & S   & \ref{fig:oglefour2}                       & IV & -                    & - & -  & semi-regular pulsations\\
PHR1757$-$2718 & S   & \ref{fig:oglefour2}                       & IV & -                    & 5634.9 & 585.0  & semi-regular pulsations $+$ orbital?\\
ShWi5          & D$'$  & \ref{fig:ogletwothreefour}, \ref{fig:macho}                        & I II III IV& 119.20354.128 & - & -  & low-amplitude orbital?\\
\hline                                        
001.33$+$01.07 & S?  & \ref{fig:oglefour1}                        & IV &  -                  &- & -  & semi-regular pulsations\\
001.37$-$01.15 & D?  & \ref{fig:oglefour1}                        & IV &  -                  &- & -  & relatively flat\\
001.71$+$01.14 & S?  & \ref{fig:oglefour1}                       & IV &  -                  &- & -  & semi-regular pulsations\\
Al2-B          & D?  & \ref{fig:oglefour2}                       & IV &  -                  &- & -  & relatively flat\\
Al2-G          & D?  & \ref{fig:oglefour2}                       & IV &  -                  &- & -  & slow variations\\
M2-24          & D?  & \ref{fig:oglethreefour3}                        & III & -                  &- & -  & slow variations\\
PHR1751$-$3349 & D?  & \ref{fig:oglethreefour3}                        & III IV & -                & - & - & slow large amplitude variations\\
PHR1803$-$2746 & D?  & \ref{fig:oglethreefour3}                        & III IV & -               &- & -  & relatively flat\\
PPA1746$-$3454 & D?  & \ref{fig:ogletwothreefour}                               & I II III IV& -           &- & -  & slow large amplitude variations\\
PPA1752$-$3542 & S?  & \ref{fig:oglethreefour3}                        & III & -                  &- & -  & semi-regular pulsations $+$ orbital?\\
PPA1807$-$3158 & D?  & \ref{fig:oglethreefour4}                      & III IV & -               & - & -  & slow variations\\
\hline
\end{tabular}
\end{table*}

\begin{table*}
\centering
\caption{Lightcurve information for other objects.}
\label{tab:lcother}
\begin{tabular}{llllllll}
\hline
Name & Type & Fig. & OGLE & MACHO FTS  & $I_\mathrm{min}$ & Period & Description \\ 
     &      &      &      &            &  (JD-2450000) & (days) & \\
\hline
003.16$-$02.31     & Nova         & \ref{fig:oglethreefour1}, \ref{fig:macho}                                      & III & 176.19611.10& - & -  & declining\\
359.88$-$03.58     & WN6             & \ref{fig:oglethreefour2}                                            & III IV & -& - & -  & dust obscuration event\\
H2-32              & Be?             & \ref{fig:H2-32}, \ref{fig:oglethreefour2}, \ref{fig:macho}   & III IV & 403.47850.5641& - & -  & slow longterm $+$ short term\\
M2-29          & HDC PN  & $(b)$                                                             & I II III & 101.21568.184 &- & -  & dust obscuration event\\
MPA1746-3412     & [WC10-11]       & \ref{fig:oglethreefour2}                                            & III IV & -& - & -  & dust obscuration events\\
PHR1803$-$2748     & B[e]            & \ref{fig:oglethreefour3}, \ref{fig:macho}                              & III IV & 104.20254.1855& - & -  & slow variations\\
PHR1805$-$2659 & dMe             & $(a)$,\ref{fig:macho}                              & III & 101.21176.282& - & -  & flare outbursts\\
PHR1806$-$2652     & HDC PN  & \ref{fig:phr1806lc}                       & I II III & -& - & -  & dust obscuration events\\
Sa3-104            & Be?             & \ref{fig:ogletwothreefour}                             & II III IV& -& - & -  & flat (spurious)\\
V4579 Sgr          & Nova         & \ref{fig:oglethreefour4}, \ref{fig:macho}                            & III IV & 104.20251.191& - & -  & declining\\
\hline
\end{tabular}
\begin{flushleft}
   $(a)$ Miszalski et al. (2009a). $(b)$ Miszalski et al. (2011a).\\
\end{flushleft}
\end{table*}

\subsection{Cool component spectral types and reddening}
\label{sec:sptype}
Table \ref{tab:spnew} lists the spectral types of red giant components that were determined for a subset of our sample. We used the indices of Kenyon \& Fernandez-Castro (1987) which measure the depth of the VO band at $\lambda$7865 ([VO]) and TiO bands at $\lambda$6180 \AA\ ([TiO]$_1$) and $\lambda$7100 \AA\ ([TiO]$_2$). Also given is the [NaI] index, which although more sensitive to luminosity, is useful when combined with the other indices. These indices are based on measurements over fairly small wavelength ranges of $\sim$400-650 \AA\ and are suitable for our AAOmega spectra. Apart from H1-45 and ShWi5, only sources with TiO and VO bands strong enough for a reliable classification were included in Tab. \ref{tab:spnew}. It is well known that these indices give earlier spectral types with decreasing wavelength in symbiotic stars (Kenyon \& Fernandez-Castro 1987; M\"urset \& Schmid 1999). We therefore give classifications for each index, noting that TiO and VO are best suited to types later than K5 and M2, respectively. The last column in Tab. \ref{tab:spnew} gives our best estimate of the spectral type which we adopt for the rest of the paper. Additionally, in the case of 355.39-02.63, 357.98+01.57, 358.46+03.54, 359.76+01.15, and PHR1757-2718, the presence of fairly strong TiO 8432 in their NIR spectra indicates the presence of an M4-5 giant or later, and such a late spectral type was adopted in the reddening estimates below. In D-types the TiO bands are usually very faint or invisible, leading to very unreliable classifications. In the case of the carbon Mira of H1-45 we used the Richer (1971) scheme to determine a spectral type of C4--7.

\begin{table*}
\centering
\caption{Spectral classifications of red giant components of new and possible symbiotic stars.}
\label{tab:spnew}
\begin{tabular}{llrrrrr}
\hline
Name           &Type& [TiO]$_1$ & [TiO]$_2$ & [VO] & [NaI]    & Adopted     \\
\hline
000.49$-$01.45 &  S &  M0       & M3.5      & M6   &  $\ga$M6  & M4--6           \\
001.70$-$03.67 &  S &  K5       & M0        & M2.5 &  K/M      & K5--M1           \\                
002.86$-$01.88 &  S &  K5       & M1        & M5.5 &  M4--5     & M4--5            \\
003.46$-$01.92 &  S &  K5.5     & M0.5      & M4.5 &  $\ga$M5  & M4--5            \\
355.28$-$03.15 & S &  M6 &  M6        & M7   &  $\ga$M6  & M6              \\
355.39$-$02.63 &  S &  K5.5     & M2        & M5   &  $\ga$M5  & M2--5           \\
356.04$+$03.20 &  S &  M4       & M5.5      & M6   &  $\ga$M6  & M6              \\                  
357.32$+$01.97 &  S &  M0.5     & M0.5      & M2:  &  K5--M3   & M0--1            \\                  
357.98$+$01.57 &  S &  M0       & M1.5      & M5   &  $\ga$M5  & M2--6            \\                  
358.46$+$03.54 &  S &  M0.5     & M3.5      & M6   &  $\ga$M5  & M4--6           \\                  
359.76$+$01.15 &  S &  K5       & K5        & M3   &  M4--5    & M3              \\                  
H1-45          &  D &  ND      & ND       & ND  &  ND      & C4--7\\                  
NSV22840       &  S &  M0       & M0        & M3.5 &  M0--3  & M0--3            \\                  
PHR1757$-$2718 &  S &  M0       & M1        & M4   &  $\ga$M5  & M1--4           \\
ShWi5          &  D$'$&  ND      & ND       & ND  &  ND      & CN              \\                  
\hline
001.33$+$01.07 & S?&  M4 &  M4        & M6   &  M4--6    & M4--6           \\
001.71$+$01.14 & S?&  ND &  M6        & M6   &  $\ga$M6  & M6              \\
001.97$+$02.41 & S?&  M1 &  M1        & M2:  &  K/M      & M1              \\
\hline
\end{tabular}
\begin{flushleft}
ND: Not detected.
\end{flushleft}
\end{table*}

We can now estimate the reddening to each object in Tab. \ref{tab:spnew} from 2MASS colours (Tables \ref{tab:newmags} and \ref{tab:othermags}) by comparing the observed $J-K_s$ colour against the intrinsic value. In the case of S-type systems, their intrinsic colours were taken from those of M giants in the Galactic Bulge (Feast, Whitelock \& Carter 1990). For the D-types with known Mira pulsation periods, their intrinsic colours were estimated using the period-colour relations from Whitelock et al. (2000, 2006) for O-rich and C-rich Miras, respectively. In the case of JaSt 2-6 and JaSt 79 the reddening estimates assume an O-rich Mira. The resultant reddenings are included in column 8 of Tab. \ref{tab:new}. The few Mira variables included in this sub-sample show reddenings derived from this method that are higher than that derived from the Balmer decrement. This is a well known fact since in D-type symbiotics, the WD companion and the surrounding ionized region is located outside the Mira's dust shell. This effect is best studied in RX Pup (Miko{\l}ajewska et al. 1999) where
the reddening towards the Mira is highly variable due to changes in dust shell opacity due to dust obscuration events, whereas the reddening towards the hot ionized region is constant and consistent with the measured interstellar reddening. See also Tab. 3 of Miko{\l}ajewska et al. (1997) and Sect. 3 of Miko{\l}ajewska (1999).

\subsection{Spectroscopic features and temperature estimates of the hot component}
Tables \ref{tab:spinfonew} and \ref{tab:spinfoother} summarise the main observed features in the continuum and the main emission lines present. For Tab. \ref{tab:spinfonew} we have included simple estimates of the hot component temperature $T_h$ based on the highest ionisation potential emission lines visible in the spectrum according to M\"urset \& Nussbaumer (1994). This technique gives a lower limit to the true temperature since we may not see all high ionisation lines. Where possible, we also give the line ratios of [O~III]$\lambda$4363/H$\gamma$, [O~III] $\lambda$5007/H$\beta$  and He~II $\lambda$4686/H$\beta$ for reference in Tables \ref{tab:newspec} and \ref{tab:otherspec}. Such ratios are useful in diagnostic diagrams (e.g. Gutierrez-Moreno et al. 1995). No reddening correction was applied to the ratios. 

\begin{table*}
\centering
\caption{Description of the continuum, including absorption line features, estimates of the hot component temperatures $T_h$ and emission line features present in new and possible symbiotic stars.}
\label{tab:spinfonew}
\begin{tabular}{lllrl}
\hline
Name & Type & Continuum/Absorption features & $T_h$ & Main emission features                                                                                                       \\
     &      &                      & (kK)  &                                                                                                                           \\
\hline                                                                                                                                                           
000.49$-$01.45 & S & TiO                                       & 100   & HI, HeI, [FeVII], OI, CaII triplet                                                     \\
001.70$-$03.67 & S & TiO, NaI D, HeI (P Cygni), CaII K         &  55   & HI, HeI, HeII, [OIII], [OII], OI, FeII, [FeII]                           \\
002.86$-$01.88 & S & TiO, NaI D                                &  114  & HI, HeI, OVI                                                                    \\
003.46$-$01.92 & S & TiO, NaI D                                &  $>$55& HI, HeI, HeII, [OIII], OI                                                              \\
354.98$-$02.87 & D & none                                      &  114  & HI, HeI, HeII, [OIII], [NeIII]                               \\
355.28$-$03.15 & S & TiO, NaI D                                &  35   & HI, HeI, [OIII]                    \\
355.39$-$02.63 & S & TiO                                       &  114  & HI, HeI, OVI, [FeVII], OI, CaII triplet        \\
356.04$+$03.20 & S & TiO, NaI D                                &  114  & HI, HeI, OVI, HeII, [FeVII]                                                \\
357.32$+$01.97 & S & TiO, CaII triplet, NaI D                  &  $>$55& HI, HeI, HeII, OI                                                                      \\
357.98$+$01.57 & S & TiO                                       &  55   & HI, HeI, HeII, [OIII], OI                                                              \\
358.46$+$03.54 & S & TiO, NaI D                                &   114 & HI, HeI, HeII, [OIII], OVI                                                 \\
359.76$+$01.15 & S & TiO, NaI D                                &   35  & HI, HeI, [OIII], OI, CaII triplet                                                      \\
H1-45          & D & CN, BaII, NaI D                           &  100  & HI, HeII, [OIII], [NeIII]                                                              \\
Hen2-375       & D & none                                      & 114   & HI, HeI, HeII, [OIII], [FeVII], OVI                                         \\
JaSt2-6        & D & -                                         &  -    & -                                                                                      \\
JaSt79         & D & TiO                                       &  100  & HI, HeI, HeII, [FeVII], [ArV], [OI], [OII], [OIII], [ArIII]                            \\
K5-33          & D & none                                      &  114  & HI, HeI, HeII, OVI, [FeVII], [OIII], [NeIII], [ArIII], [ArIV], [ArV]                   \\
NSV22840       & S & TiO, NaI D                                & $>$55 & HI, HeI, HeII, [OIII], OI                                                              \\
PHR1757$-$2718 & S & TiO                                       &  114  & HI, HeI, HeII, [FeVII], OVI, OI, CaII triplet                                         \\
ShWi5          & D$'$&                CN, CH, NaI D, BaII 6495   &  100  & HI, HeI, HeII, [FeVII], [OIII], [NeIII], [ArIII], [ArIV], [ArV], [NII]                 \\
\hline                                                          
001.33$+$01.07 & S?& TiO, NaI D                                &  25   & HI, HeI, OI, CaII triplet                               \\
001.37$-$01.15 & D?& red                                       &  25   & HI, HeI, [NII], OI, [OI], [OII], CaII triplet, [SIII]   \\
001.71$+$01.14 & S?& TiO                                       &  25   & HI, HeI, CaII triplet                            \\
001.97$+$02.41 & S?& TiO, CaII triplet, NaI D                  &  25   & HI, HeI                                                 \\
355.12$+$03.82 & S?& TiO, NaI D                                &  25   & HI, HeI?, [OI]                                        \\
357.12$+$01.66 & D? & NaI D                                    &  14   & OI, HI                                               \\
Al2-B          & D?& $(a)$                                     &  41   & HI, HeI, HeII, [OIII], [NeIII], [ArIII]                 \\
Al2-G          & D?& red                                       &  60   & HI, HeI, HeII, [OIII], [ArV], [NII], [NeIII], [ArIII]   \\
M2-24          & D?& none                                      &  46   & HI, HeI, [NII], [OIII], [NeIII], [ArIV]                   \\
PHR1751$-$3349 & D?& weak                                      &  25   & HI, [NII], OI, [OI], [OII], HeI, [SII]                                     \\
PHR1803$-$2746 & D?             & $(b)$                        &  41   & HI, HeI, [OIII], [NeIII], [ArIII], [OII], [NII], [OI]   \\
PPA1746$-$3454 & D?& weak                                      &  35   & HI, [OIII], [NII], [SII], HeI, OI              \\
PPA1752$-$3542 & S?& -                                         &  35   & HI,[OIII]                                                \\
PPA1807$-$3158 & D?& -                                         &   -   & HI,[NII]                                                 \\
Th3-9          & S?& red, CaII triplet, NaI D                  &  25   & HI, HeI                                          \\
\hline
\end{tabular}
\begin{flushleft}                                                                                         
   $(a)$ Continuum is not real and comes from neighbouring fibre.\\
   $(b)$ Continuum is not real and comes from nearby star. \\
\end{flushleft}                                                                                           
\end{table*}

\begin{table*}                                                                                            
\centering
\caption{Description of the continuum, including absorption line features, and emission line features present in other objects.}
\label{tab:spinfoother}
\begin{tabular}{llll}
\hline
Name &  Type &  Continuum/Absorption features & Main emission features                                                                                                          \\
\hline
003.16$-$02.31     & Nova        & NaI D                                            & HeII, NIII, [FeVII], CIV                                           \\
359.88$-$03.58     & WN6            & NaI D                                            & HeII, NIII, CIV                                                            \\
H2-32              & Be?            & NaI D, CaII H \& K & HI, OI, [SII], [OI], [OII], [OIII], FeII                                         \\
K2-17              & HDC PN & blue                                             & HI, HeI, HeII, [OIII], CaII triplet, [FeVII], CIV, [ArV]  \\
M1-44              & PN (superposition) & -                                                & -                                                                          \\
M2-11              & HDC PN & NaI D                                            & [FeVII]                                                                    \\
M2-29              & HDC PN & Of(H)                                            & HI, HeI, [OIII], [NeIII], [ArIII], [NII], [SII], [OII]                                                                 \\
M3-8               & PN (superposition)  & -                                                & -                                                                          \\ 
M3-38              & HDC PN & none                                             & [FeVII]                                                                    \\
M4-4               & PN (superposition)  & -                                                & -                                                                          \\
MPA1746-3412   & [WC10-11] & $(a)$                                       & HI, [NII], [SII], CII, CIII                                                       \\
PHR1803$-$2748     & B[e]           & red, NaI D                                       & HI, OI, [OII], [NiII], [OIII], [FeII], FeII, [NII]\\
PHR1805$-$2659     & dMe            & dM                                             & HI, CaII H\& K                                                             \\
PHR1806$-$2652     & HDC PN & weak                                             & [FeVII]                                                                    \\
PPA1758$-$2628     & UCHII          & red                                              & HI, HeI, [OIII], [OII], [OI], [NII]                                        \\
PPA1808$-$2700     & PN (superposition)  & -                                                & -                                                                          \\
Sa3-104            & Be?            & weak, CaII H\& K, NaI D, H$\beta$            & HI, HeI, [OIII], [OII], [ArIII], [NII], [SIII], [SII]                      \\
Th3-28             & WN6             & red                                              & HeII, NIII, CIV, NIV, HeI                                                  \\
V4579 Sgr          & Nova        & blue, CaII H\& K, NaI D                          & HeII, NIII, [FeVII], CIV                                           \\
\hline
\end{tabular}
\begin{flushleft}
   $(a)$ Continuum is not real and comes from nearby star. \\
\end{flushleft}
\end{table*}

\begin{table*}
\centering
\caption{Line ratios of new and possible symbiotic stars.}
\label{tab:newspec}
\begin{tabular}{lllll}
\hline
Name & Type & $\lambda$4363/H$\gamma$ & $\lambda$5007/H$\beta$ & $\lambda$4686/H$\beta$ \\
\hline
001.70$-$03.67 & S & 0.03 & 0.15 & 0.06    \\
003.46$-$01.92 & S & 0.67 & 1.22 & 0.18    \\
354.98$-$02.87 & D & 2.04 & 2.28 & 0.07    \\
355.28$-$03.15 & S & 0.70 & 1.50 & -       \\
355.39$-$02.63 & S & 0.11 & 0.45 & 0.50    \\
356.04$+$03.20 & S & - & - & 0.87          \\ 
357.32$+$01.97 & S & 0.24 & 0.14 & 0.26    \\
357.98$+$01.57 & S & 0.42 & - & $(a)$      \\
358.46$+$03.54 & S & 1.36 & 1.61 & 0.30    \\
359.76$+$01.15 & S & - & 0.40 & -          \\
H1-45 & D & 1.26 & 6.88 & 0.30             \\
Hen2-375 & D & 2.49 & 5.37 & 0.27          \\
JaSt79 & D & 1.38 & 11.15 & 0.53           \\
K5-33 & D & 2.11 & 8.35 & 0.27             \\
NSV22840 & S & 0.30 & 0.51 & 0.18          \\
PHR1757$-$2718 & S & - & - & 1.18          \\
ShWi5 & D$'$& 1.49 & 9.81 & 0.35            \\
\hline                                     
Al2-B & D? & 1.56 & 11.43 & 0.10           \\
Al2-G & D? & 2.70 & 8.27 & 0.20            \\
M2-24 & D? & 1.41 & 3.69 & 0.01            \\
PPA1746$-$3454 & D? & 0.55 & 0.93 & -      \\
PPA1752$-$3542 & S? & 1.80 & 2.70 & -      \\
PHR1803$-$2746 & D? & 3.02 & 8.07 & 0.04   \\
\hline
\end{tabular}
\begin{flushleft}
   $(a)$ He~II $\lambda$4686 detected, but H$\beta$ is affected by bad column on CCD.
\end{flushleft}
\end{table*}

\begin{table*}
\centering
\caption{Line ratios of other objects.}
\label{tab:otherspec}
\begin{tabular}{lllll}
\hline
Name & Type & $\lambda$4363/H$\gamma$ & $\lambda$5007/H$\beta$ & $\lambda$4686/H$\beta$ \\
\hline
H2-32 & Be? & - & 0.07 & -                           \\
K2-17 & HDC PN& 0.91 & 0.55 & 0.68                        \\
M2-11 & HDC PN& 0.69 & 13.94 & 0.46                       \\
M2-29 & HDC PN& 1.00 & 4.00 & -                           \\
M3-38 & HDC PN& 0.59 & 18.09 & 0.33                       \\
PHR1806$-$2652 & HDC PN & 0.10 & 4.05 & 0.36   \\
PPA1758$-$2628 & UCHII & - & 18.37 & -                 \\
Sa3-104 & Be? & 0.47 & 1.63 & -                      \\
\hline                                                    
\end{tabular}                                            
\end{table*}

\section{New S-type symbiotic stars}
\label{sec:stypes}
\begin{figure*}
\begin{center}
\includegraphics[angle=270,scale=0.36]{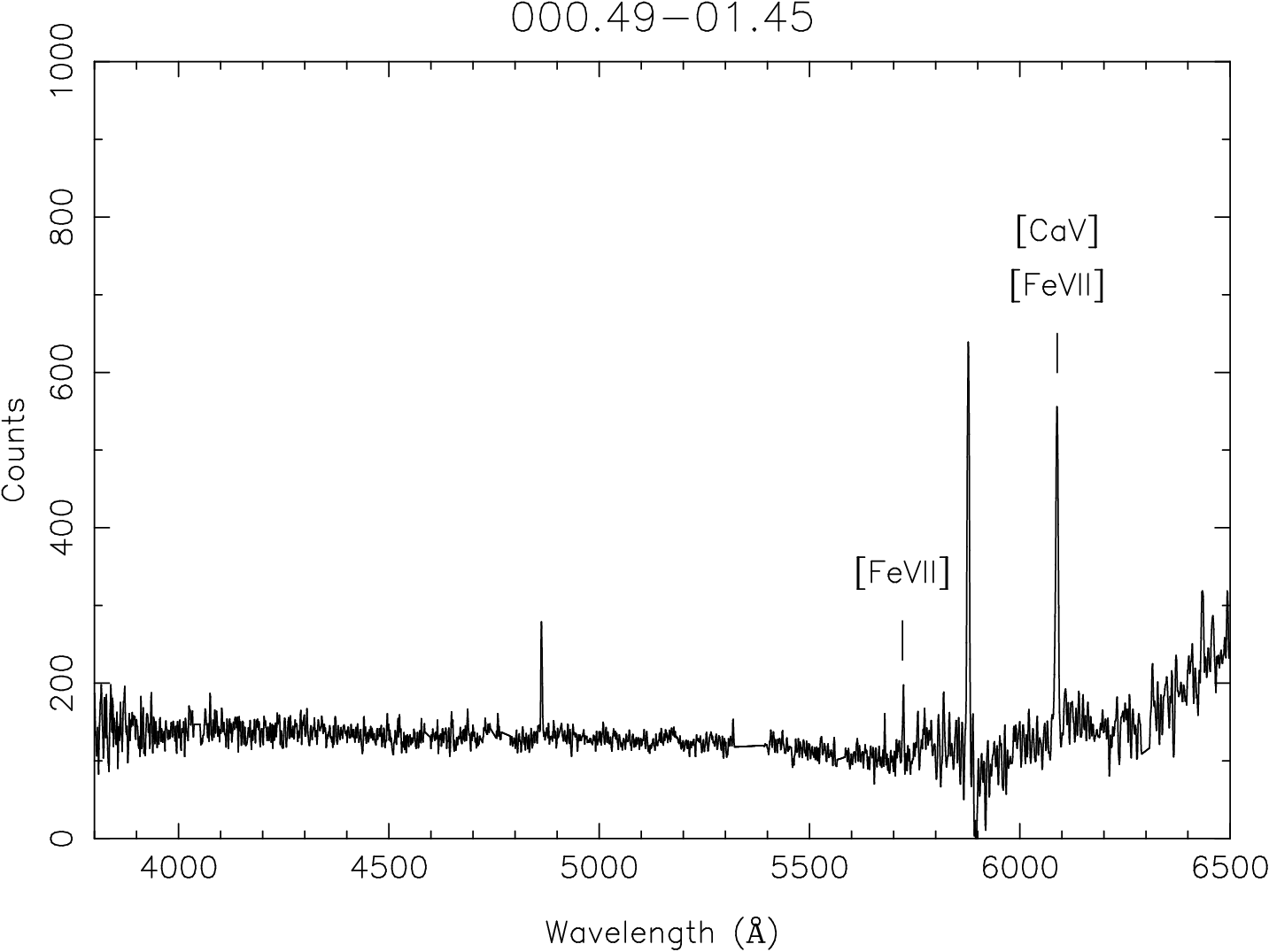}
\includegraphics[angle=270,scale=0.36]{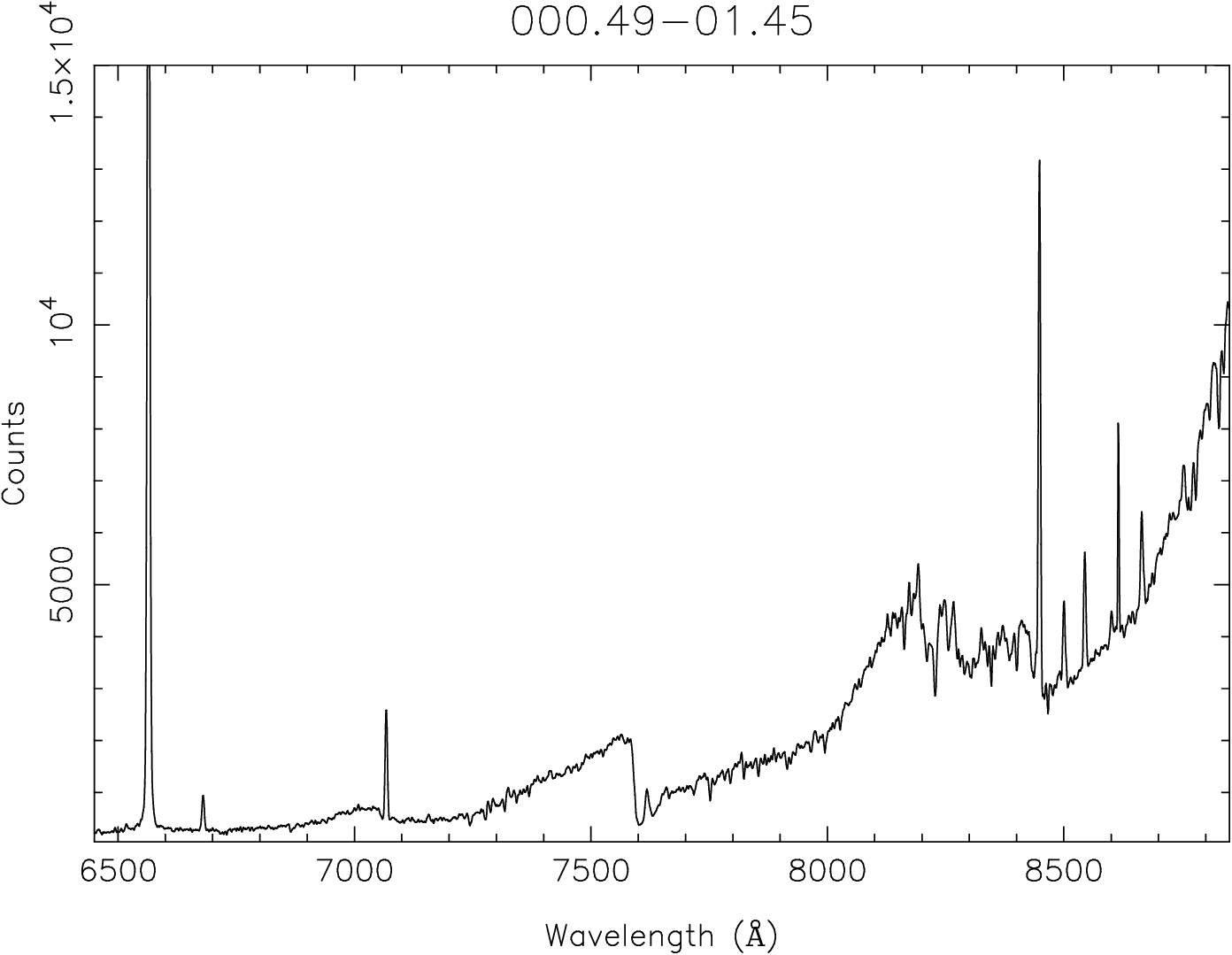} \\
\includegraphics[angle=270,scale=0.36]{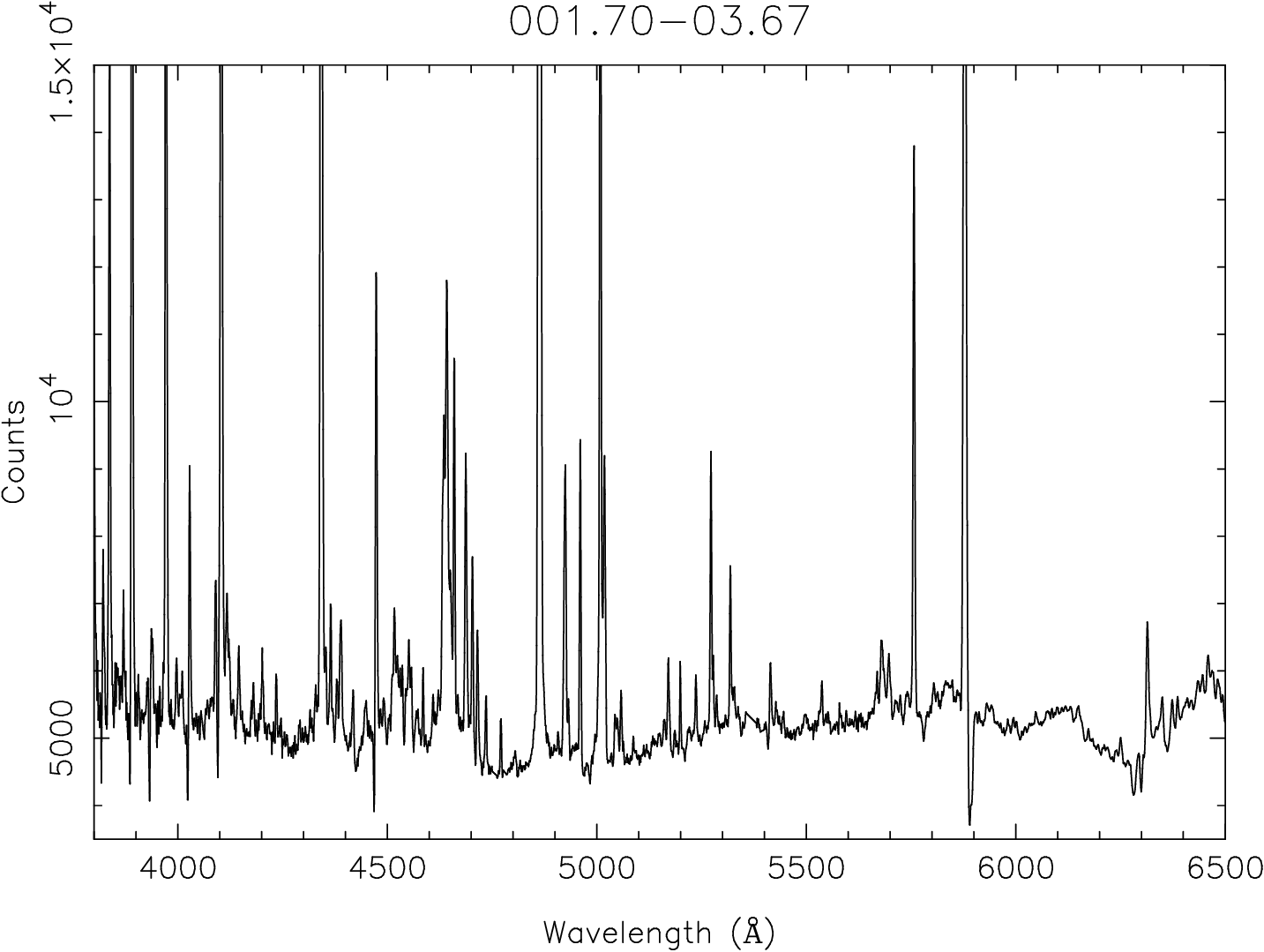}
\includegraphics[angle=270,scale=0.36]{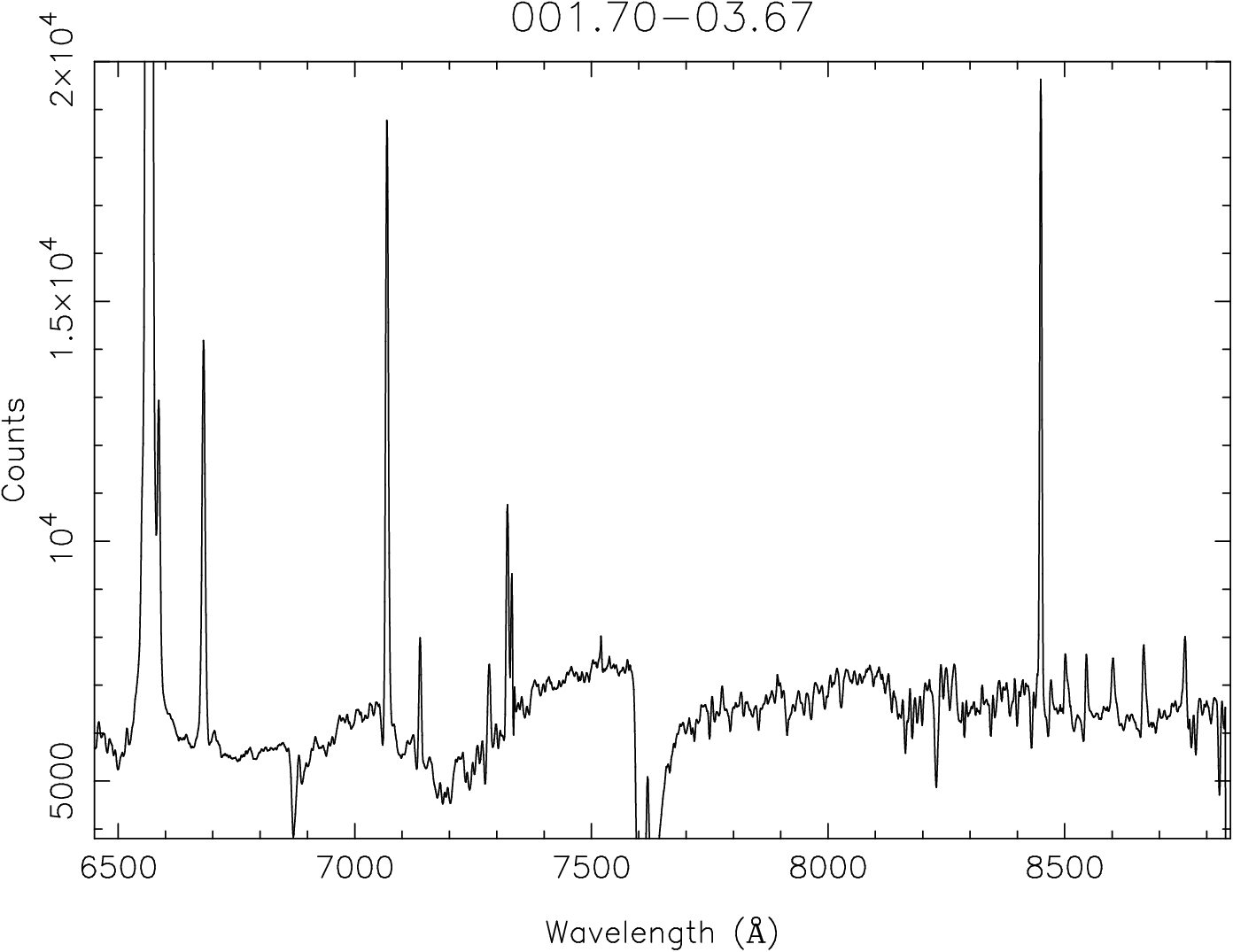}\\
\includegraphics[angle=270,scale=0.36]{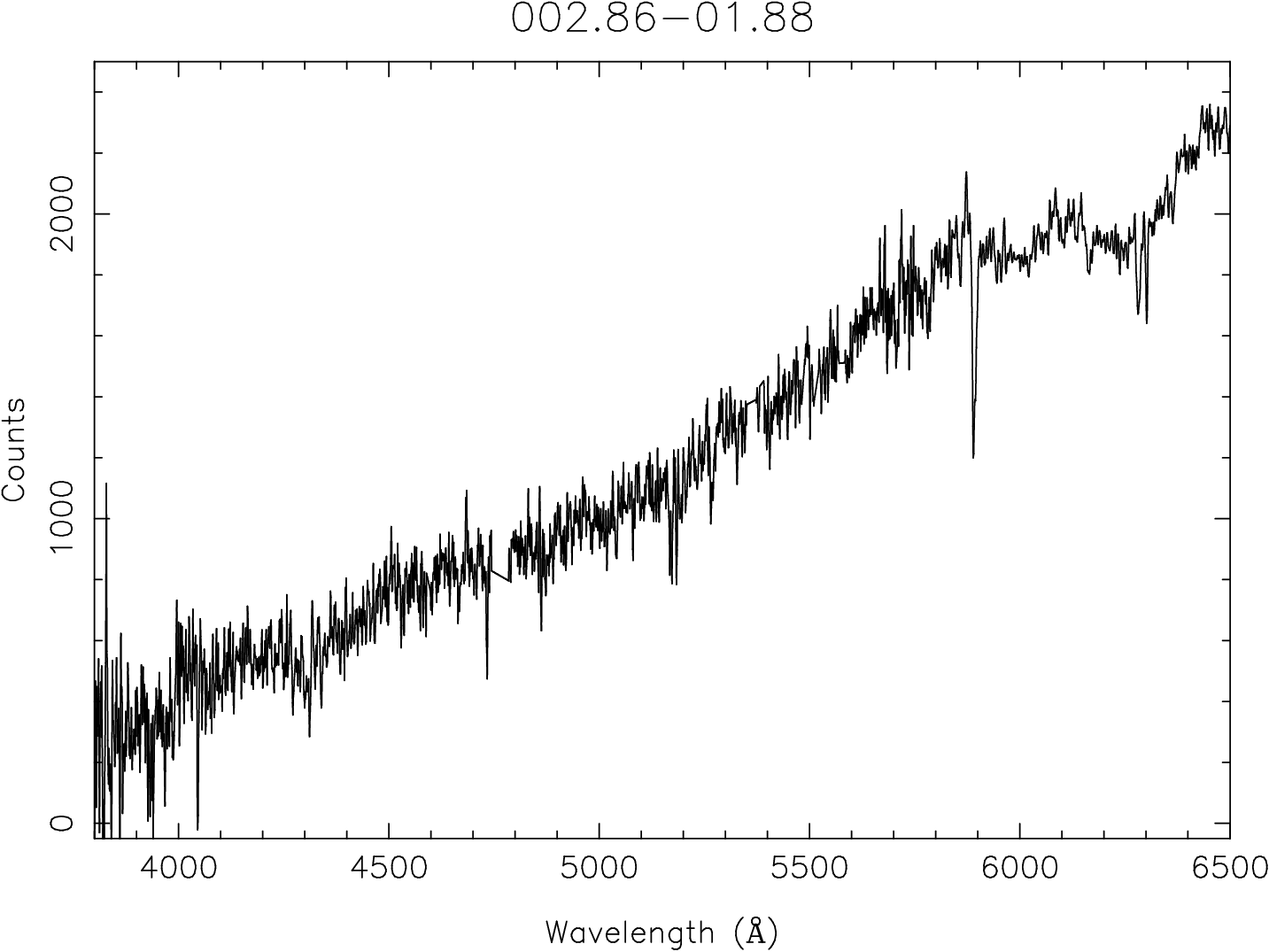}
\includegraphics[angle=270,scale=0.36]{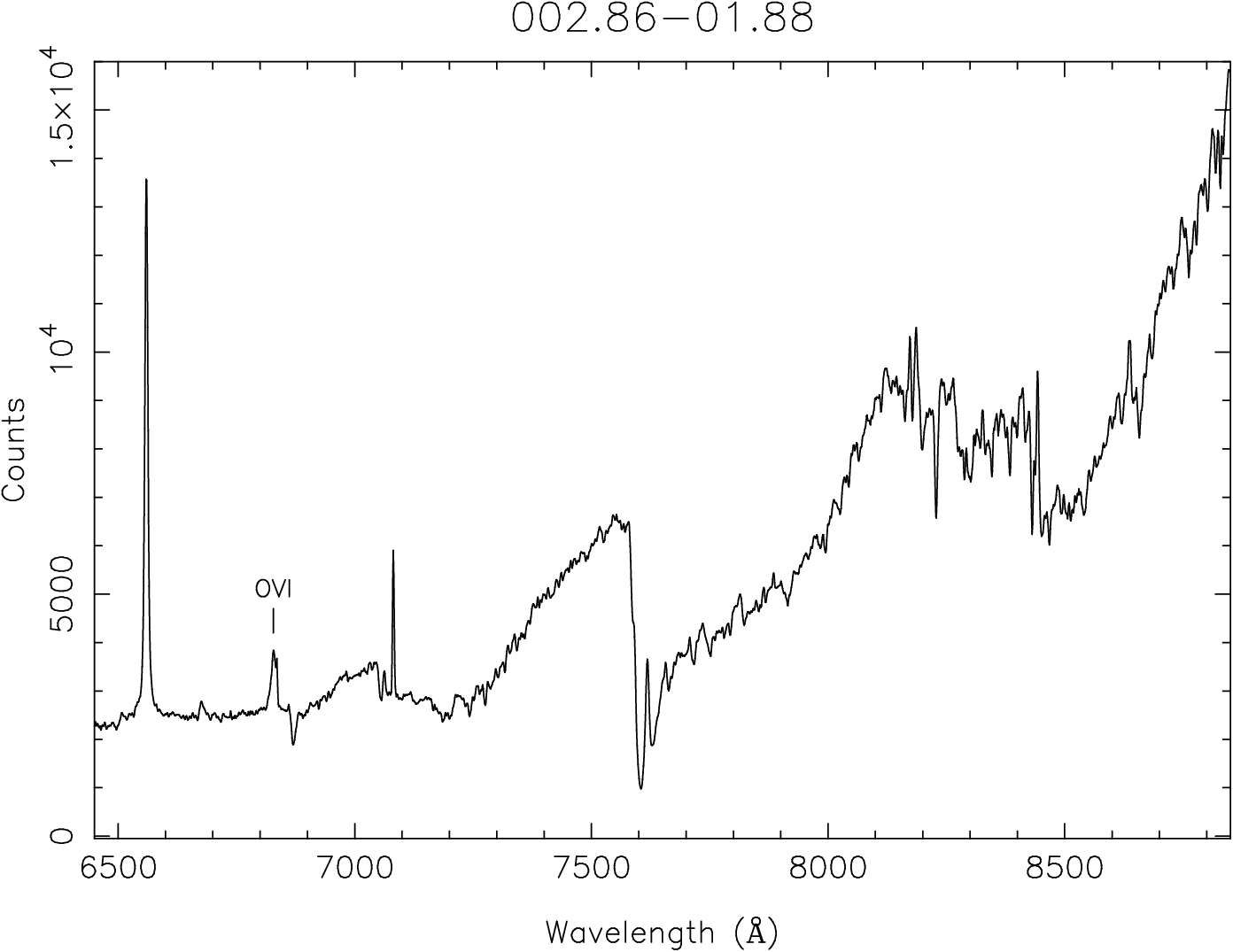} \\
\end{center}
\caption{AAOmega spectra of new S-type symbiotic stars.}
\label{fig:s1}
\end{figure*}

\begin{figure*}
\begin{center}
\includegraphics[angle=270,scale=0.36]{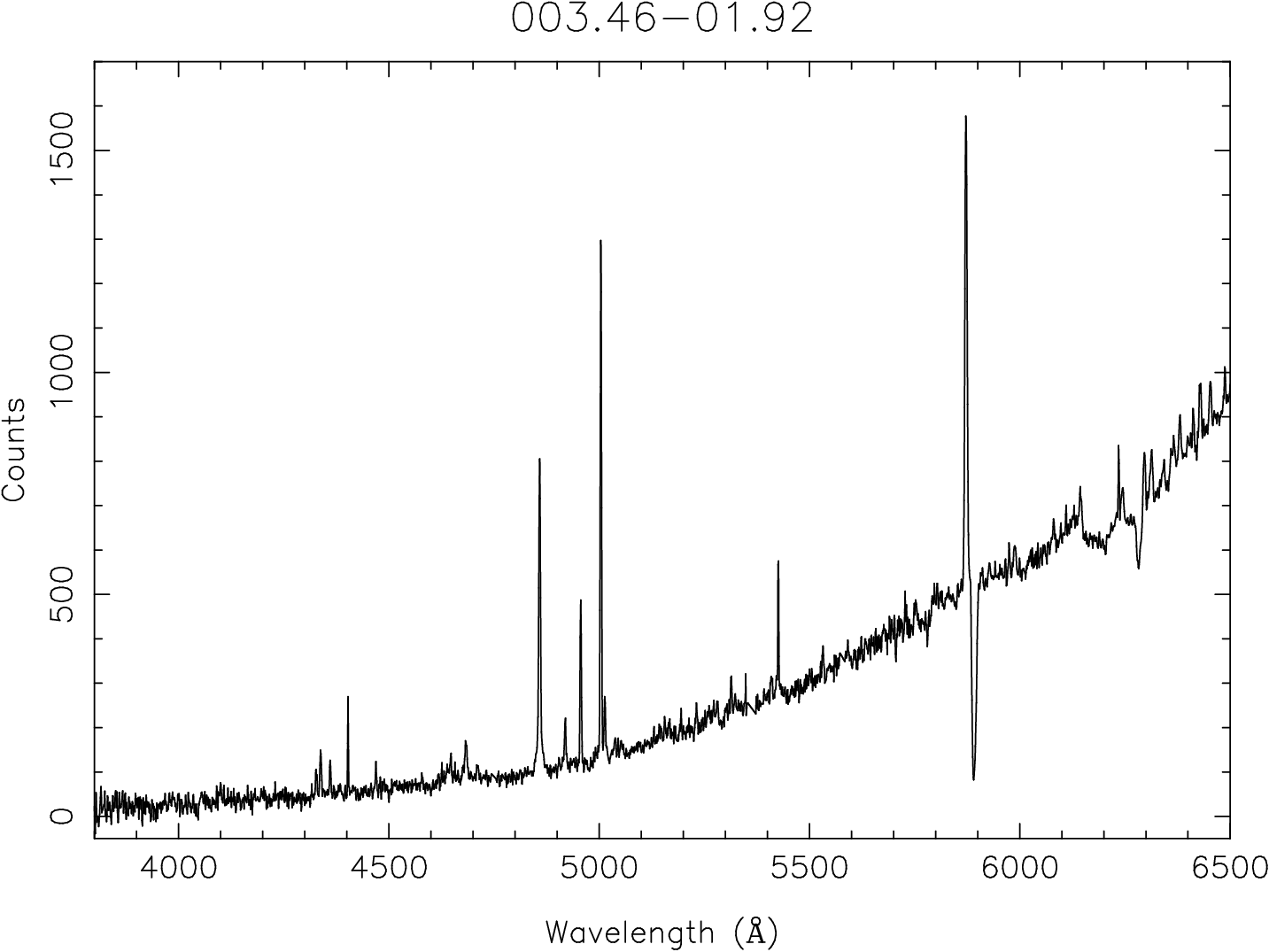}
\includegraphics[angle=270,scale=0.36]{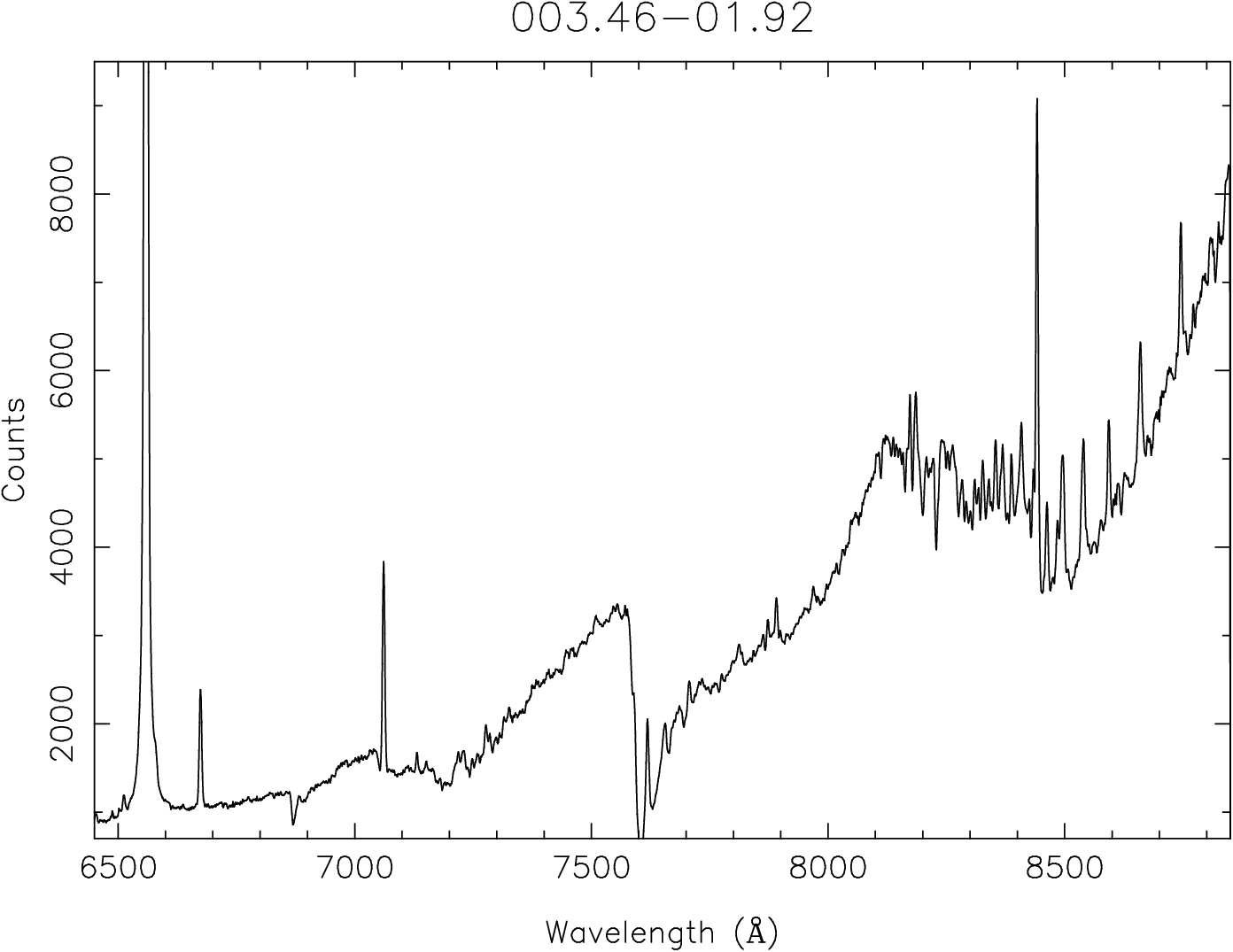}\\
\includegraphics[angle=270,scale=0.36]{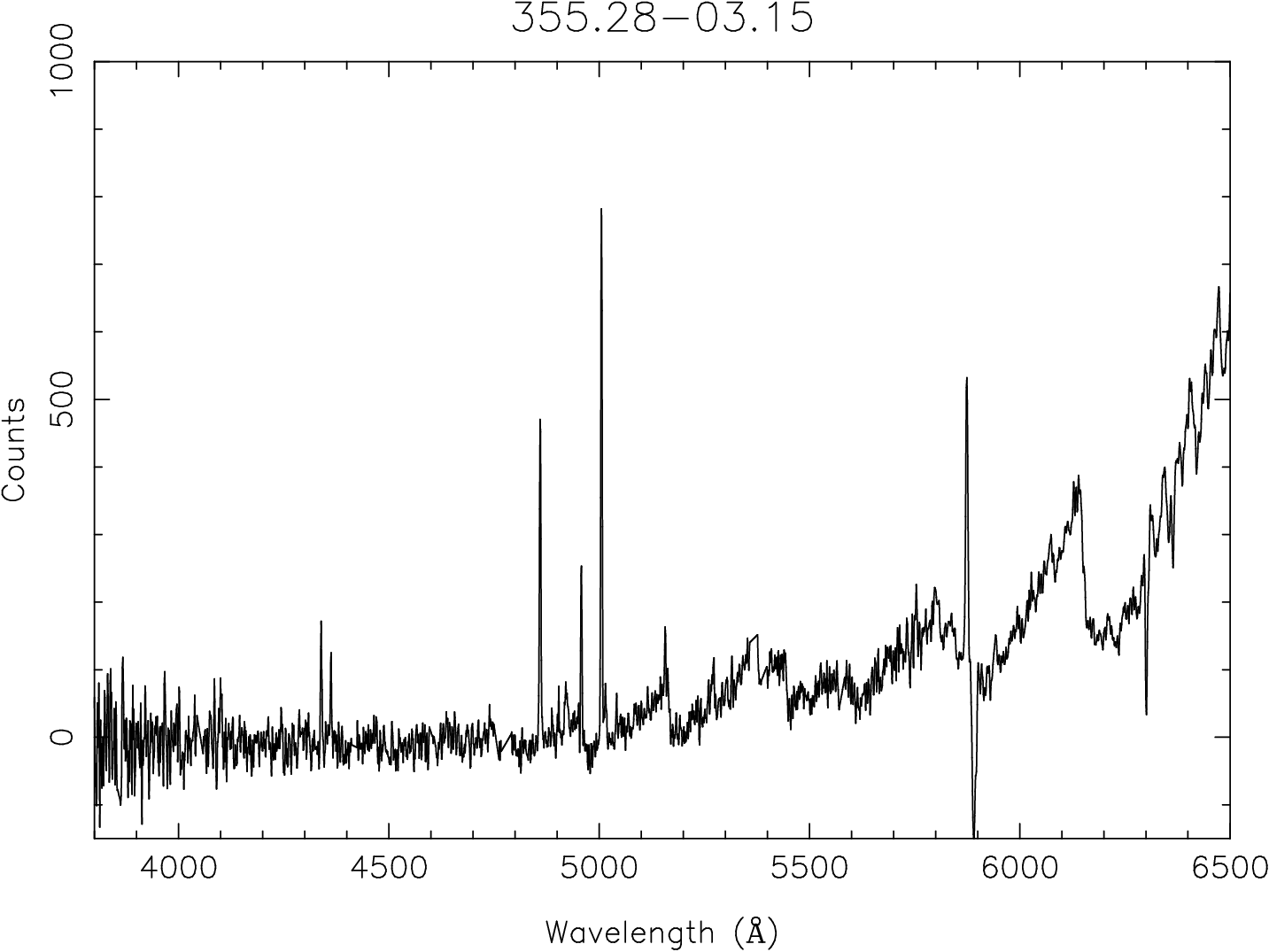}
\includegraphics[angle=270,scale=0.36]{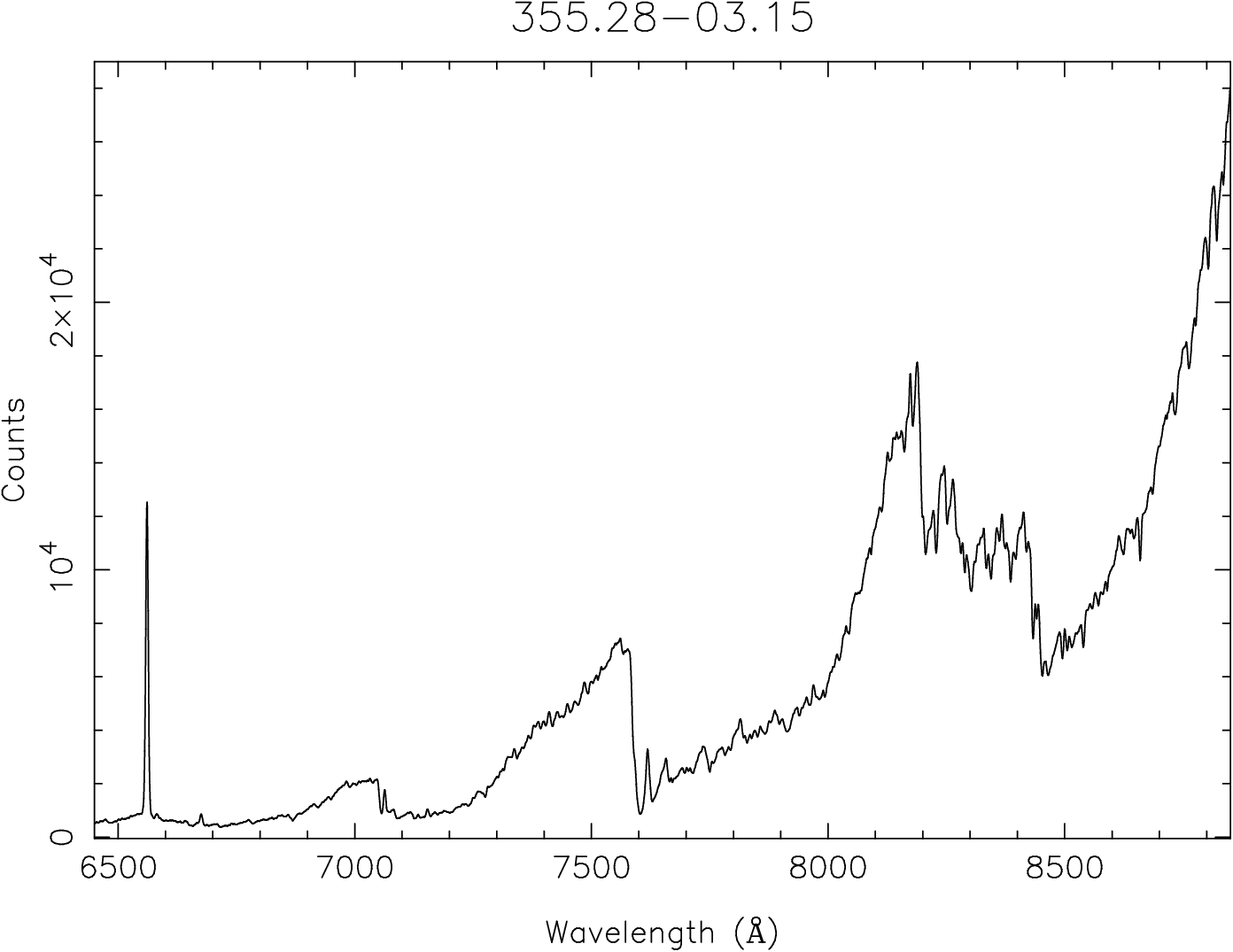} \\
\includegraphics[angle=270,scale=0.36]{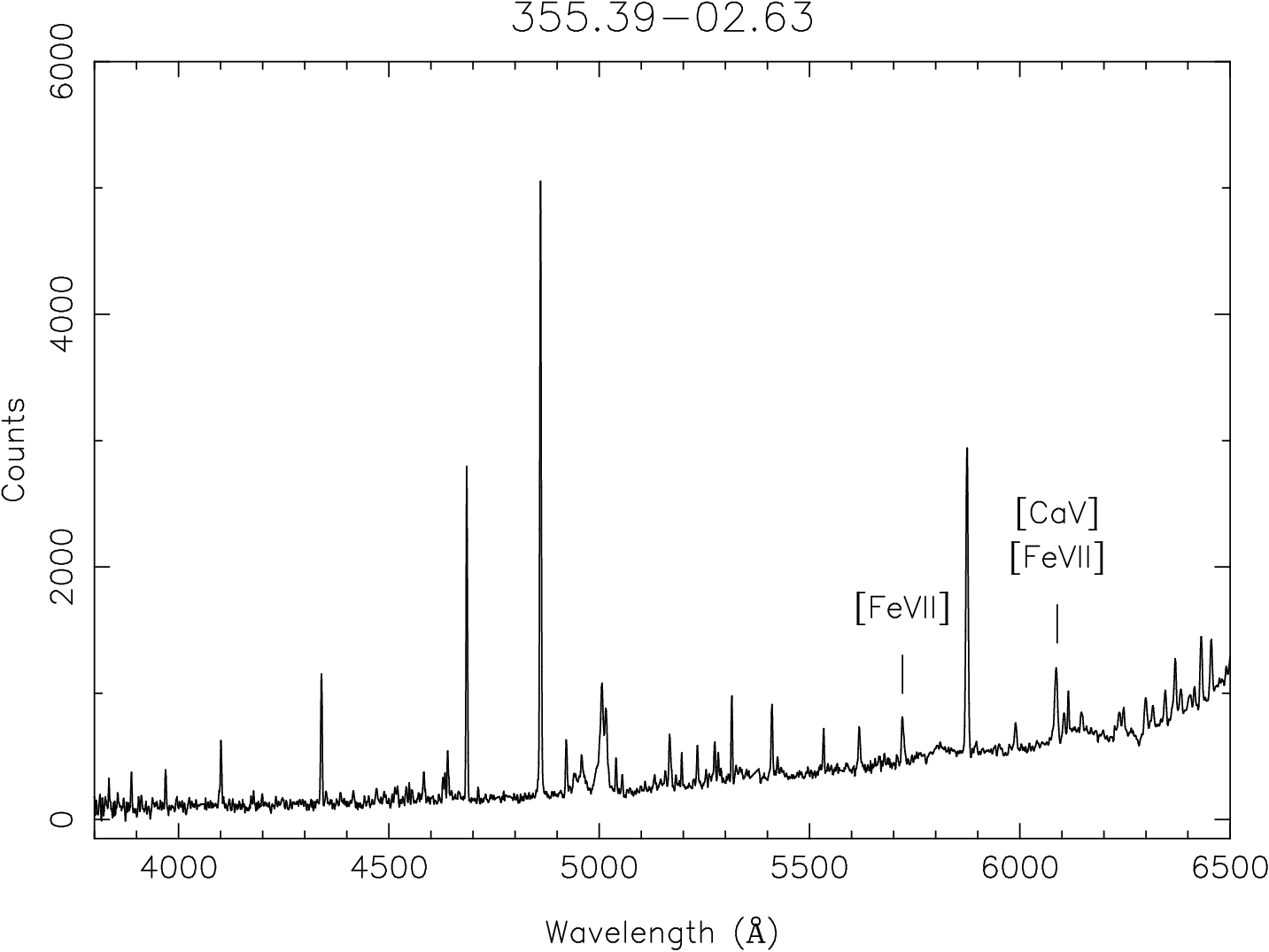}
\includegraphics[angle=270,scale=0.36]{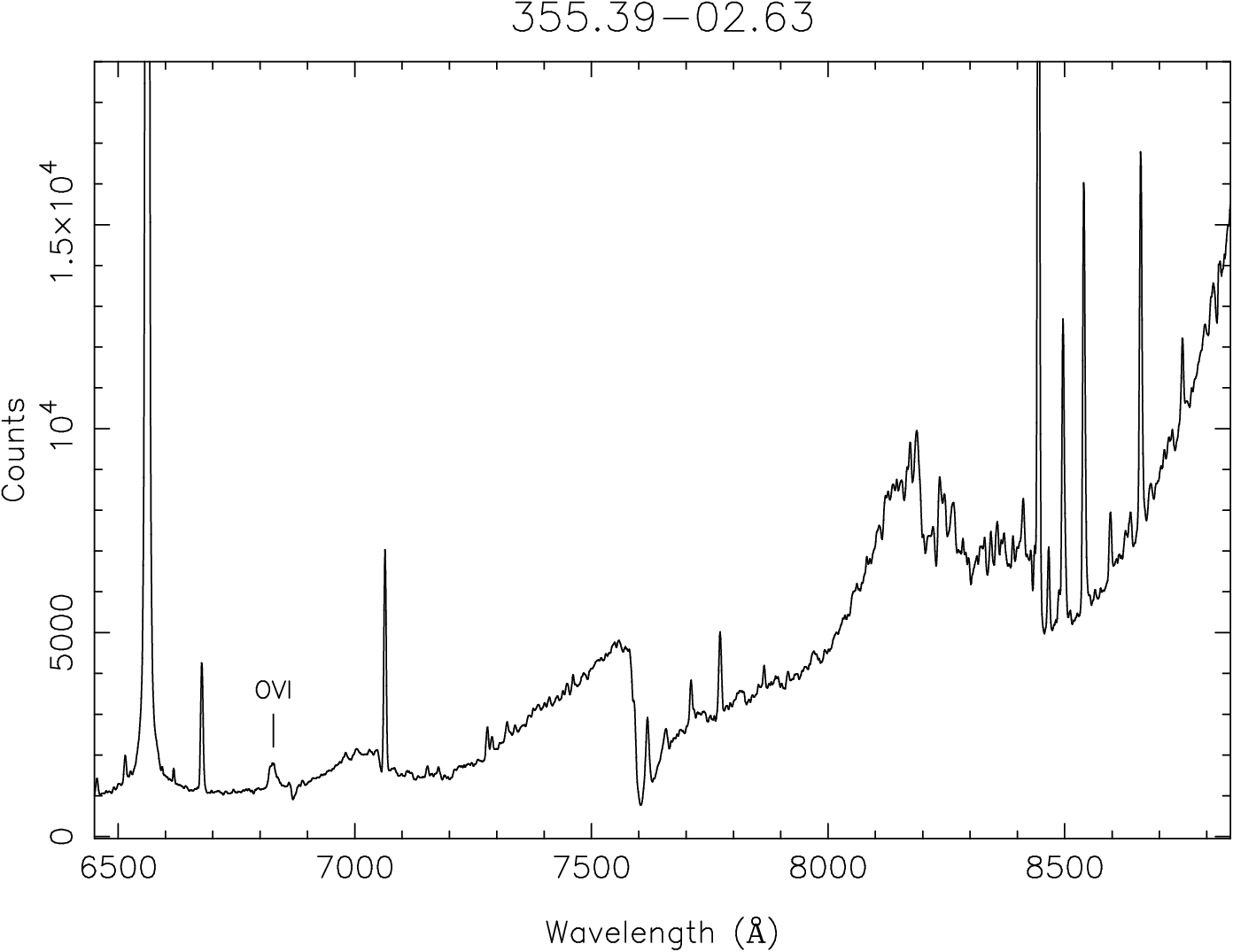} \\
\end{center}
\caption{AAOmega spectra of new S-type symbiotic stars (continued).}
\label{fig:s2}
\end{figure*}

\begin{figure*}
\begin{center}
\includegraphics[angle=270,scale=0.36]{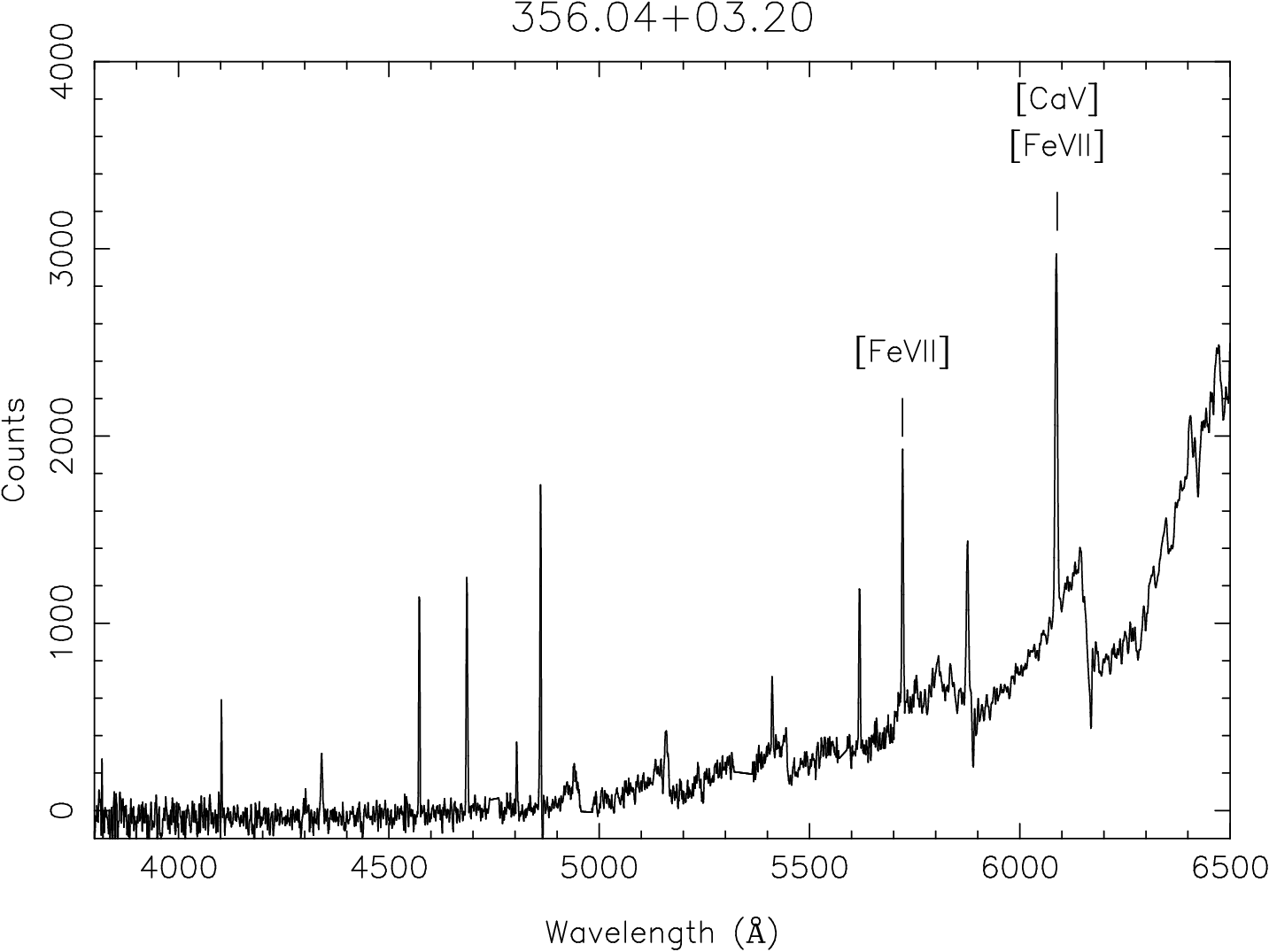}
\includegraphics[angle=270,scale=0.36]{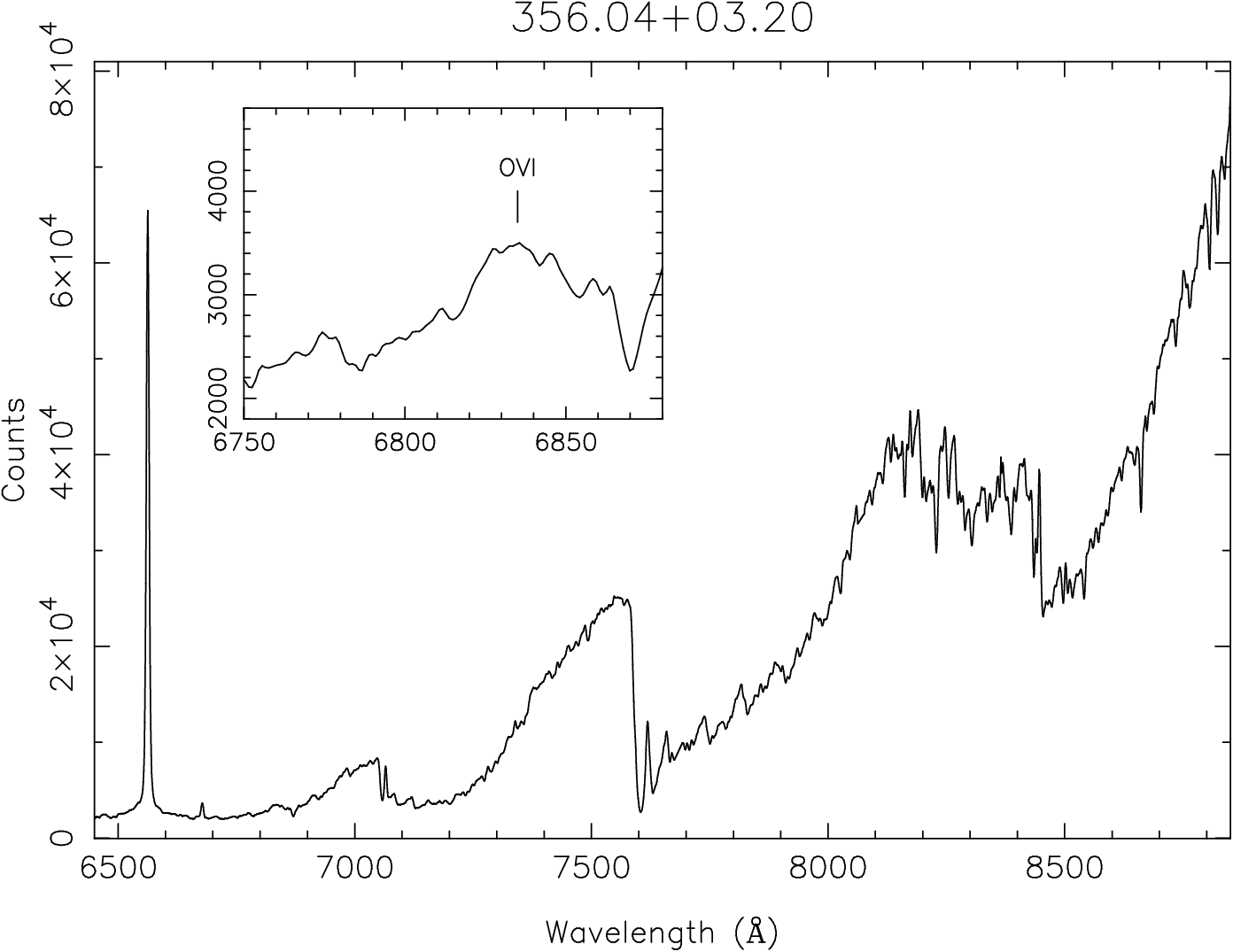} \\
\includegraphics[angle=270,scale=0.36]{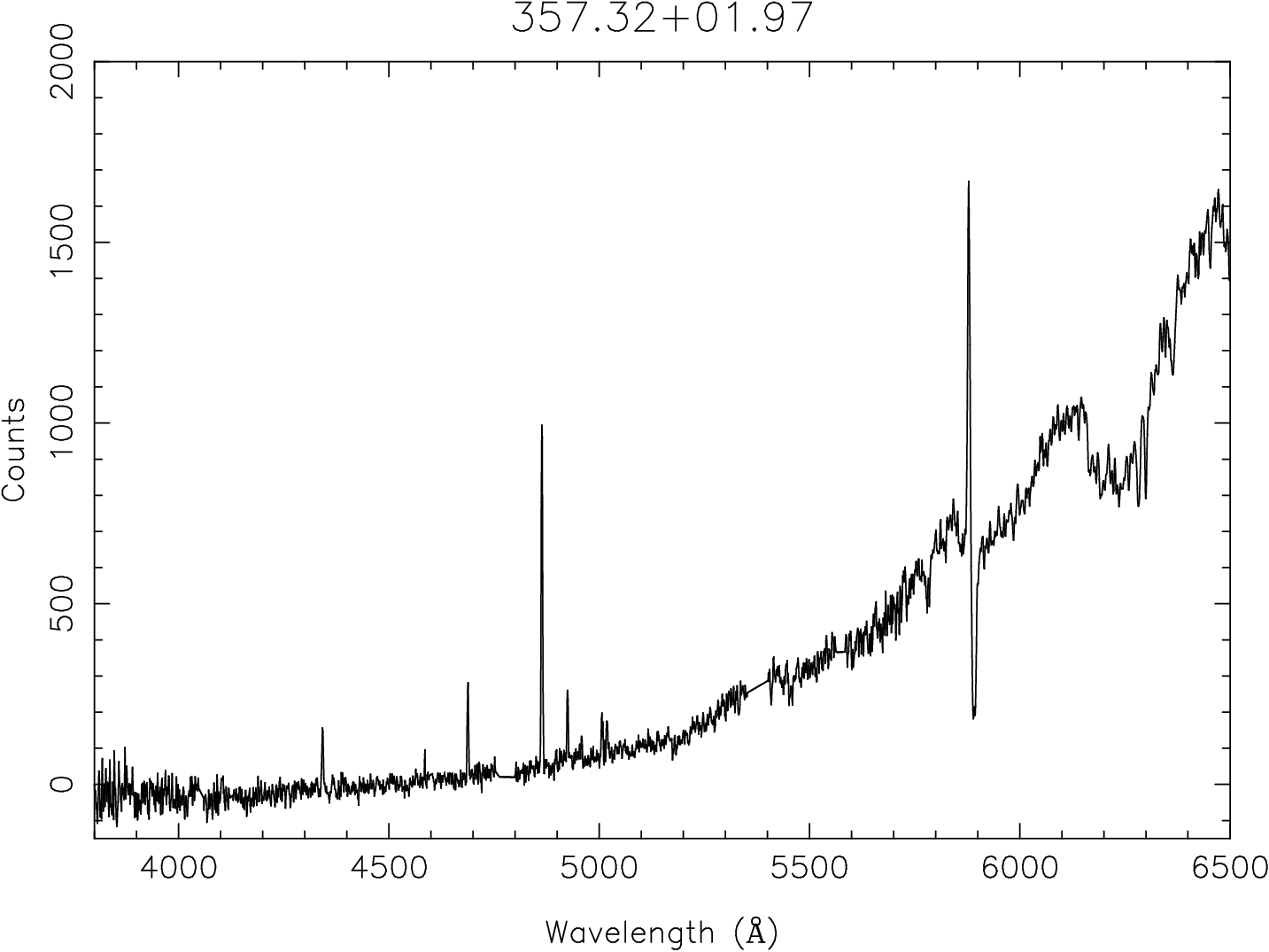}
\includegraphics[angle=270,scale=0.36]{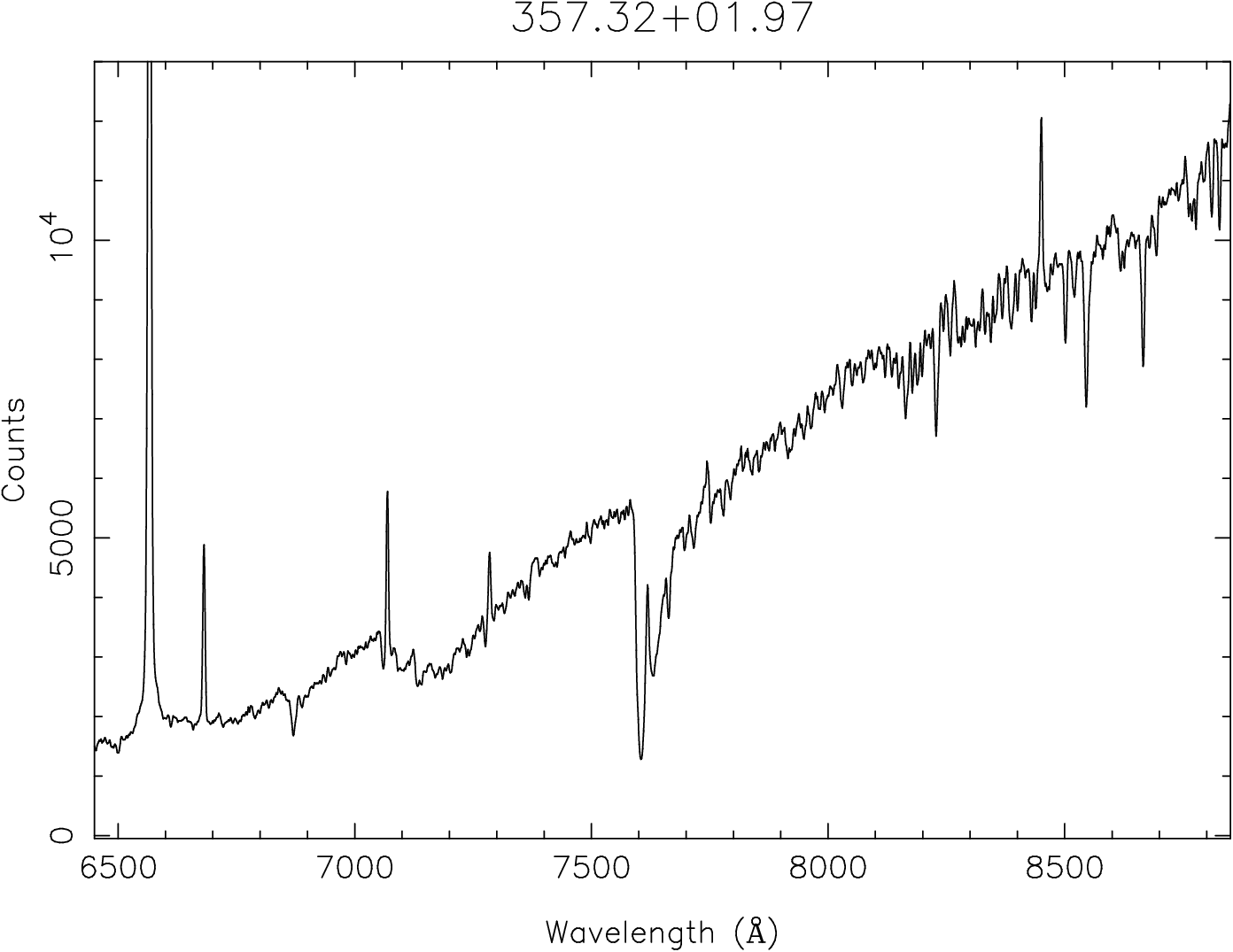} \\
\includegraphics[angle=270,scale=0.36]{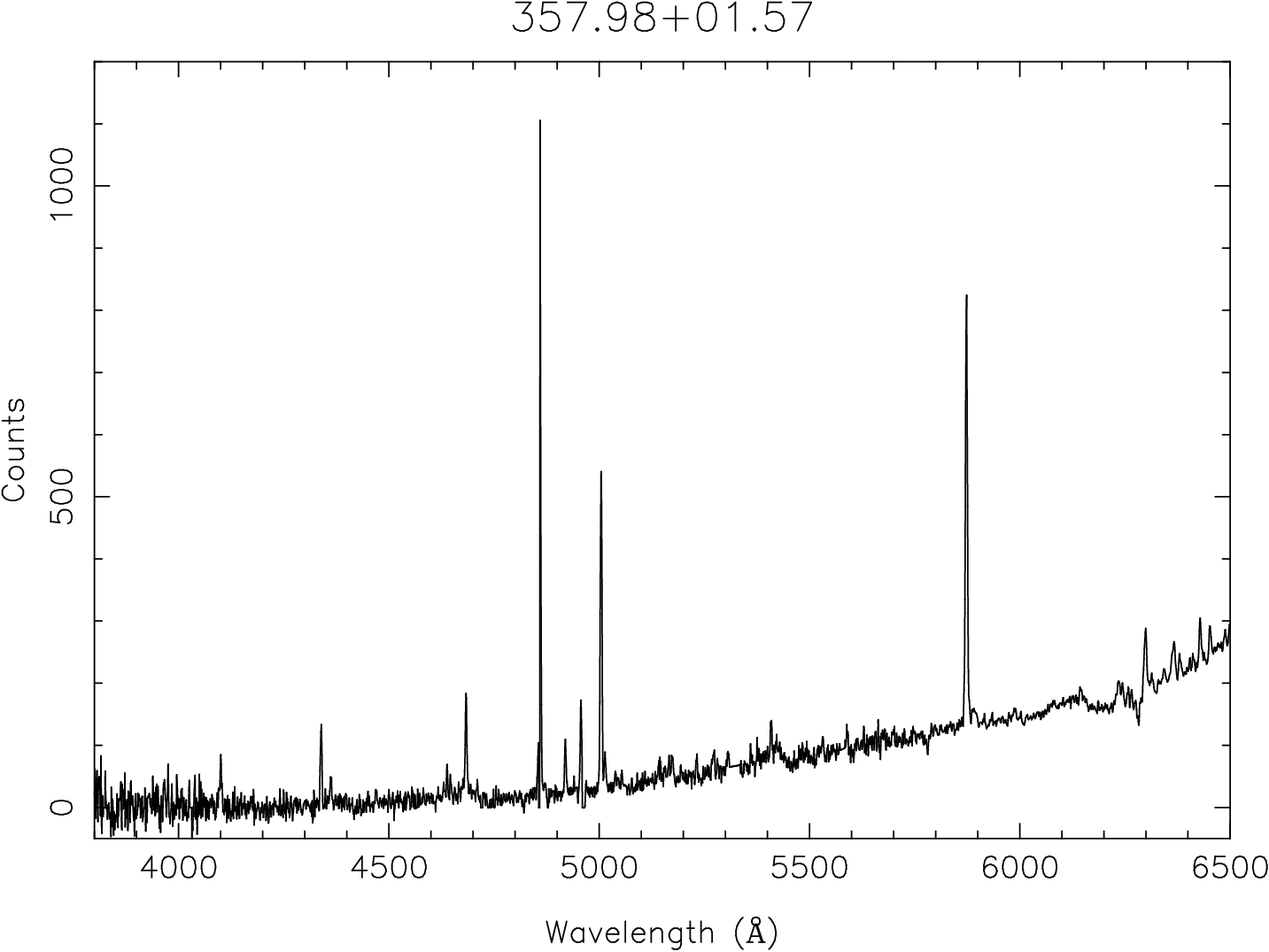}
\includegraphics[angle=270,scale=0.36]{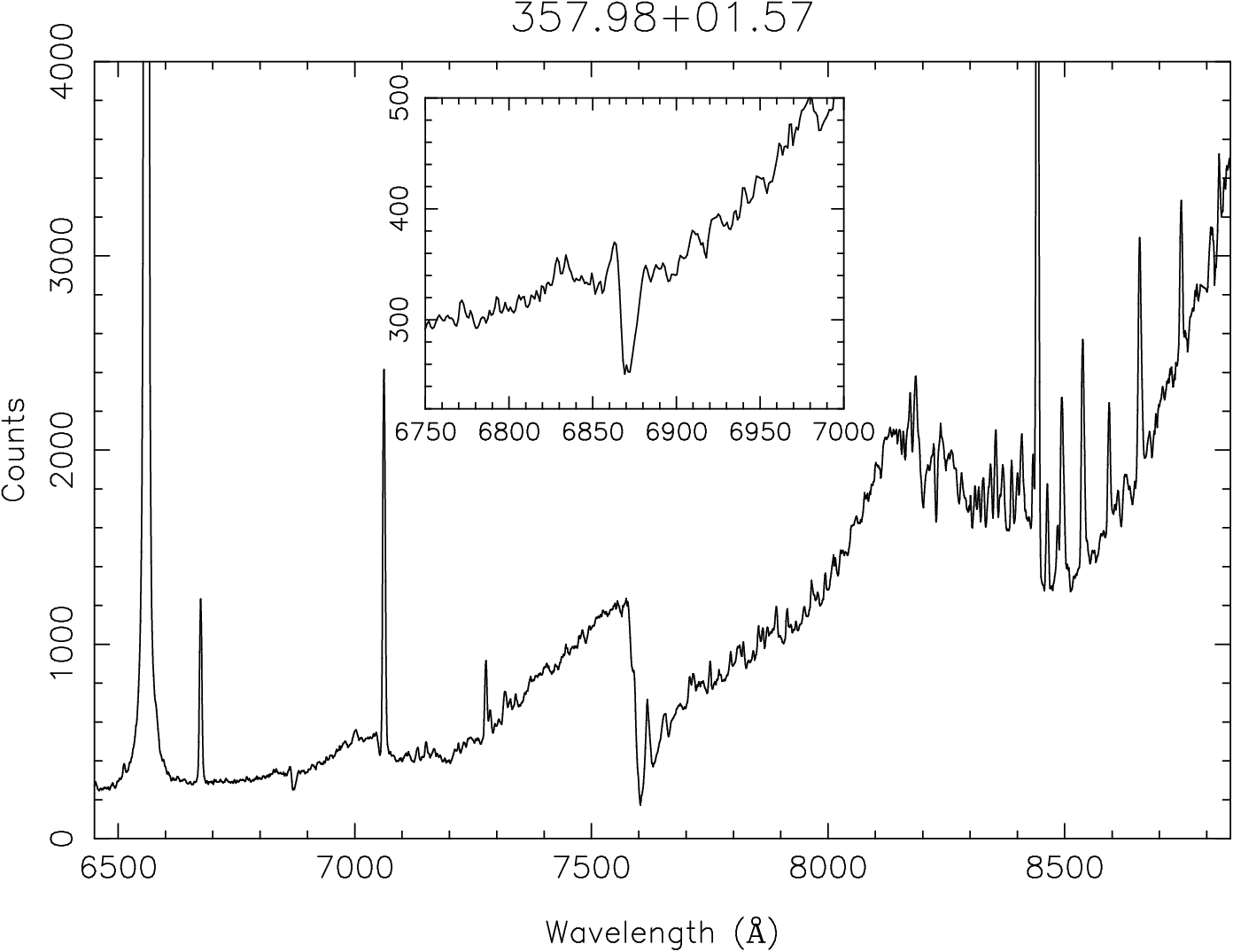}\\
\end{center}
\caption{AAOmega spectra of new S-type symbiotic stars (continued).}
\label{fig:s3}
\end{figure*}

\begin{figure*}
\begin{center}
\includegraphics[angle=270,scale=0.36]{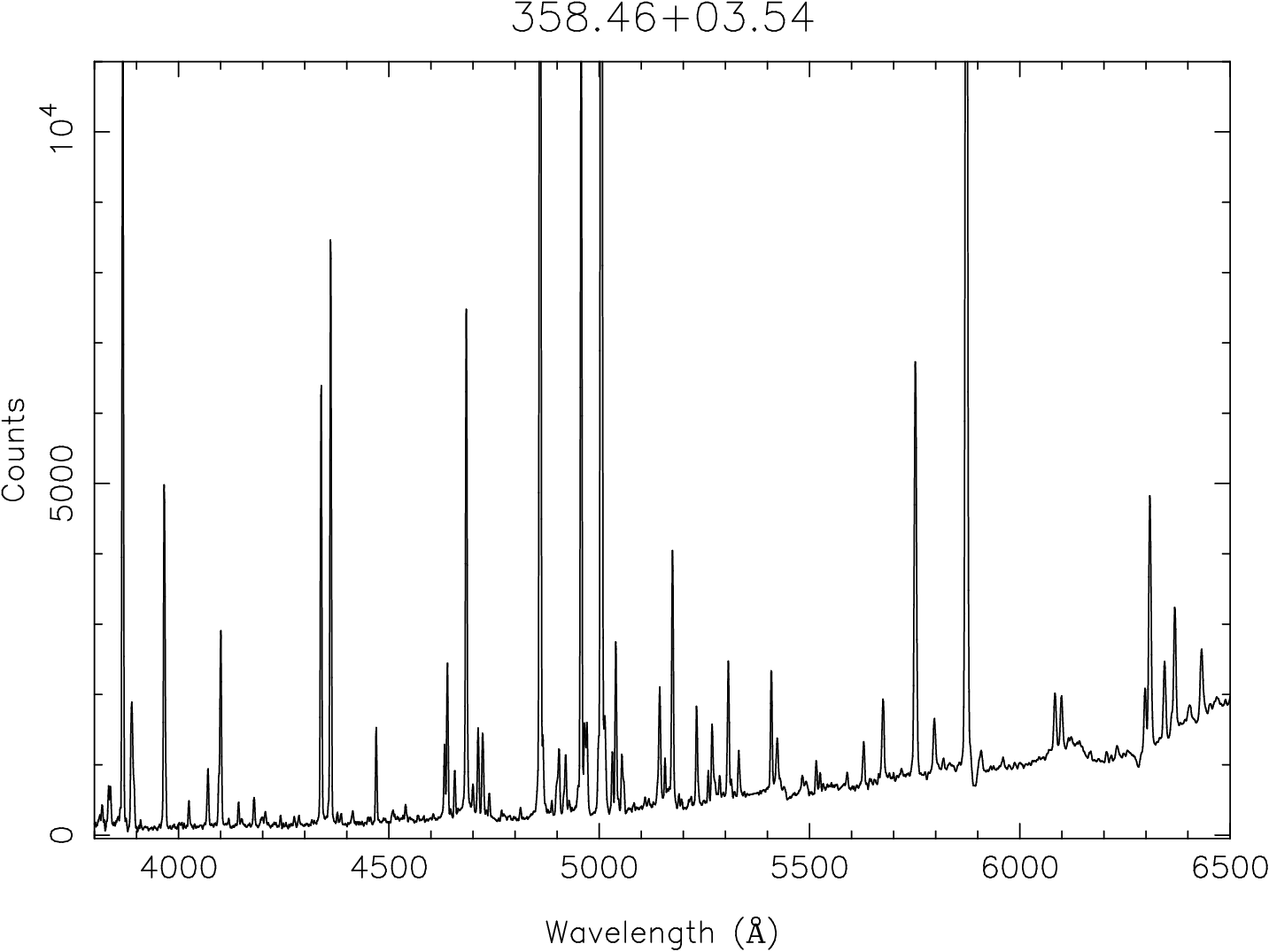}
\includegraphics[angle=270,scale=0.36]{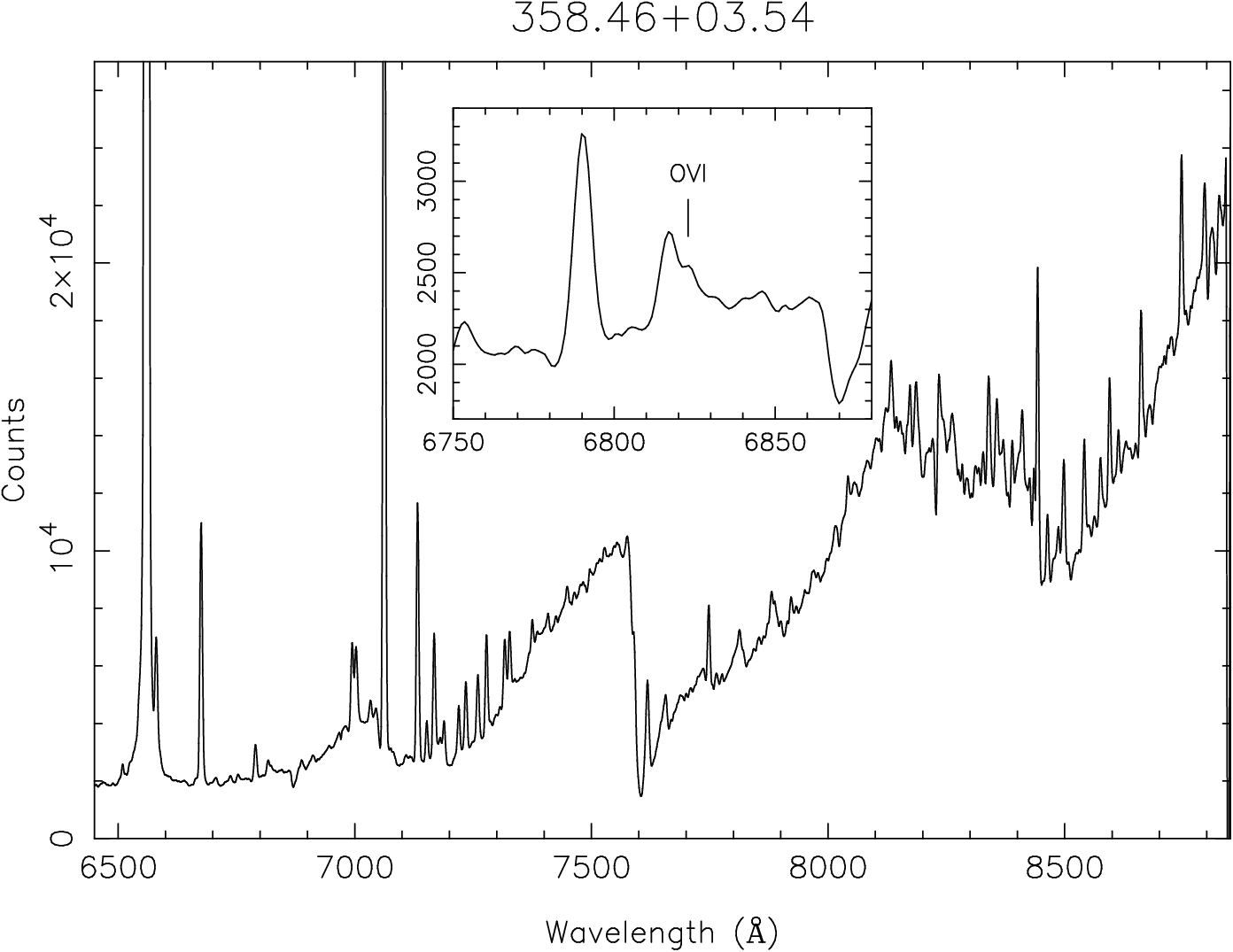}\\
\includegraphics[angle=270,scale=0.36]{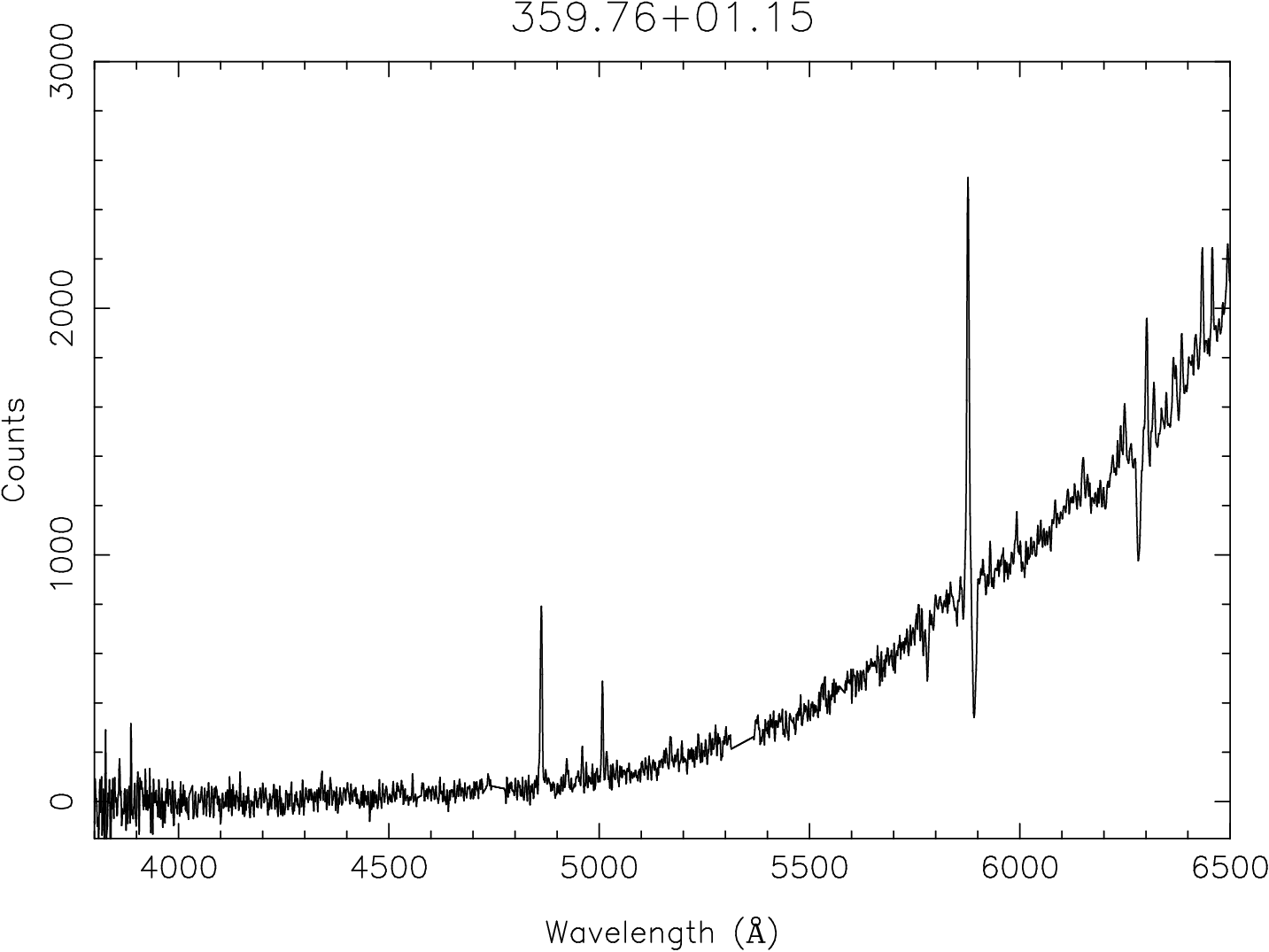}
\includegraphics[angle=270,scale=0.36]{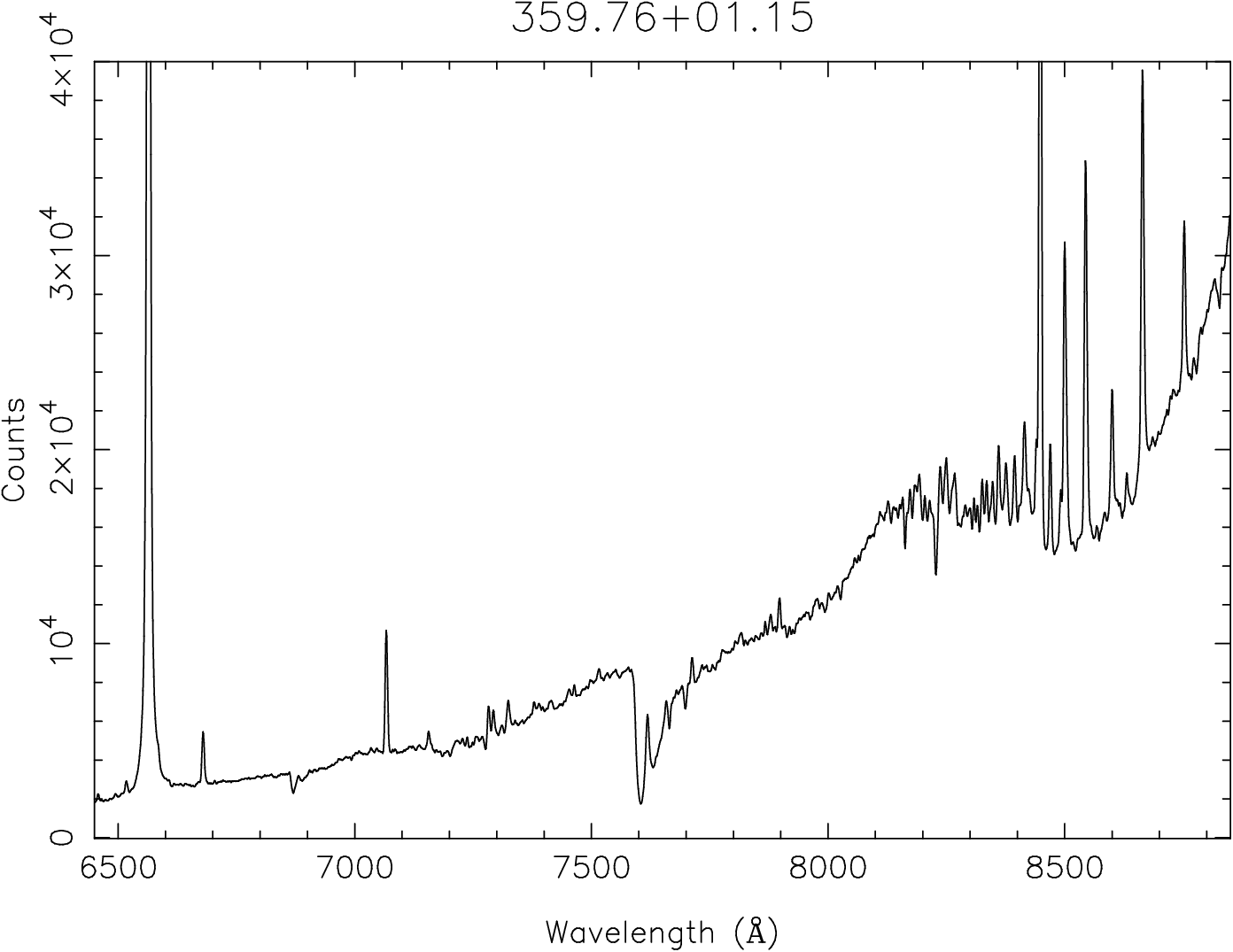} \\
\includegraphics[angle=270,scale=0.36]{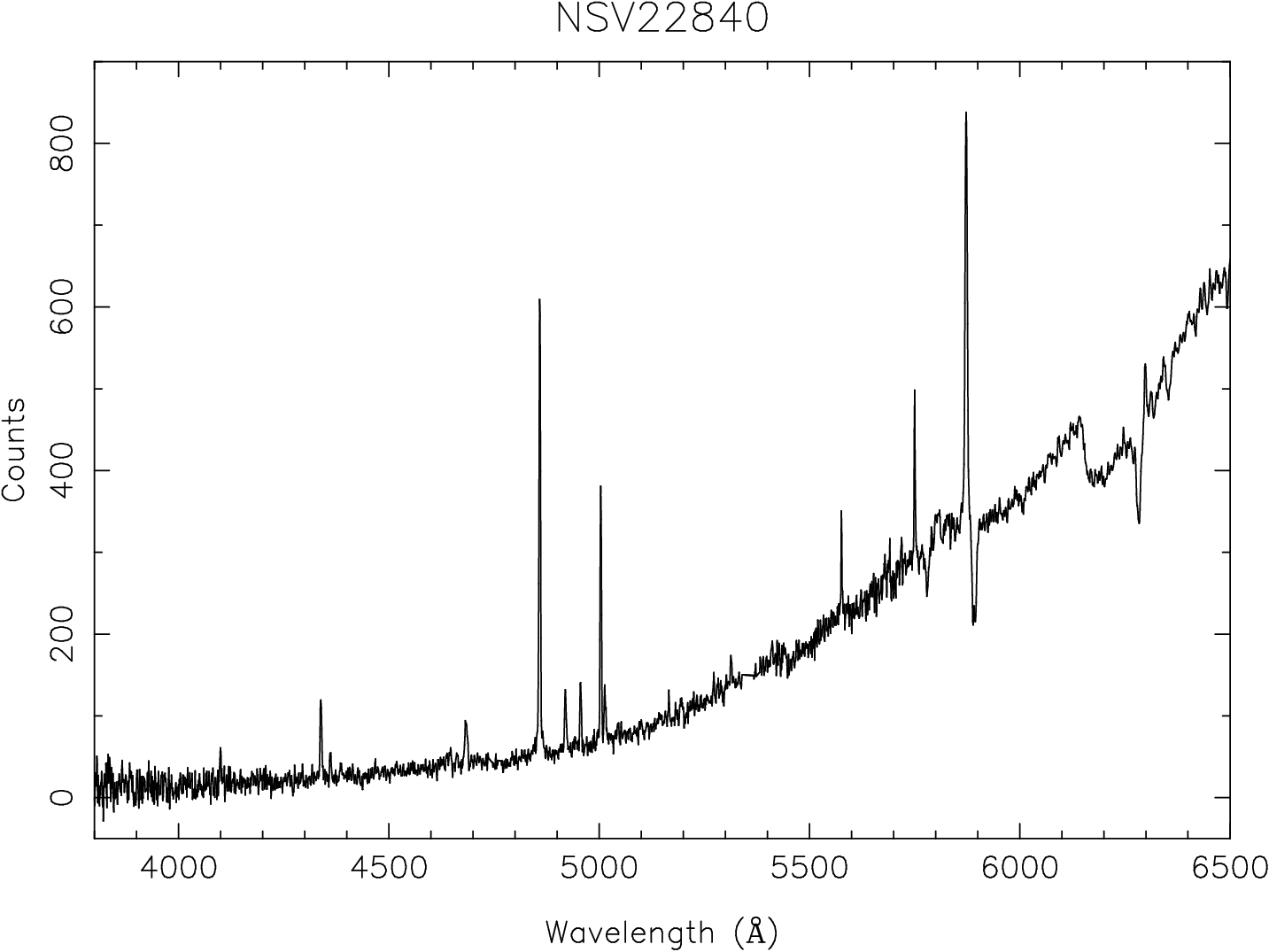}
\includegraphics[angle=270,scale=0.36]{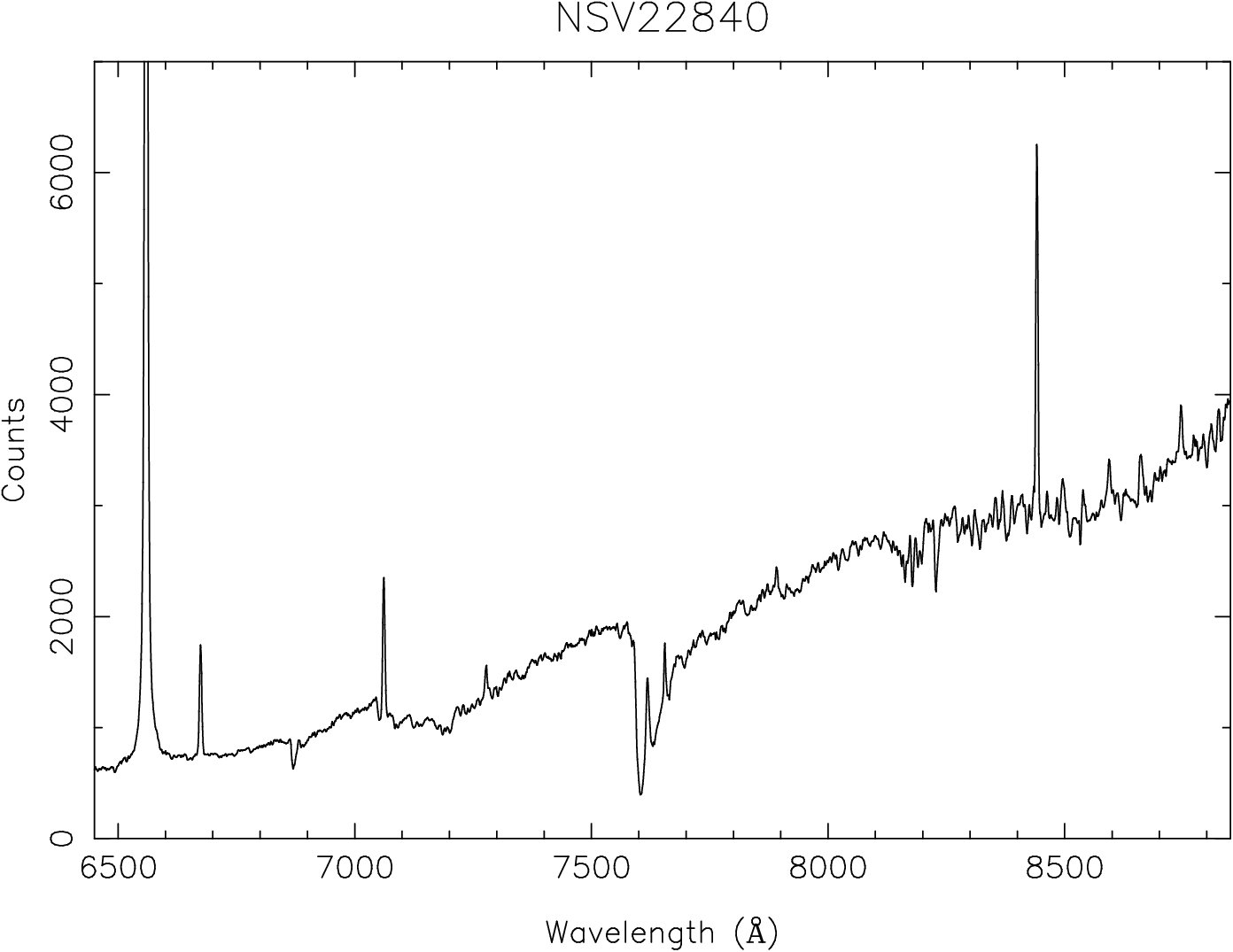}\\
\end{center}
\caption{AAOmega spectra of new S-type symbiotic stars (continued).}
\label{fig:s4}
\end{figure*}

\begin{figure*}
\begin{center}
\includegraphics[angle=270,scale=0.36]{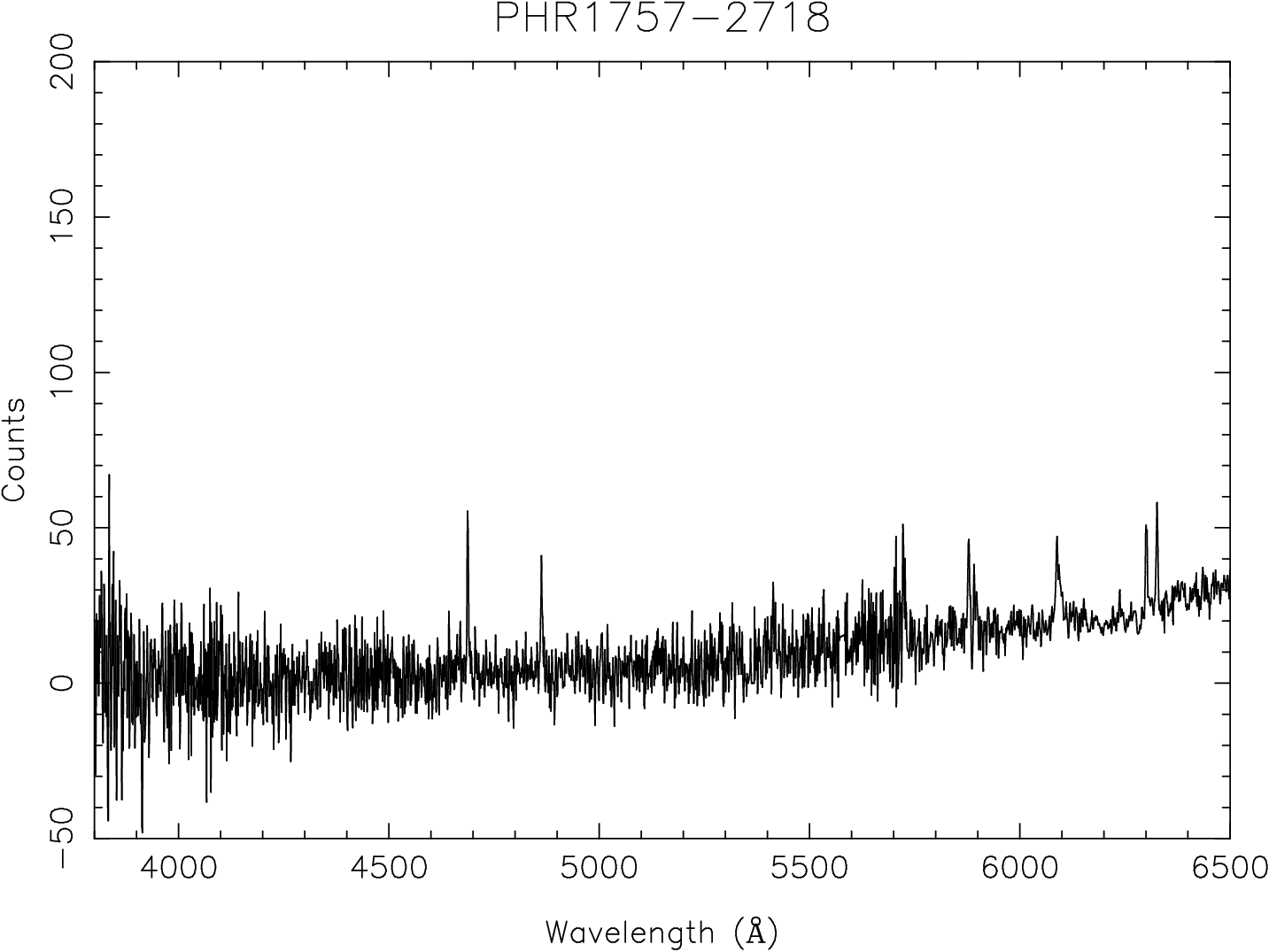}
\includegraphics[angle=270,scale=0.36]{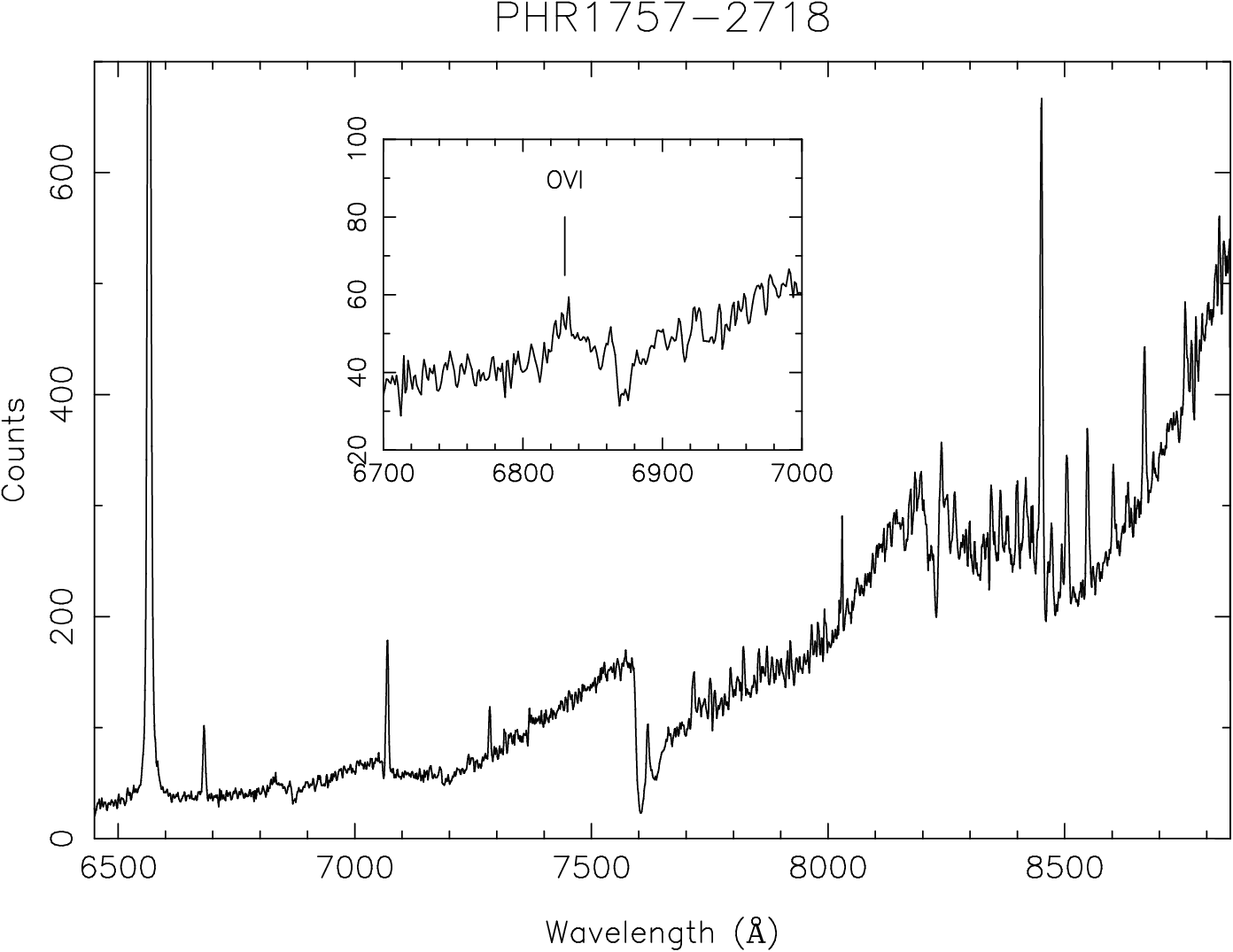}\\
\end{center}
\caption{AAOmega spectra of new S-type symbiotic stars (continued).}
\label{fig:s5}
\end{figure*}

Figures \ref{fig:s1}, \ref{fig:s2}, \ref{fig:s3}, \ref{fig:s4} and \ref{fig:s5} present spectra of the new S-type symbiotic systems.

\subsection{000.49-01.45}
The presence of high ionisation [FeVII] emission lines in the AAOmega spectrum (Fig. \ref{fig:s1}) ensures the symbiotic nature of this object. 
The I-band lightcurve in Fig. \ref{fig:oglethreefour1} is typical of an S-type system with an M5--6 giant, namely semi-regular pulsations on top of a probable orbital variability of 1014.5 days. 

\subsection{001.70-03.67}
One of the brightest symbiotics in our sample ($J=9.0$ mag) with a rich emission line spectrum on top of an K5--M1 red giant (Tab. \ref{tab:spnew}). It is saturated in the OGLE photometry and the MACHO lightcurves are rather noisy, making it difficult to derive any conclusion about variability.

\subsection{002.86-01.88}
This is another significantly reddened S-type symbiotic as proven by the presence of the Raman scattered OVI 6825 feature. 
As in the other S-type symbiotics with late M-type red giants the OGLE lightcurve (Fig. \ref{fig:oglethreefour1}) is dominated by low amplitude semi-regular pulsations. A longer scale orbital variation of $\sim$1068.4 days seems to be present. The spectrum shown in Fig. \ref{fig:s1} corresponds to the first AAOmega observation which shows the Raman-scattered OVI feature, albeit with a lack of corresponding high ionisation features (e.g. [O~III], [Fe~VII]). The OVI feature subsequently vanished in the second AAOmega observation 324 days later.

\subsection{003.46-01.92}
We find an orbital period of 801.0 days from the OGLE lightcurve that closely resembles that of AR Pav (Rutkowski, Miko{\l}ajewska \& Whitelock 2007). The eclipsing binary shows a probable secondary minimum indicative of ellipsoidal variability.

\subsection{355.28-03.15}
There is no He~II $\lambda$4686 in the AAOmega spectrum, however strong [O~III] and He~I emission lines support its symbiotic nature. 
The combined OGLE lightcurve is dominated by low amplitude semi-regular pulsations, but no periodic long term signal is present. 

\subsection{355.39-02.63}
An excellent example of an S-type symbiotic with a very rich, high-ionisation emission line spectrum with all features present that would be expected including Raman scattered OVI 6825. The OGLE lightcurve is typical for an S-type, resembling closely e.g. 003.46-01.92, with semi-regular pulsations from the red giant and an orbital variation of 854.5 days.

\subsection{356.04+03.20}
The emission line spectrum is consistent with a highly ionized S-type symbiotic such as CI Cyg (Kenyon et al. 1991) and AX Per (Miko{\l}ajewska \& Kenyon 1992). The Raman scattered OVI 6825 feature is weak but its presence is consistent with the detected [FeVII] emission.

\subsection{357.98+01.57}
The emission line spectrum of 357.98+01.57 is typical of an S-type symbiotic. There may be a very weak detection of OVI 6825, but without any [FeVII] emission lines its presence is unlikely. Strong TiO $\lambda$8432 in the IR part of the spectrum suggests that the spectral type is as late as indicated by the VO band, i.e. most likely M5--6, where the weakness of the TiO bands at shorter wavelengths is caused by the symbiotic nebular continuum veiling the M giant. The OGLE-IV lightcurve in Fig. \ref{fig:oglefour1} shows low amplitude semi-regular pulsations of the red giant on top of longer term changes over 400--500 days, possibly due to orbital motion. The overall shape and timescales are typical of S-type symbiotics, but it is too early to make any conclusions about an orbital period.

\subsection{358.46+03.54}
This is one of the brightest objects in our sample at $J=9.19$ mag. As in 356.04+03.20, the rich emission line spectrum is typical of a highly ionized symbiotic star. The OVI emission line appears to be blended with a nebular emission line, most likely [Kr~III] $\lambda$6827 as seen in NGC~7027 (Zhang et al. 2005). 

\subsection{NSV22840}
NSV22840 is a little studied variable star which Terzan \& Gosset (1991) found to vary from $R=17.7$ to $R=16.5$ mag. Kohoutek \& Wehmeyer (2003) noted strong H$\alpha$ emission and a faint visible red continuum in their objective prism survey. The AAOmega spectrum proves its S-type symbiotic nature and the OGLE-IV lightcurve shows what may be an eclipse in addition to semi-regular pulsations.

\subsection{PHR1757-2718}
Listed as a possible PN by Parker et al. (2006) based on weak H$\alpha$ in their published spectrum. In Tab. \ref{tab:spnew} we estimated a spectral type of M1--4 for the red giant, however as in the case of 357.98+01.57, we suspect the symbiotic nebular continuum is masking the TiO bands, suggesting a later spectral type of M5--6. The emission line spectrum is typical for a highly ionized S-type symbiotic. Although the Raman scattered OVI 6825 feature is very weak, its presence is probably real as there is also [FeVII] emission. The OGLE-IV lightcurve shows longterm changes over 500--600 days that are most likely due to orbital motion. Continued photometric monitoring is necessary to confirm our preliminary orbital period of 585 days (see Fig. \ref{fig:oglefour2}).

\section{New D-type symbiotic stars}
\label{sec:dtypes}
\begin{figure*}
\begin{center}
\includegraphics[angle=270,scale=0.36]{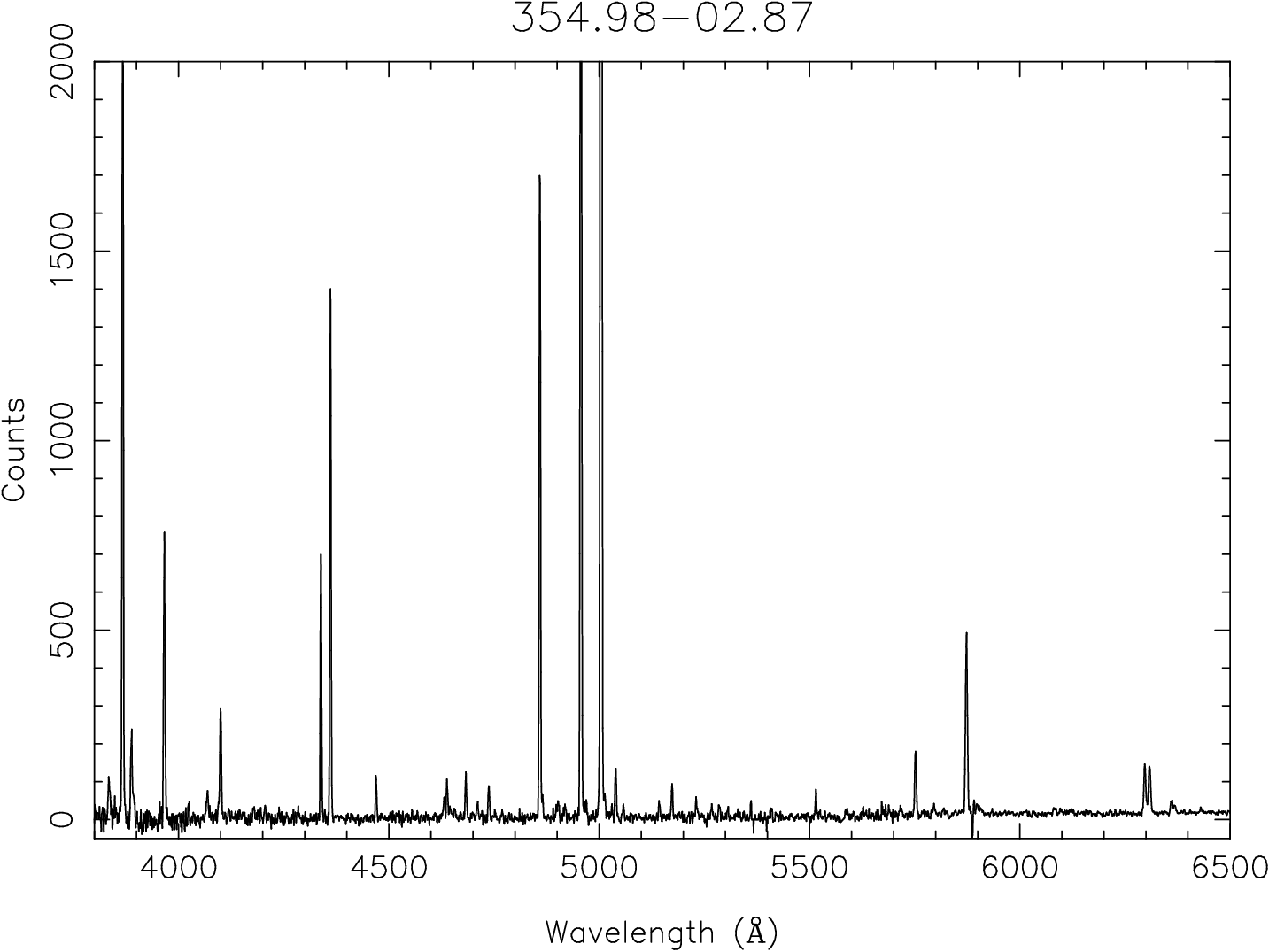}
\includegraphics[angle=270,scale=0.36]{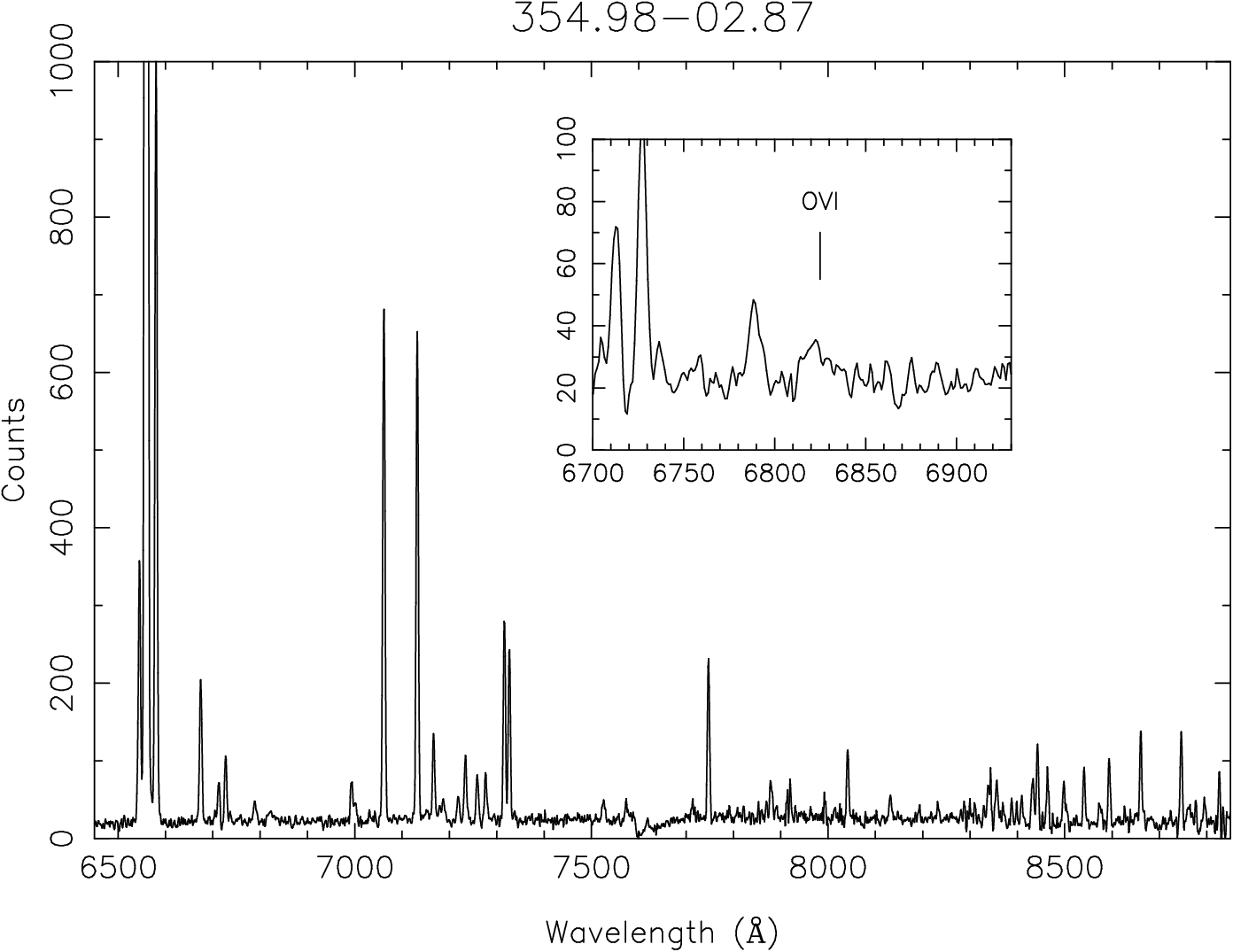}\\
\includegraphics[angle=270,scale=0.36]{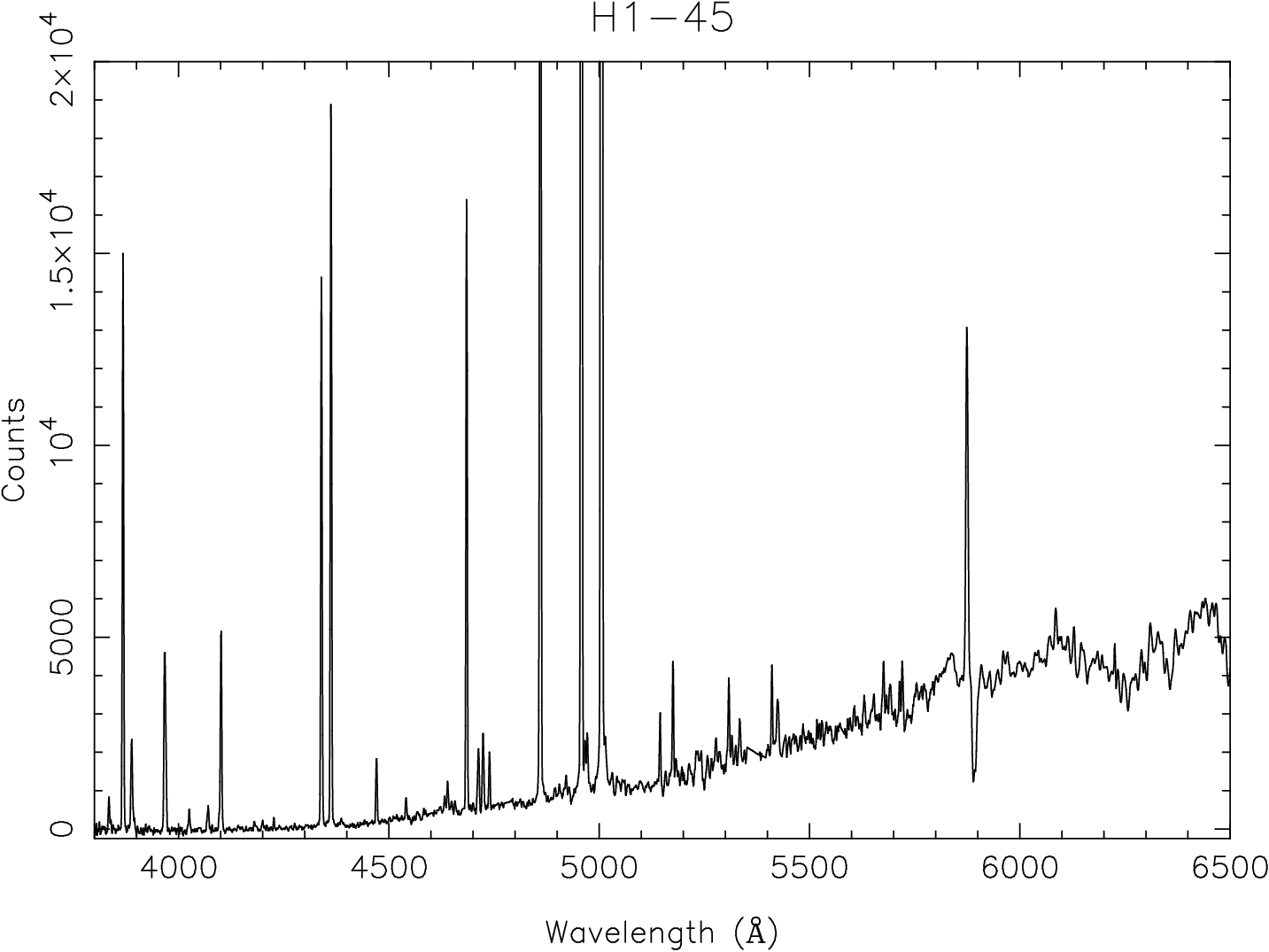}
\includegraphics[angle=270,scale=0.36]{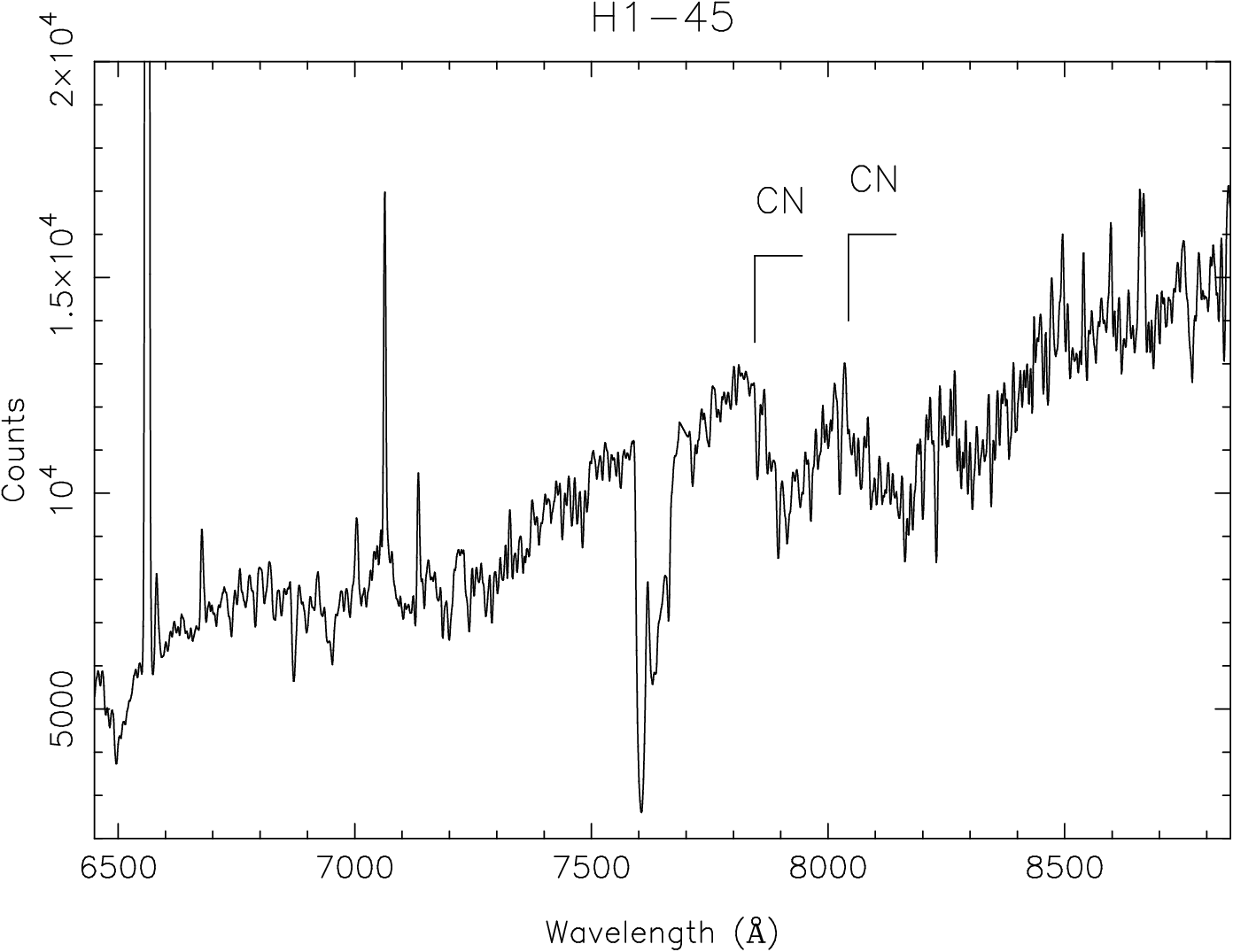}
\includegraphics[angle=270,scale=0.36]{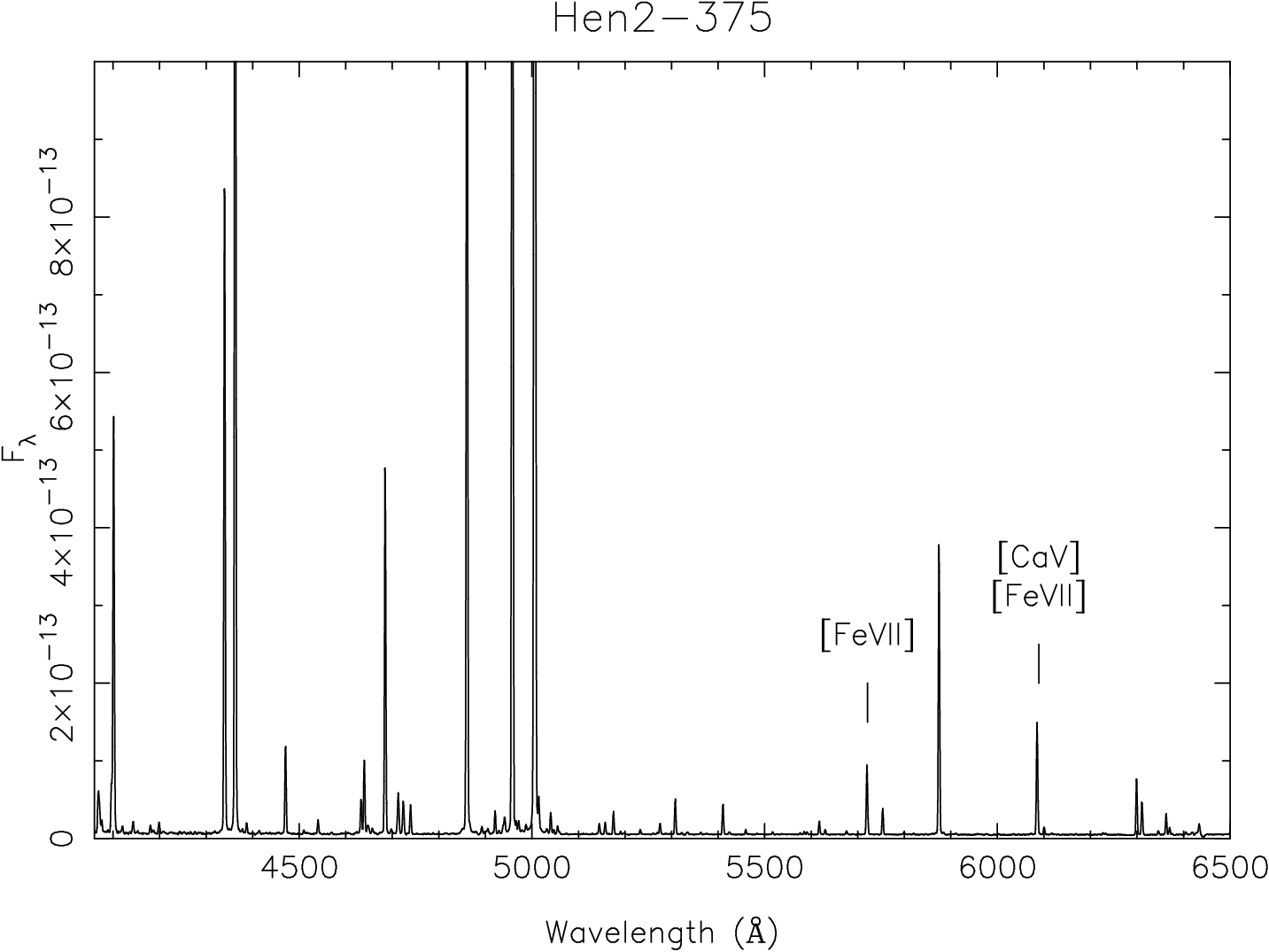}
\includegraphics[angle=270,scale=0.36]{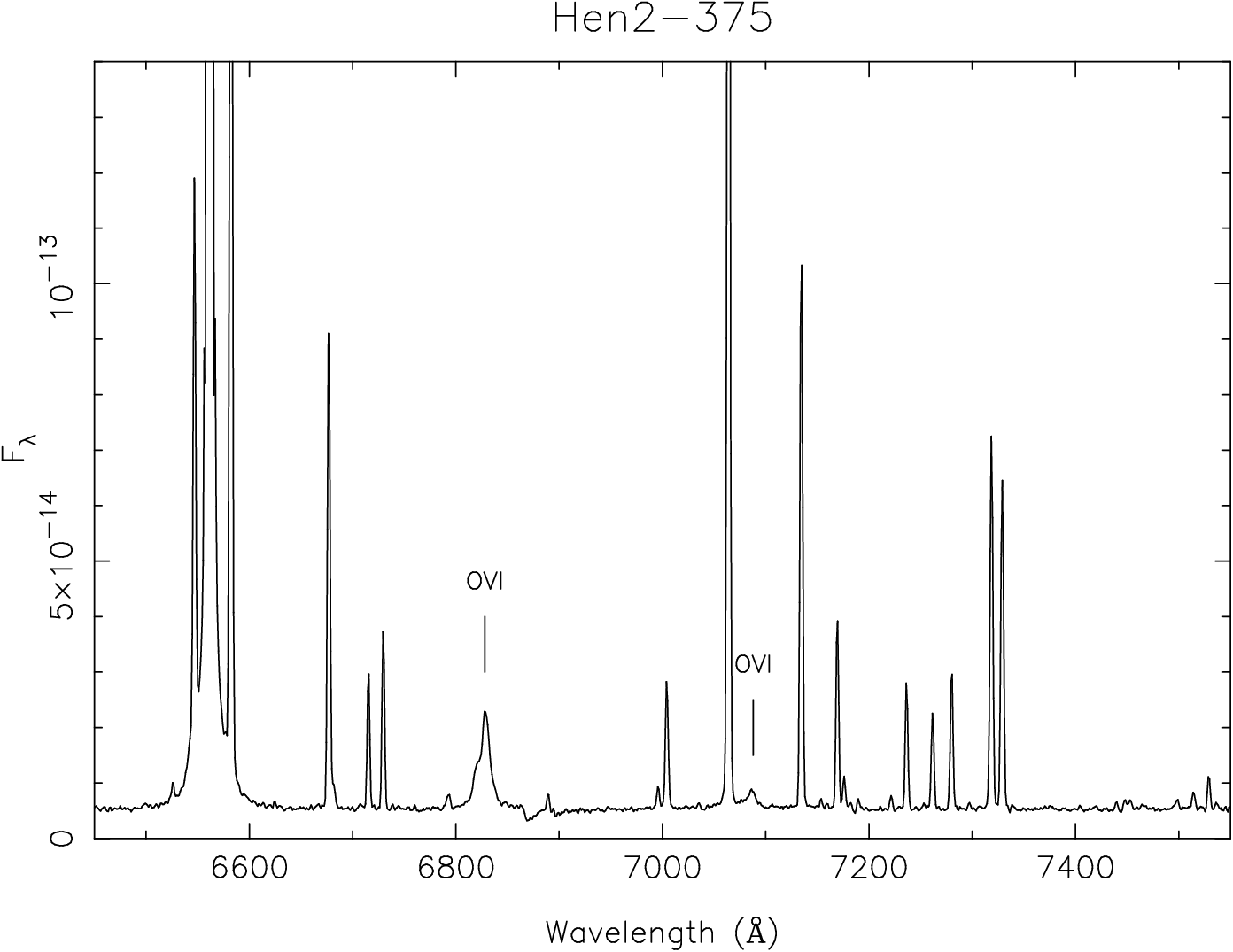}
\end{center}
\caption{AAOmega spectra of new D-type symbiotic stars.}
\label{fig:d1}
\end{figure*}

\begin{figure*}
\begin{center}
\includegraphics[angle=270,scale=0.36]{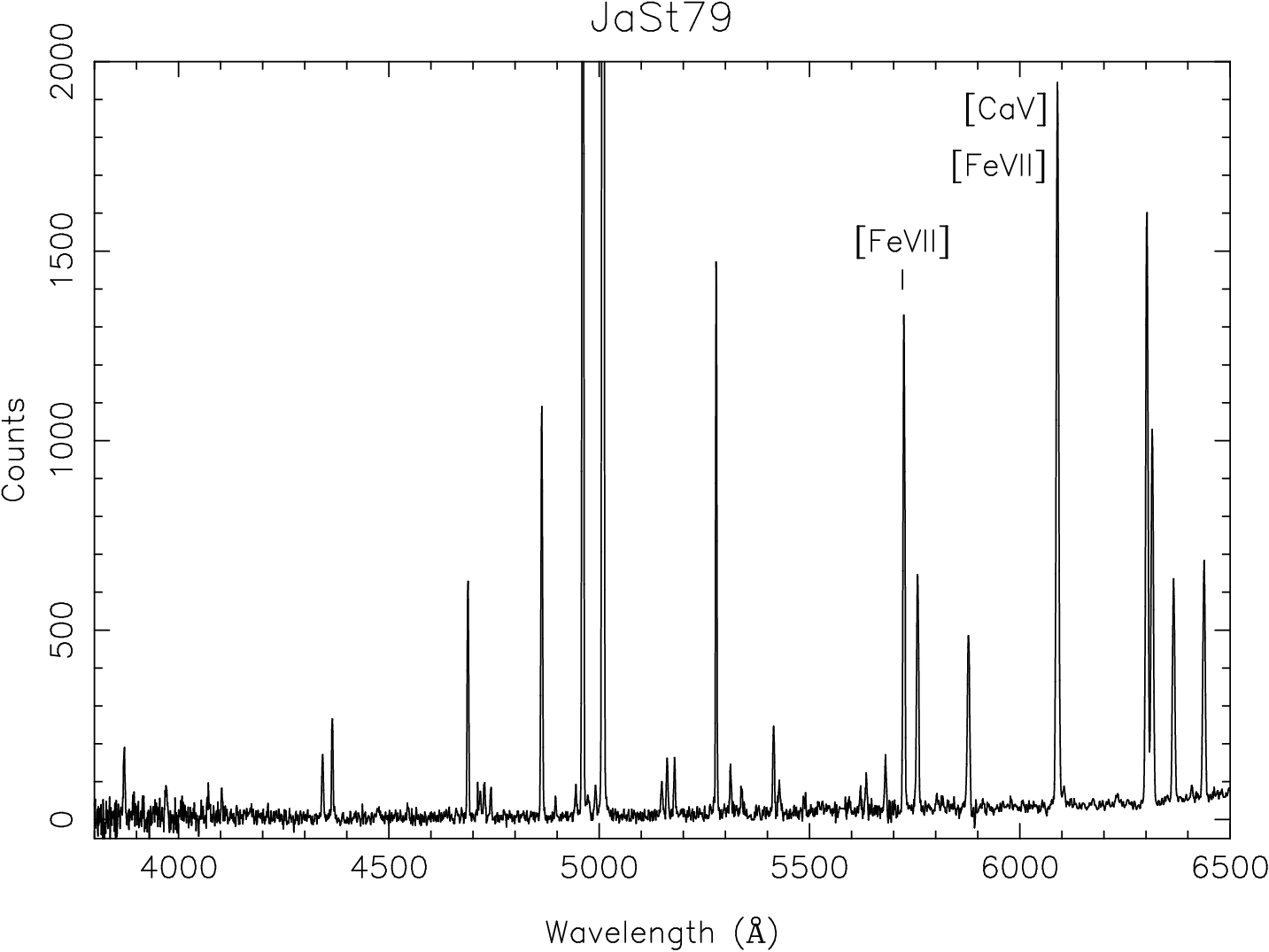}
\includegraphics[angle=270,scale=0.36]{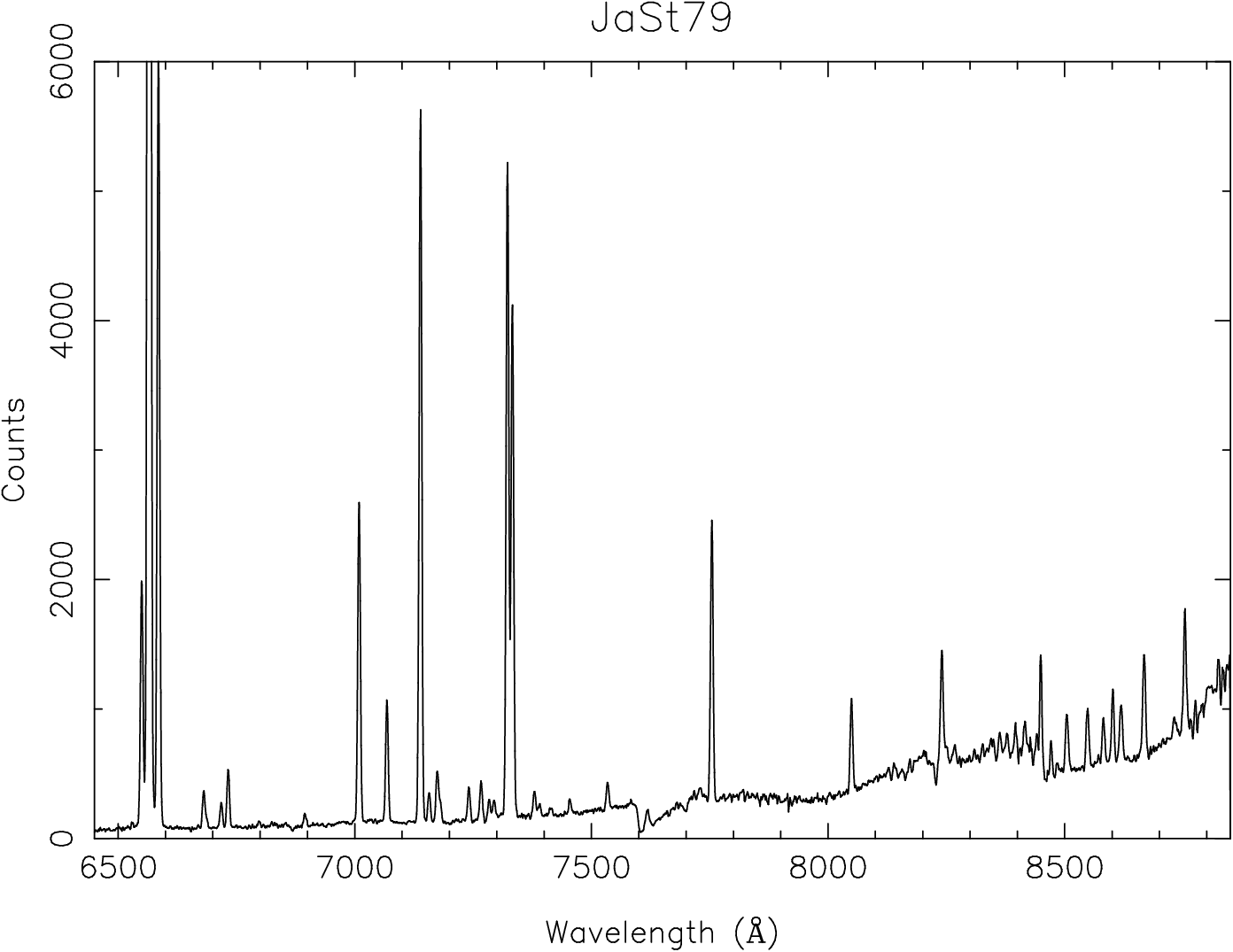}\\
\includegraphics[angle=270,scale=0.36]{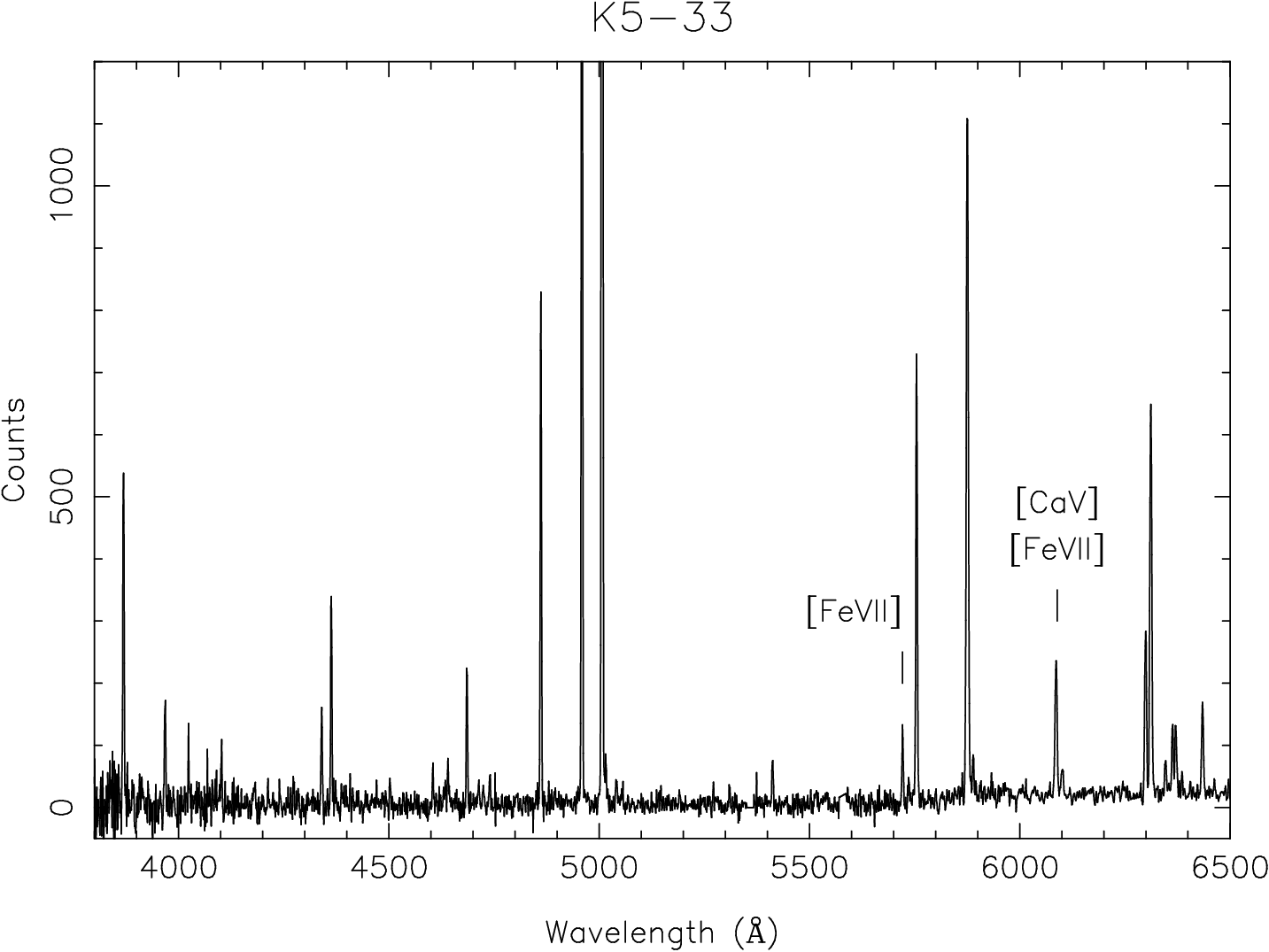}
\includegraphics[angle=270,scale=0.36]{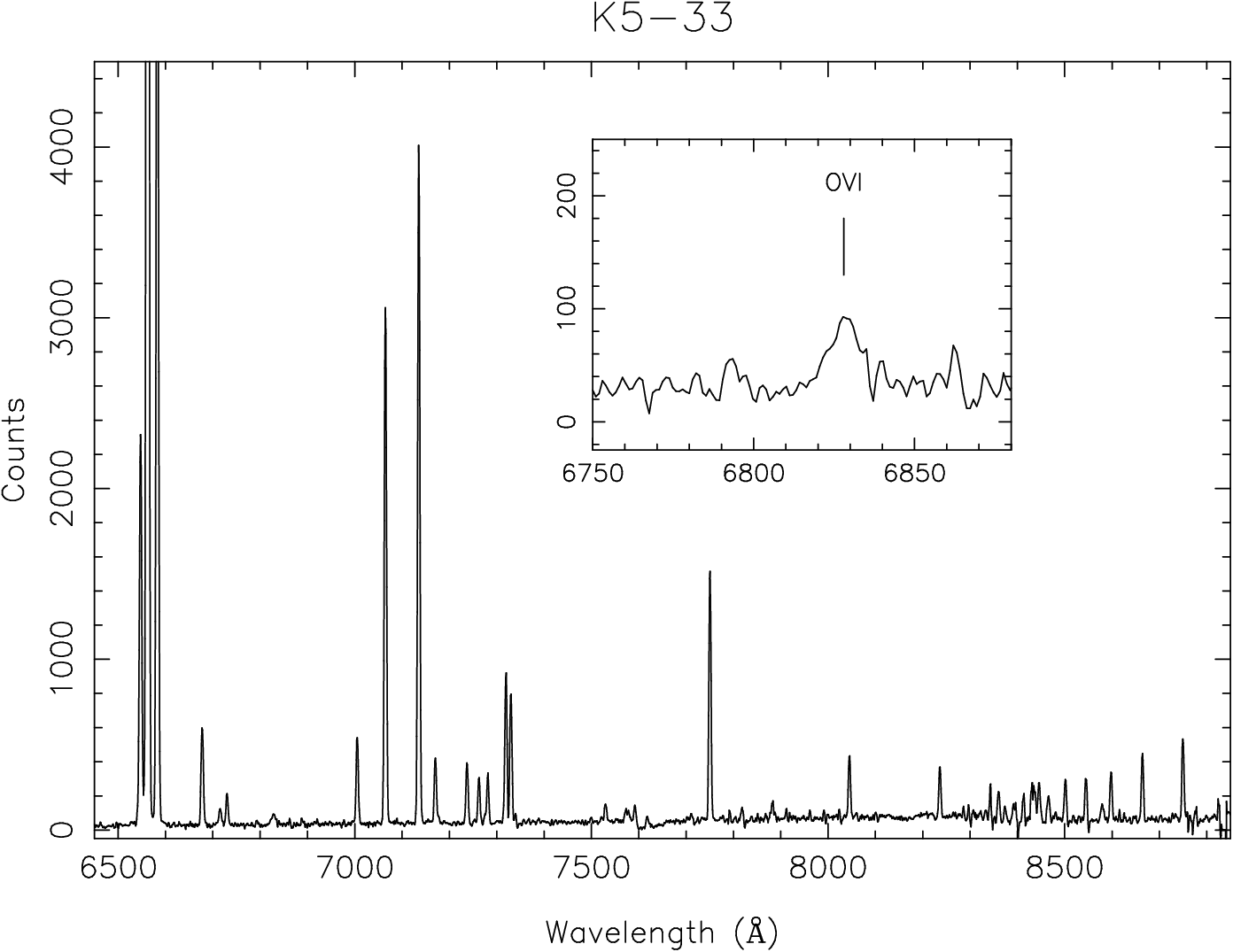}
\end{center}
\caption{AAOmega spectra of new D-type symbiotic stars (continued).}
\label{fig:d2}
\end{figure*}

Figures \ref{fig:d1} and \ref{fig:d2} present spectra of the new D-type symbiotic systems (except for JaSt2-6).

\subsection{354.98-02.87}
A typical D-type emission line spectrum complete with a weak detection of the Raman scattered OVI 6825 feature. The very red 2MASS colours agree well with the known symbiotic Miras.  The OGLE-IV lightcurve (Fig. \ref{fig:oglefour1}) is the redder of two sources in Fig. \ref{fig:ogleim1}, the one with $V-I=1.59$ (slightly to NE) compared to $V-I=0.281$ (slightly to SW).
As would be expected for Miras embedded in their dust shell, the OGLE lightcurve detects only small variations and the Mira is not observed in the AAOmega spectrum. 

\subsection{H1-45}
\label{sec:h145}
The AAOmega spectrum of H1-45 reveals strong emission lines typical of D-type symbiotic stars on top of a red continuum with strong CN bands and strong Ba~II $\lambda\lambda$ 4554, 4934 and 6495 absorption lines. These properties are characteristic of a cool carbon star enhanced in s-process elements for which we assign a tentative spectral type of C4--C7 according to the Richer (1971) classification scheme. Furthermore, H1-45 is remarkable because of the 408.6 day pulsation period uncovered by the OGLE lightcurve, making H1-45 only the fourth known Galactic carbon-rich symbiotic Mira and the first luminous carbon star with likely Galactic Bulge membership (Sect. \ref{sec:cmira}). The three other Galactic systems known are SS38 and AS 210, with pulsation periods of 463 d and 423 d, respectively (Gromadzki et al. 2009), and IPHAS J205836.43$+$503307.2, which is variable but does not yet have a determined pulsation period (Corradi et al. 2011). Another carbon-rich symbiotic in our sample may be ShWi5 (Sect. \ref{sec:shwi5}). There may also be some evidence for a dust obscuration event at the start of the OGLE lightcurve. 

We can estimate the distance to H1-45 using the period-luminosity relationship for carbon Miras (Whitelock et al. 2008). The 2MASS observations were taken at a pulsation phase of $\sim$0.79, implying the observed $K_s$ magnitude and colours may differ by $\sim$0.15 mag from their average values. The pulsation period of 408.6 days corresponds to an absolute magnitude of $M_K=-8.06\pm0.12$ (Whitelock et al. 2008), that when combined with the corrected 2MASS photometry, results in $A_K=0.78\pm0.2$ mag and $K_0=5.90\pm0.35$ mag. This results in a distance of $d=6.2\pm1.4$ kpc suggesting either an intervening Galactic Disk or near-side Galactic Bulge membership. We discuss this crucial point further in Sect. \ref{sec:cmira}.

H1-45 was also observed by the deep H$\alpha$, Sloan $r'$ and $i'$ imaging survey taken with the CTIO Blanco MOSAIC II instrument (Jonker et al. 2011). We retrieved these images from the NOAO archive\footnote{http://portal-nvo.noao.edu} and created a quotient image H$\alpha$/$r'$ to search for an extended H$\alpha$ nebula. Despite the inner core (2--3\arcsec) being saturated in both H$\alpha$ and $r'$ images, the image quality is sufficient to rule out the presence of a larger extended nebula, in contrast to IPHAS J205836.43$+$503307.2 which appears to have a nebula (Corradi et al. 2011). 

\subsection{Hen2-375}
\begin{figure}
   \begin{center}
      \includegraphics[scale=0.3]{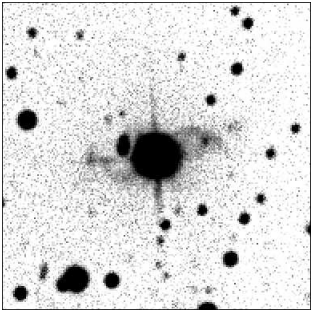}
   \end{center}
   \caption{SAAO 1.0-m [OIII] image of Hen2-375 with an unsharp mask applied. The image is $120\times120$ arcsec$^2$ with North up and East to left. The vertically elongated source NE of Hen2-375 is a filter artefact.}
   \label{fig:Hen2-375}
\end{figure}

The SAAO 1.9-m spectrum clearly proves Hen2-375 to be a D-type symbiotic star with both OVI Raman features detected amongst a rich emission line spectrum. A giant companion is not detected in our spectroscopy and we found no photometric variations in the ASAS lightcurve (Pojmanski 2002). In the colour-composite image of Hen2-375 made from SSS $I_N$ (red), $R_F$ (green) and $B_J$ (blue) images (Hambly et al. 2001) we noticed an extended blue nebula (Fig. \ref{fig:c1}). To check whether this was a photographic flaw or not, we performed imaging in a variety of filters on the SAAO 1.0-m telescope. The STE4 SAAO CCD was used with $2\times2$ binning to give a pixel scale of 0.62\arcsec\ pixel$^{-1}$. Figure \ref{fig:Hen2-375} shows the nebulosity that was strongest in the [OIII] filter ($\lambda_\mathrm{cen}=503.2$ nm, FWHM=21 nm). Only a marginal detection in an H$\alpha$ filter was recorded, consistent with previous findings that bipolar nebulae around symbiotic stars are very strong in [OIII] emission (e.g. Solf 1984). Balmer line emission in such systems is strongly dependent on the density contrast of the nebula which in the polar direction is an order of magnitude less than in the orbital plane. The bipolar nebula measures 48\arcsec\ tip-to-tip and appears to be point-symmetric (i.e. $S$-shaped). The SAAO 1.9-m spectrum was oriented at PA=90, however it does not show any velocity structure in the extended [OIII] emission. No spatially resolved [NII] emission lines from the nebula were detected in our spectrum, making it rather unusual amongst symbiotic nebulae that often show strong [NII] emission (e.g. Solf 1984). A Balmer decrement of H$\alpha$/H$\beta$=4.1 was measured from the shorter exposures which corresponds to $E(B-V)=0.33$ mag under Case B nebular conditions. This is a reasonable assumption since the line emission regions of D-types are usually located outside the thick dust shell and are only affected by interstellar reddening (e.g. Miko{\l}ajewska et al. 1997, 1999).

\subsection{JaSt2-6}
A pulsation period of 599.3 days was measured from OGLE-II data by Matsunaga et al. (2005). With now more OGLE observations we have measured a similar 605.4 days. Surrounding the object is resolved H$\alpha$ nebulosity, explaining its inclusion in the PN catalogue of Jacoby \& Van de Steene (2004). Even though we lack deep spectroscopy of this system, these two facts combined allow us to confidently classify JaSt2-6 as a symbiotic Mira. It is also detected by the Rosat All-Sky Survey Faint Source Catalogue (see Voges et al. 2000; Sect. \ref{sec:xray}).

\subsection{JaSt79}
Discovered by Van de Steene \& Jacoby (2001) who remarked JaSt79 as a point source and also studied by Jacoby \& Van de Steene (2004). The OGLE lightcurve shows a large $I$-band amplitude ($>2$ mag) and a period of 424.8 days, that when combined with the rich emission line spectrum taken by 2dF/AAOmega, clearly classifies JaSt79 as a symbiotic Mira. The Mira is detected in the AAOmega spectrum and does not appear to be carbon rich. We estimate a late M type based on the very strong TiO 8432 band, but the precise spectral type of the Mira is difficult to quantify because of the emission lines. As done for H1-45, we can also estimate the distance to JaSt79 using the period-luminosity relation for Galactic O-rich Miras (Whitelock et al. 2008). The 424.8 day period corresponds to $M_K=-8.12\pm0.12$ mag and the 2MASS photometry was taken at a pulsation phase of $\sim$0.36, requiring a correction of $\sim$0.15 mag to the $K_s$ magnitude and colours. This yields $A_K=1.1\pm0.2$ mag and $K_0=6.00\pm0.35$ mag, to give a distance of $6.7\pm1.6$ kpc. Slower variations in the OGLE lightcurve are probably due to dust obscuration (see e.g. Gromadzki et al. 2009) and JaSt79 may also be an X-ray source (see Hong et al. 2009; Sect. \ref{sec:xray}). 

\subsection{K5-33}
PN candidate PBOZ10 (Pottasch et al. 1988) and later rediscovered by Kohoutek (2002) as K5-33.
It is however a clear D-type symbiotic with the Raman scattered OVI 6825 feature detected and a typical D-type emission line pattern. As in 354.98-02.87, the OGLE-IV lightcurve (Fig. \ref{fig:oglefour2}) shows minimal variability, consistent with the D-type classification having an obscured Mira. The source identified in the OGLE image (Fig. \ref{fig:ogleim2}) is the redder of the two in $V-I$.

\section{New D$'$-type symbiotic stars}
\label{sec:ddtypes}

\subsection{ShWi5}
\label{sec:shwi5}
\begin{figure*}
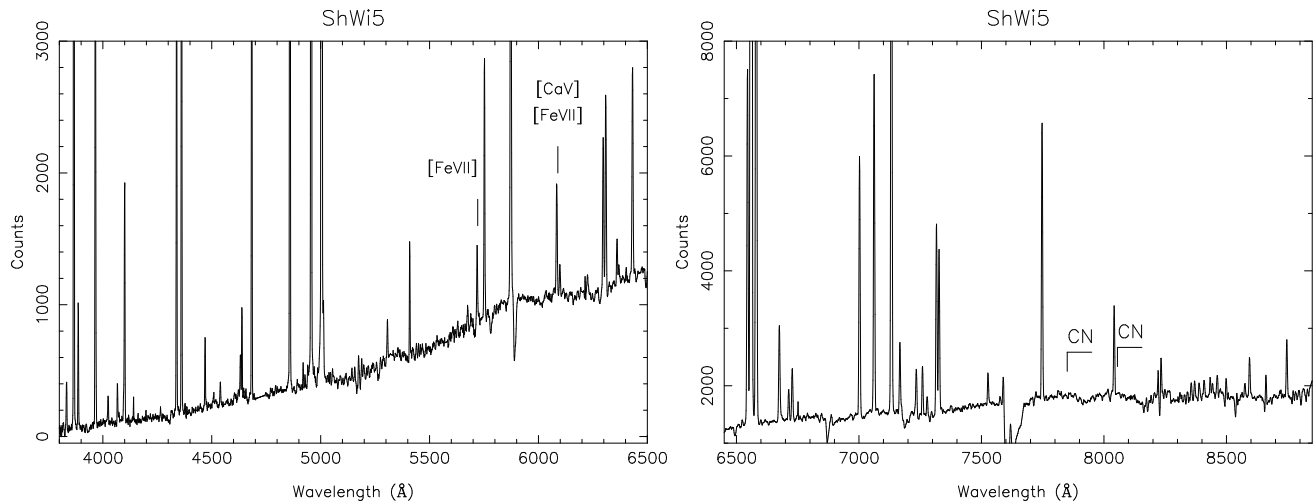

\begin{center}
\includegraphics[angle=270,scale=0.36]{syfigs/ShWi5b.ps}
\includegraphics[angle=270,scale=0.36]{syfigs/ShWi5r.ps}
\end{center}
\caption{AAOmega spectrum of the new D$'$-type symbiotic star ShWi5.}
\label{fig:dd1}
\end{figure*}

Discovered by Shaw \& Wirth (1985), ShWi5 shows a rich emission line spectrum on top of a red stellar continuum (Fig. \ref{fig:dd1}). The 2MASS colours are consistent with a D$'$-type symbiotic and this classification is supported by the warm carbon-rich giant detected in the AAOmega spectrum (equivalent to a late G-type). There are no TiO bands, but the CH band and weak CN bands are present, typical of other D$'$-types (M\"urset \& Schmid 1999) that often have solar metallicities and are overabundant in s-process elements (e.g. Pereira et al. 2005 and ref. therein). The continuum is complemented by [FeVII] emission and an emission line pattern typical of D$'$-type symbiotics. The lightcurve is relatively flat as is the case for most D$'$-type symbiotics (Gromadzki et al. in preparation), however there may be some low-amplitude variability that could be ascribed to orbital motion. 

\section{Possible S-type symbiotic stars}
\label{sec:pstypes}
The spectra of possible S-types are presented in Figures \ref{fig:ps1} and \ref{fig:ps2}.
\subsection{PPA1752-3542}
The spectrum published online by Parker et al. (2006) displays [O~III] $\lambda$4363/H$\gamma$ $>$ 1, H$\alpha$, H$\beta$ and [O~III] $\lambda$5007,4959. Also present in the spectrum is an unrelated nearby late-type star that falls in the large 6.7\arcsec\ 6dF fibre diameter. The OGLE lightcurve resembles an S-type symbiotic star, but the system is too faint to be detected in 2MASS. Given the lack of better observations we leave this as a probable system that requires more observations to check our suspected S-type classification.

\section{Possible D-type symbiotic stars}
\label{sec:pdtypes}
The spectra of possible D-types are presented in Figures \ref{fig:pd1}, \ref{fig:pd2} and \ref{fig:pd3}.

\subsection{357.12+01.66 and PHR1751-3349}
Both objects have NIR colours consistent with D-type symbiotics, however they lack the high ionisation emission lines that would be required for a definite symbiotic star status. We note that two well-studied symbiotic novae, RX Pup and V407 Cyg (Miko{\l}ajewska et al. 1999; Munari et al. 2011; Shore et al. 2011, 2012), both associated with D-type symbiotics, only show high ionisation lines during their nova outburst. Otherwise their optical spectra only show low ionisation emission lines (e.g. HI, ionized metals) and they may mimic B[e] stars (e.g. Lamers et al. 1998). Spectra of PHR1751-3349 were taken during the OGLE lightcurve minimum and the lightcurve shows a large-amplitude slow variation typical of some D-type symbiotic stars. PHR1751-3349 also has a small bipolar nebula that measures 13.5\arcsec\ tip-to-tip (0.5 pc at 8 kpc).

\subsection{Al2-B}
The broad wings on the [O~III] $\lambda$4959, 5007 and H$\alpha$ emission lines are not physical and are caused by inter-fibre contamination by a brighter PN observed in an adjacent fibre on the spectrograph pseudo-slit. A weak stellar continuum is also imprinted on the spectrum by the adjacent spectrum. Fainter features are however real, e.g. the $\lambda$4363/H$\gamma$ ratio. The possible D-type classification is based mostly on the emission line spectrum since the 2MASS detection is too faint to allow for a meaningful classification. 

\subsection{Al2-G}
The AAOmega spectrum shows an emission line pattern similar to a D-type symbiotic. The most notable features being [O~III] $\lambda$4363 $>$ H$\gamma$ and strong [ArIV] and [ArV] emission lines. There may be a possible detection of [Fe~VII] emission lines, but this requires further confirmation. During the OGLE-IV lightcurve Al2-G slowly brightened by $\sim$0.1 mag, consistent with a slowly variable D-type system, which is also hinted at by the weak red stellar continuum in the AAOmega spectrum. As for Al2-B, the source is at the limit of what 2MASS can detect, so the colours are not meaningful in this case.

\subsection{M2-24}
Zhang \& Liu (2003) presented spectroscopy of this very red source ($J-K_s=3.61$ mag) which resembles other D-types in our sample with a highly ionized emission line spectrum. Some peculiar features were detected including a high [O~III] $\lambda$4363/H$\gamma$ ratio and the presence of Mg~II. The OGLE finder chart reveals a bipolar nebula and although a slowly brightening trend is seen in the OGLE lightcurve, presumably due to dust, we lack evidence for a solid D-type classification. 

\subsection{PHR1803-2746}
The emission lines detected by AAOmega are typical of a D-type symbiotic star but the stellar component in the spectrum is spurious, coming from the nearby unrelated star. The OGLE-III $I$-band image shows the fibre diameter which in typical seeing would bring in light from this star. 

\subsection{PPA1746-3454}
The OGLE lightcurve is similar to PHR1751-3349 and resembles that of some D-type symbiotic stars. The red 2MASS colours and AAOmega spectrum also resemble D-type symbiotics, but we lack the critical features necessary to secure a D-type classification.

\subsection{PPA1807-3158}
The shallow spectrum published online by Parker et al. (2006) shows H$\alpha$, [N~II] $\lambda$6548, 6583 as well as weak He~I $\lambda$6678, [O~I] $\lambda$6300,6363 and [S~III] $\lambda$6312. The red 2MASS colours and OGLE lightcurve are typical of D-type symbiotics, but as in the case of PPA1746-3454 we lack the critical features to secure a D-type classification.

\section{Planetary nebulae with high density cores}
\label{sec:egb6}
At least five objects in our sample demonstrate nuclei or `cores' that have properties more consistent with a symbiotic star, i.e. [FeVII]/[CaV] emission lines (requiring a dense environment with a hot ionising source) and/or emission lines indicating densities $n_e\ga$10$^6$--10$^7$ cm$^{-3}$, but crucially do not show any evidence for a red giant companion or the Raman-scattered OVI emission bands. It is expected that a high mass loss rate wind from a red giant is necessary to create the high densities found in symbiotic stars, so it is surprising that no red giants are present. Another source of high density material may be dust remaining from the AGB phase of the WD. Except for M2-29 (that we previously studied in Miszalski et al. 2011a), these objects are discussed in the following subsections after we summarise the few extant examples known. 

The first such example to be studied was the central star of the evolved PN EGB6 (Ellis, Grayson \& Bond 1984; Liebert et al. 1989; Fulbright \& Liebert 1993; Bond 2009; Su et al. 2011). Since then only a few other examples have been identified including M2-29 (Torres-Peimbert et al. 1997; Miszalski et al. 2011a), Abell 57 and PHR 1553-5738 (Miszalski et al. 2011c), and PHR1641-5302 (Parker \& Morgan 2003; Frew \& Parker 2010). Such peculiar central stars of otherwise normal PNe are now referred to as EGB6-like (Frew \& Parker 2010), though this may be a diverse group so in this work we call them high density core (HDC) PNe. An excellent example of this diversity is Hen2-428 for which Rodr\'iguez et al. (2001) found a dense core ($n_e>10^{10}$ cm$^{-3}$), but which was subsequently found to be a double-degenerate close binary central star (possibly a contact binary) with an orbital period of 0.18 days (Santander-Garc\'ia et al. 2011).

\begin{figure*}
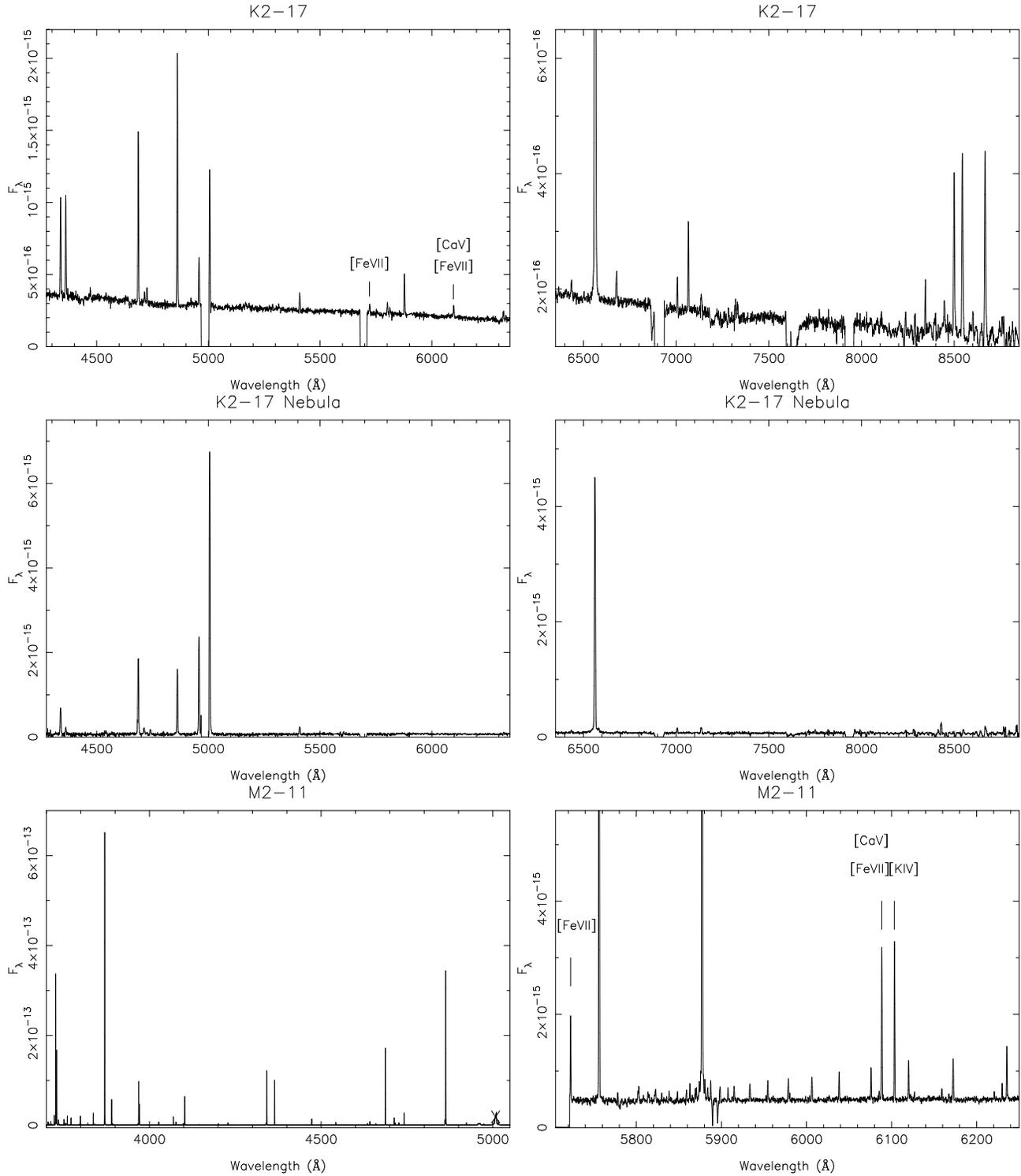

\begin{center}
\includegraphics[angle=270,scale=0.36]{syfigs/K2-17b.ps}
\includegraphics[angle=270,scale=0.36]{syfigs/K2-17r.ps}\\
\includegraphics[angle=270,scale=0.36]{syfigs/K2-17Nebulab.ps}
\includegraphics[angle=270,scale=0.36]{syfigs/K2-17Nebular.ps}\\
\includegraphics[angle=270,scale=0.36]{syfigs/M2-11b.ps}
\includegraphics[angle=270,scale=0.36]{syfigs/M2-11r.ps}\\
\end{center}
\caption{Spectra of high density core (HDC) PNe.}
\label{fig:egb61}
\end{figure*}

\begin{figure*}
\begin{center}
\includegraphics[angle=270,scale=0.36]{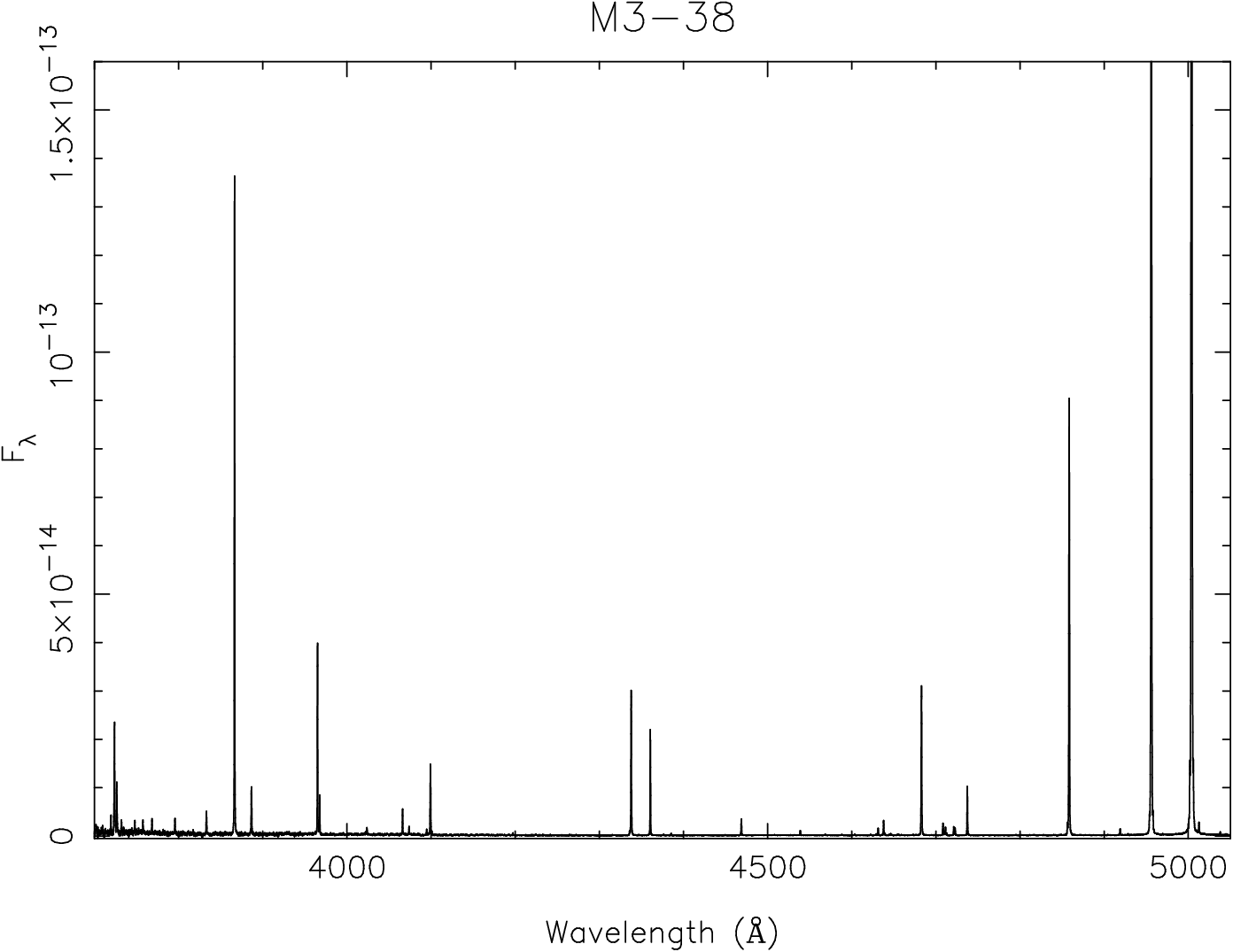}
\includegraphics[angle=270,scale=0.36]{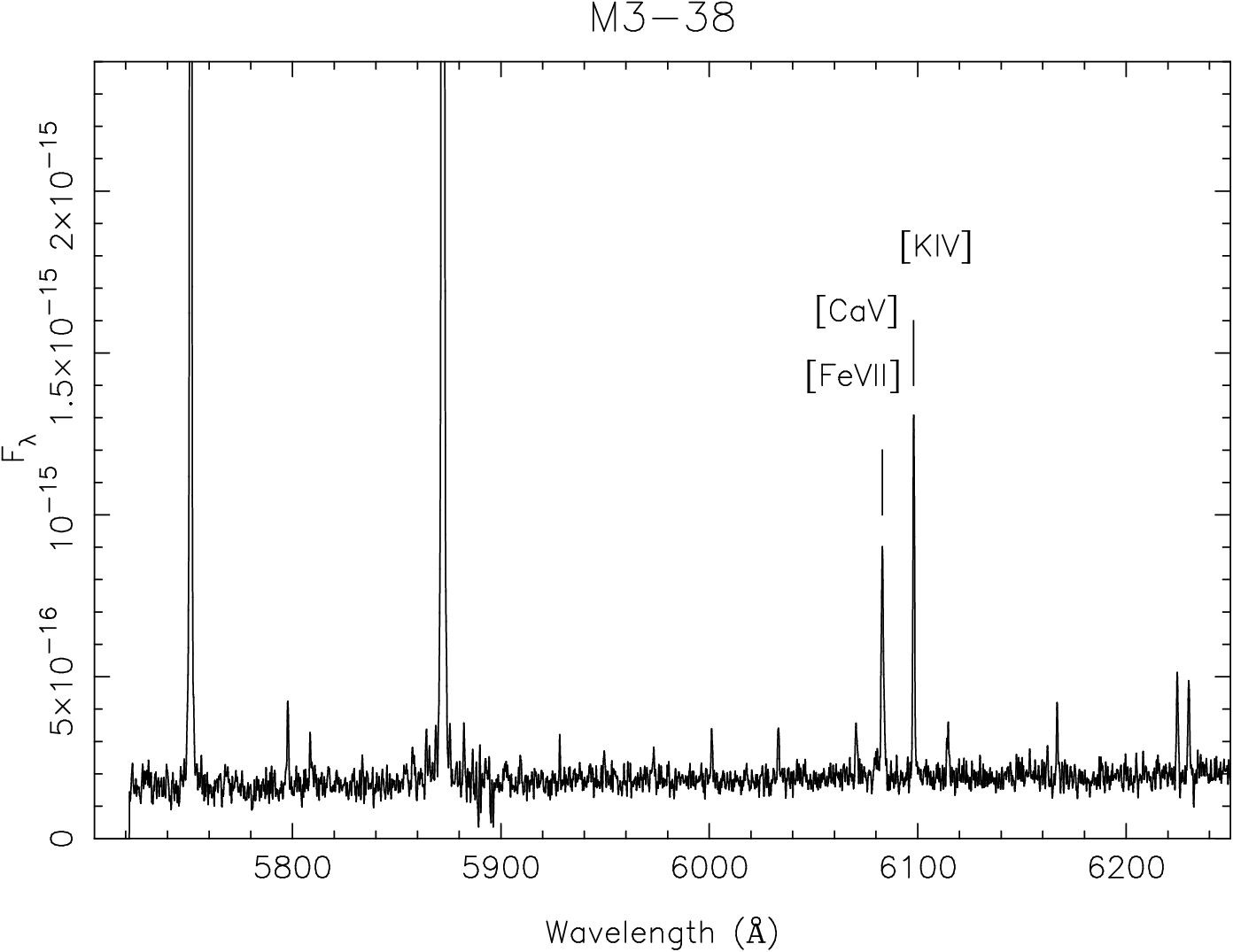}\\
\includegraphics[angle=270,scale=0.36]{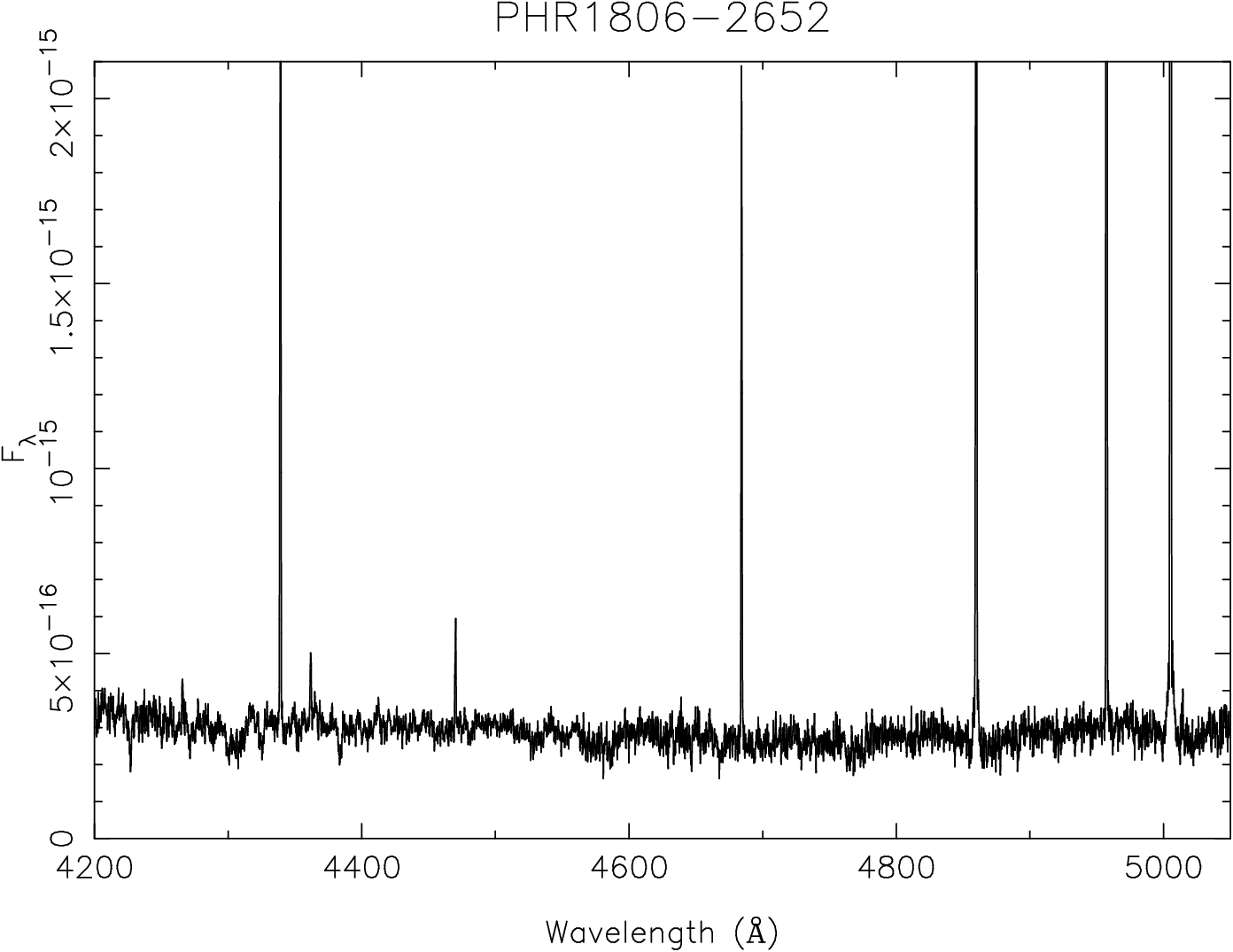}
\includegraphics[angle=270,scale=0.36]{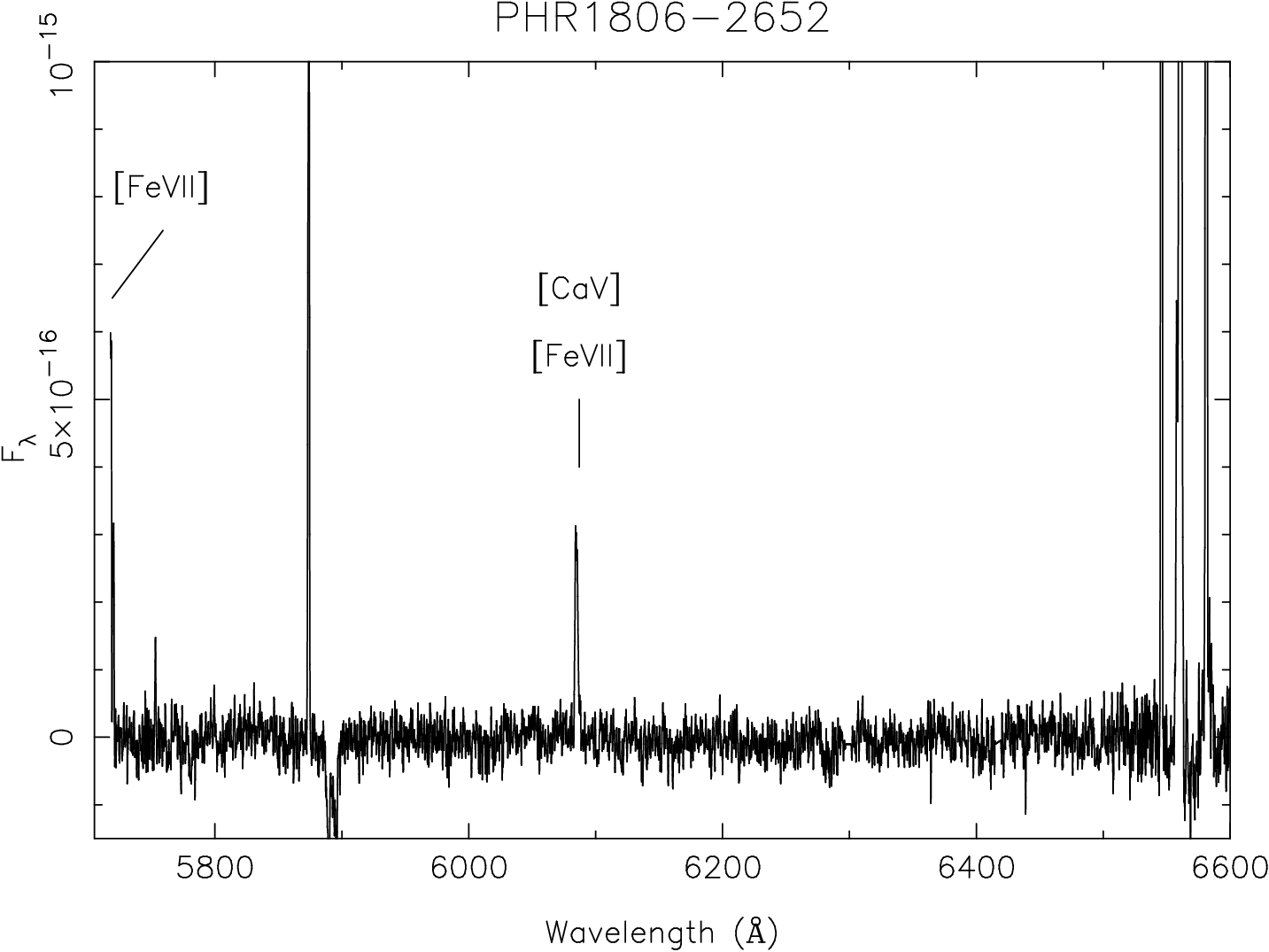}\\
\includegraphics[angle=270,scale=0.36]{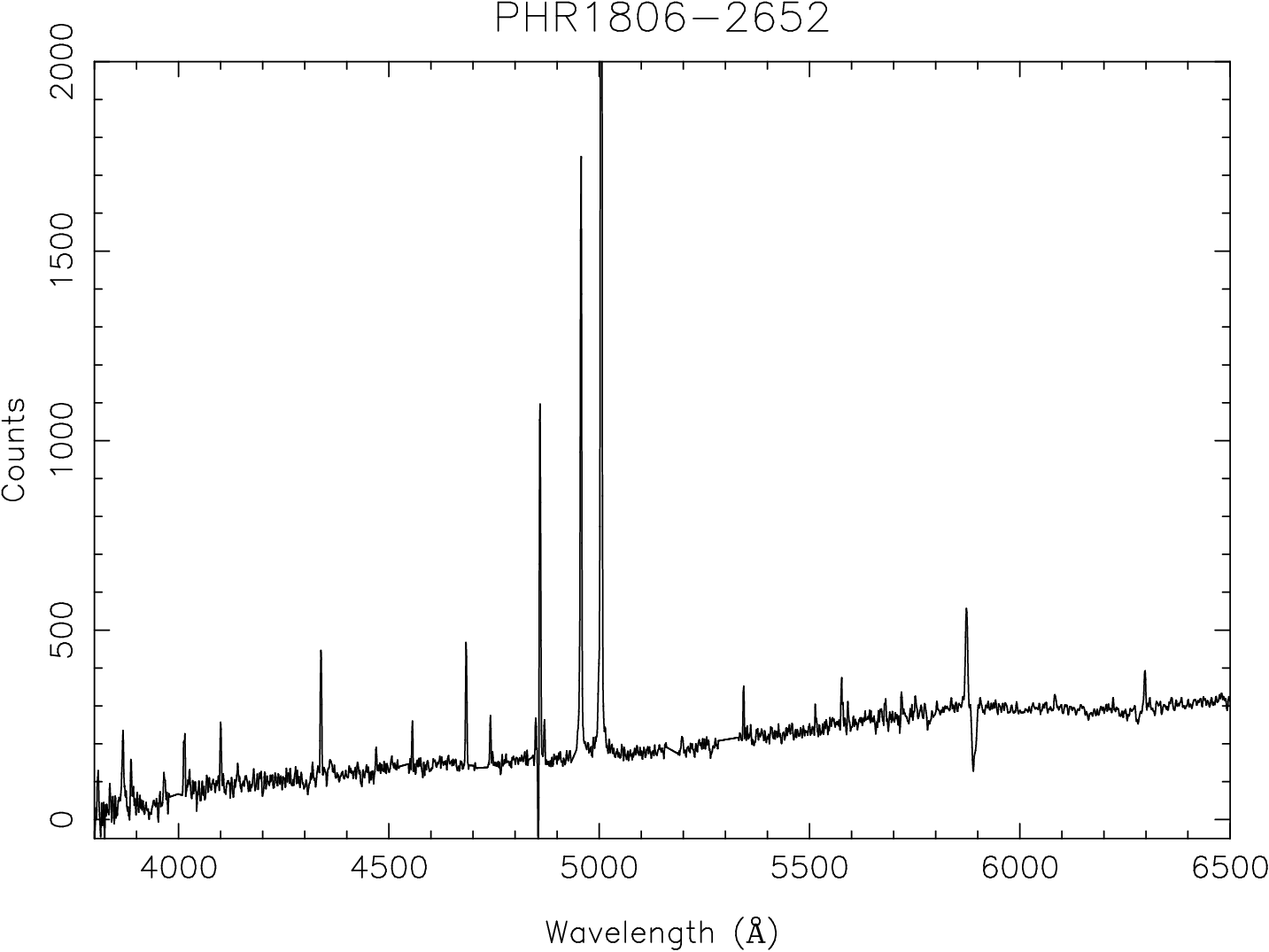}
\includegraphics[angle=270,scale=0.36]{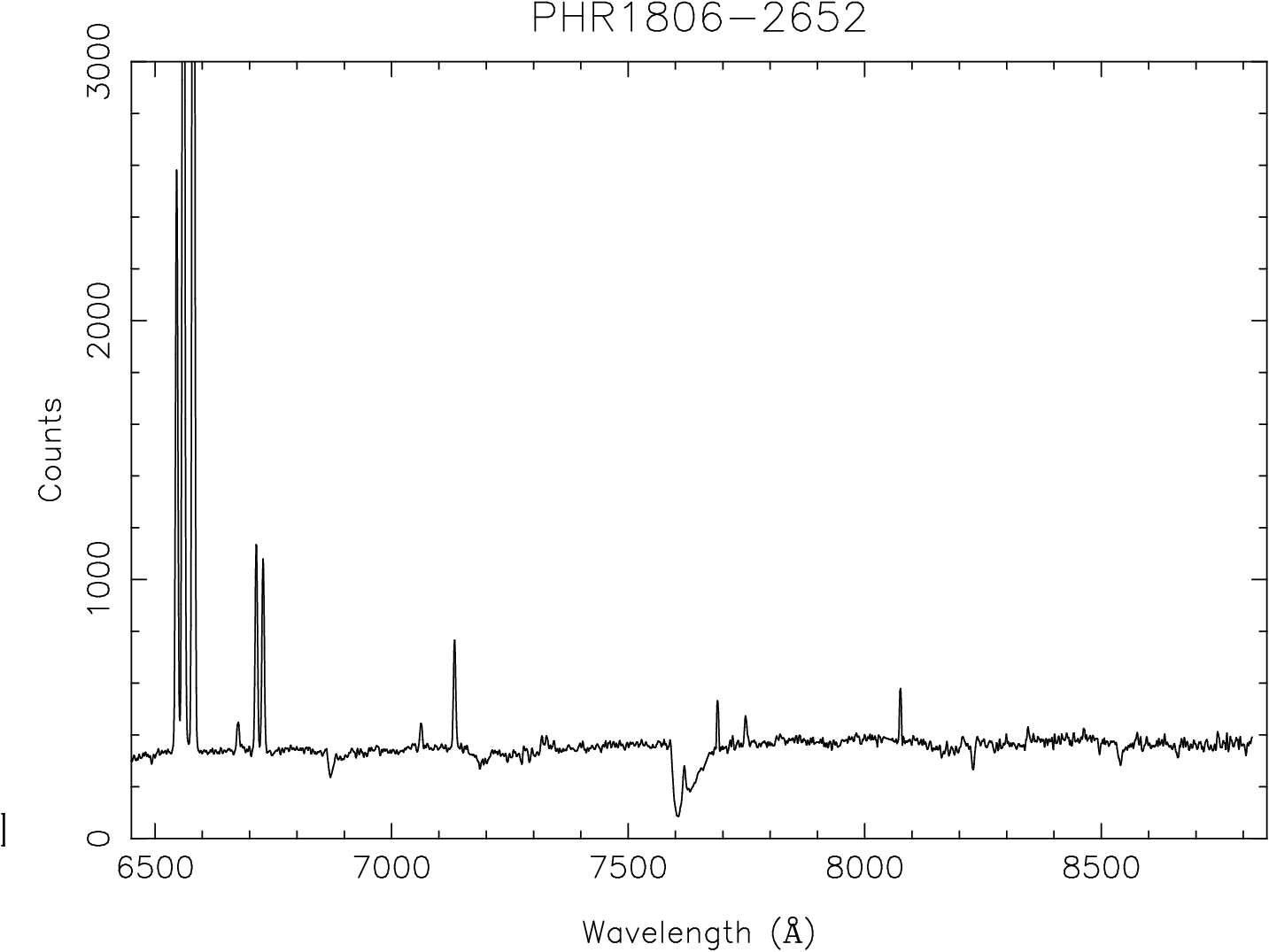}\\
\end{center}
\caption{Spectra of high density core (HDC) PNe (continued).}
\label{fig:egb62}
\end{figure*}

Figures \ref{fig:egb61} and \ref{fig:egb62} present spectroscopy of the five HDC PNe in our sample. Members of this class are suspected to share a similar binary configuration to EGB6, i.e. to have an M dwarf surrounded by a dense emission nebula (Bond 2009) located in a wide orbit ($\sim$100 AU) around the WD. Improving our understanding of dust obscuration events in HDC PNe may help us understand the origin of long secondary periods in some AGB stars (see e.g. Nicholls et al. 2009).

A dust disk is also anticipated to be present but its exact location is not known (Su et al. 2011). The only other HDC PN to show evidence for a dust disk is M2-29 which experienced a remarkable dust obscuration event in its lightcurve (Hajduk et al. 2008; Gesicki et al. 2010; Miszalski et al. 2011a). Miszalski et al. (2011a) applied a simple R Coronae Borealis star model (Goeres \& Sedlmayr 1992) to the M2-29 lightcurve to determine that the dust responsible for these events forms at $\sim$70 AU from the WD. This separation is comparable to the distance of the M-dwarf companion from the WD in EGB6. One possible explanation is that there is a dust disk around the M dwarf companion rather than around the WD (e.g. Su et al. 2011). Another possibility is that the dust, leftover from the AGB phase in a surrounding circumbinary disk, is formed under the disturbing influence (wake) of the M dwarf companion. A similar pinwheel like effect is seen in AGB stars with companions (Maercker et al. 2012) and massive Wolf-Rayet stars (e.g. Tuthill et al. 1999).

\subsection{K2-17}
K2-17 is a bipolar PN measuring $35\times44$\arcsec\ across with lobes visible to the East and West (Fig. \ref{fig:c1}). It resembles somewhat Abell 57 which also has a HDC (Miszalski et al. 2011c) and NGC2371-2. The central star shows a clear H$\alpha$ excess in the SHS data and is located in the centre of an apparent torus which is brighter on the SE side. Fig. \ref{fig:egb62} shows the typical high-excitation PN nebular spectrum (HeII $\lambda$4686/H$\beta$=1.2) which has a Balmer decrement of H$\alpha$/H$\beta$=4.15 corresponding to $E(B-V)=0.50$ mag. The nucleus notably shows a blue continuum with a S/N of $\sim$35 at $\lambda$5500\AA\ and $\sim$45 at $\lambda$6800\AA, the Ca II triplet in emission ($n_e\ga10^{10}$ cm$^{-3}$), [Fe~VII] $\lambda$5721 and [Ca~V]/[Fe~VII] $\lambda$6087, C~IV $\lambda\lambda$5801,5812 and [ArV] $\lambda\lambda$6435, 7005. We note that the [Fe~VII] emission is substantially weaker than seen in the symbiotic stars of our sample. The spectrum somewhat resembles that of Hen~2-428 (Rodr\'iguez et al. 2001), however it is unlikely to be a close binary as unpublished $I$-band photometry with the SAAO 1.0-m telescope found minimal variability. The asymmetric bipolar morphology and unusual spectrum strongly suggest a main-sequence companion in a wide-orbit is present. The 2MASS colours are rather red, suggesting the presence of warm dust as in some other HDC PNe, but this requires confirmation as the detection is near the limit for 2MASS.

\subsection{M2-11 and M3-38}
\begin{figure*}
   \begin{center}
      \includegraphics[scale=0.444]{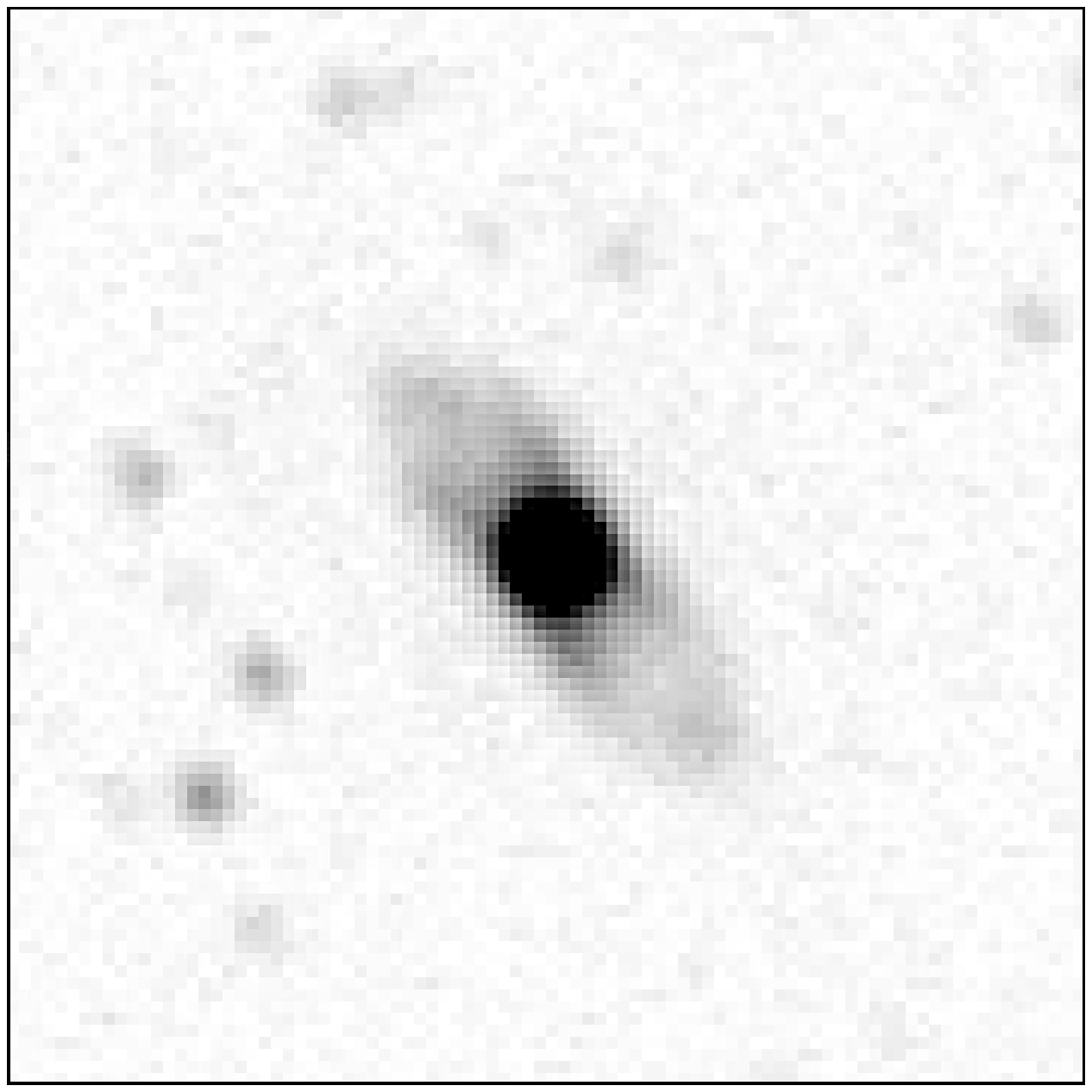}
      \includegraphics[scale=0.4]{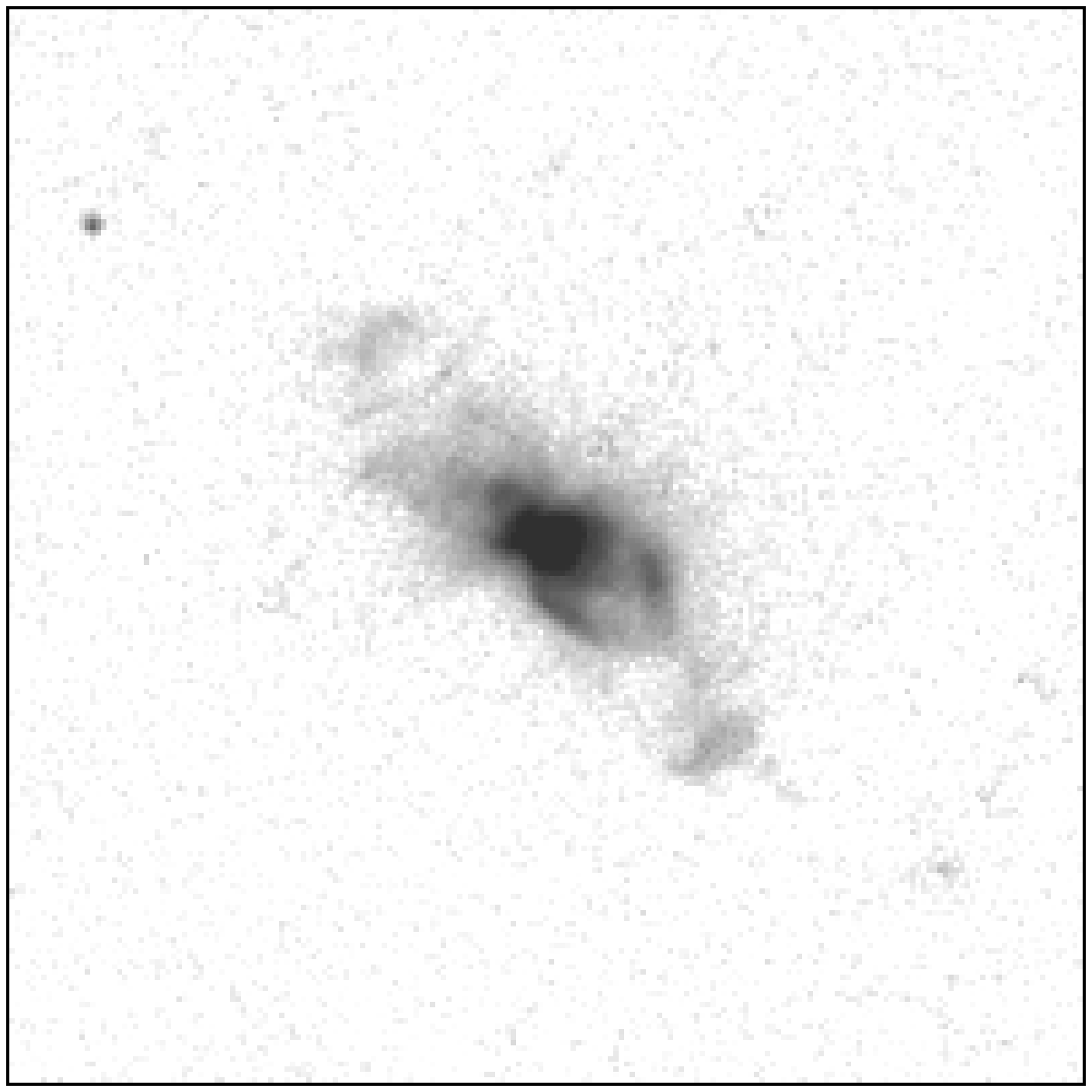}
   \end{center}
   \caption{Images of the PNe M2-11 (left, ESO NTT/EMMI [O~III] from programme 079.D-0764(B), $30\times30$ arcsec$^2$) and M3-38 (right, \emph{HST} WFPC2 $F656N$, $10\times10$ arcsec$^2$). Note the similarity in the highly collimated outflows surrounding both HDC PNe. North is up and East to left.}
   \label{fig:m211m338}
\end{figure*}

Figure \ref{fig:m211m338} shows the highly collimated outflows of these similar PNe. Each shows [Fe~VII] emission from their nuclei. The average [O~III] $\lambda$4363/H$\gamma$ ratios of $0.67\pm0.05$ and $0.55\pm0.07$ for M2-11 and M3-38, respectively, were calculated from the 20 spectra in the mini-IFUs. The errors represent the standard deviation $\sigma$ of all 20 measurements and the maximum value in each case did not exceed 2$\sigma$, suggesting there is no real variation over the nebula that could be attributed to a dense nucleus. The weak emission lines at $\lambda$5801,5812 may be members of the Pfund He~II series (rather than C~IV), of which several lines are visible just to the red of these lines in each object. Unfortunately, neither object has OGLE photometry. 

\subsection{PHR1806-2652}
PHR1806-2652 was observed in the same FLAMES field as M2-29 (see Miszalski et al. 2011a).
The mini-IFU for PHR1806-2652 was centred on 18$^\mathrm{h}$06$^\mathrm{m}$56\fs07 $-$26$^\circ$52$'$53.4 at a PA of 135.24 deg (see Fig. \ref{fig:ogleim2}) which includes a central star (CS) whose 2MASS and GLIMPSE magnitudes are recorded in Tab. \ref{tab:othermags}. The OGLE $I$- and $V$-band images show that the CS is slightly brighter in $V$ and surrounded on either side by nebulosity in an $S$-shaped (bipolar) configuration. The nebula is visible in the SHS (Fig. \ref{fig:c4}), which also reveals an extra patch of H$\alpha$ nebulosity 34.1\arcsec to the SW which may be related at a separation of 1.3 pc at 8 kpc. The nebular morphology resembles the unusual He-rich object IPHASJ195935.55$+$283830.3 (Corradi et al. 2010a). 

The emission line ratios of the nebula are consistent with a PN after averaging all fibres. On the other hand, the [Fe~VII] $\lambda$5721 and [Fe~VII]/[Ca~V] $\lambda$6087 emission originate from the CS and are not typical of PNe. The radial velocities of the [Fe~VII] emission lines match those of the He I $\lambda$5876 emission line from the nebula. The S/N is too low to tell if there is stellar [O~III] $\lambda$4363 emission, but the CS shows slightly broader H$\alpha$ emission and stronger He~I emission lines. No features of a companion are found, although we note that the apparent G-band in Fig. \ref{fig:egb62} is spurious, originating from the integrated background light of the Galactic Bulge. We estimated the reddening of $E(B-V)=1.0$ mag from an average of estimates made from AAOmega and FLAMES spectra, the former estimate made use of the [OIII] $\lambda$5007/H$\beta$ ratio from the FLAMES spectrum. This is rather high compared to $E(B-V)=0.65$ mag of the very nearby M~2-29 (Miszalski et al. 2011a), suggesting the 0.3--0.5 mag uncertainty of this method is realistic and that the reddening may be closer to that of M~2-29. 

\begin{figure*}
   \begin{center}
      \includegraphics[scale=0.7,angle=270]{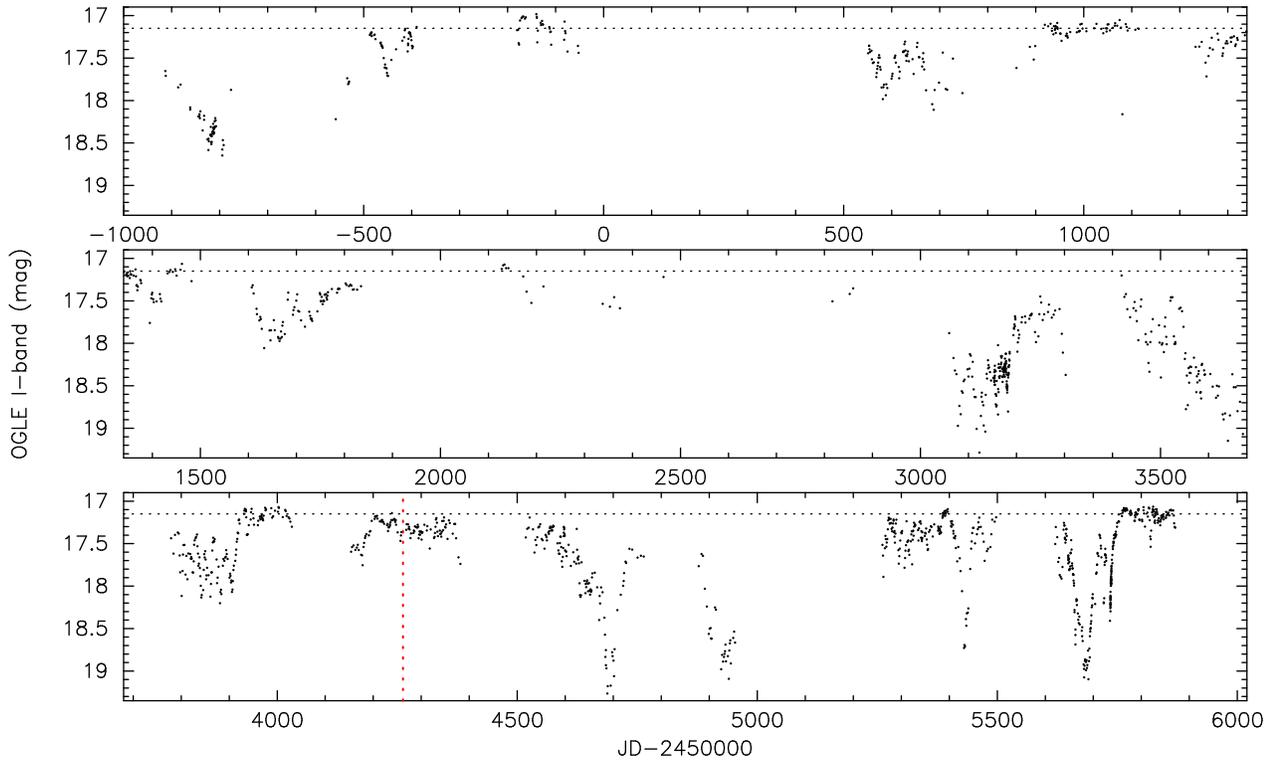}
   \end{center}
   \caption{Full OGLE $I$-band lightcurve of the nucleus of PHR1806-2652. The horizontal dotted line at $I=17.15$ mag marks an arbitrary light maximum and the vertical dotted red line marks the epoch of the VLT FLAMES observations.}
   \label{fig:phr1806lc}
\end{figure*}

Figure \ref{fig:phr1806lc} presents the full OGLE $I$-band lightcurve of PHR1806-2652. The lightcurve shows several dust obscuration events of up to $\sim$2.0 mag in depth that are considerably more frequent and irregularly spaced than in NGC2346. A periodogram analysis of the lightcurve did not find any clear periodicity, consistent with the known pseudo-periodic nature of dust obscuration events (e.g. NGC2346). Together with M2-29, PHR1806-2652 suggests dust obscuration events may be a common occurrence in the HDC PN class. Another PN to show dust obscuration events in our sample is MPA1746-3412 which hosts a [WC10-11] central star (Sect. \ref{sec:MPA1746}). 

\section{Other objects}
\label{sec:other}

\subsection{Novae: 003.16-02.31 and V4579 Sgr}
The strong HeII, NIII, [FeVII] and CIV emission lines present in both spectra are typical of He/N-type classical novae in the nebular phase (e.g. Warner 2003). The MACHO lightcurves and, to a lesser extent, OGLE lightcurves show the decline of the novae over a span of 14.1 (003.16-02.31) and 18.7 (V4579 Sgr) years. Fitting a line to the MACHO lightcurves we find decline rates of 1.2$\times$10$^{-3}$ mag day$^{-1}$ (003.16-02.31) and 1.9$\times$10$^{-4}$ mag day$^{-1}$ (V4579 Sgr). In the OGLE lightcurves the corresponding rates are 9.6$\times$10$^{-5}$ mag day$^{-1}$ and 2.8$\times$10$^{-5}$ mag day$^{-1}$, respectively. Discovery of V4579 Sgr is attributed to R. McNaught in 1986, however we could not locate the corresponding IAU circular and it is listed as a slow nova in Samus et al. (2012). Woudt, Warner \& Spark (2005) found 003.16-02.31 to have an eclipsing lightcurve with a period of 0.117 days and give further background information on the object. Some recent systematic studies of novae at similar stages in their evolution are Tappert et al. (2012) and Walter et al. (2012).

\subsection{A B[e] star: PHR1803-2748}
\label{sec:phr1803}
Following Corradi et al. (2010a), the AAOmega spectrum of PHR1803-2748 detects [NiII] $\lambda$6666, 7378 and 7412, which combined with the multiple [FeII] lines and 2MASS colours indicative of hot dust, classifies PHR1803-2748 as a B[e] star (Lamers et al. 1998). See also Miroshnichenko (2006) and the discussion by Frew \& Parker (2010). A bipolar nebula is present and measures $\sim$18\arcsec\ tip-to-tip (0.7 pc at 8 kpc). Its presence suggests that PHR1803-2748 may be a symbiotic B[e] star, but further observations are required to examine this possibility. The OGLE lightcurve shows slow changes similar to those seen in D-type symbiotic stars that are probably due to dust. 

\subsection{The [WC10-11] Wolf-Rayet central star of MPA1746-3412}
\label{sec:MPA1746}
Figure \ref{fig:ogleim1} identifies the object correctly which corresponds to object 57 from KW2003 (see later) which is clearly an H$\alpha$ emission source (Fig. \ref{fig:c1}). The AAOmega spectrum is contaminated by the unrelated red giant (2MASS J17461818$-$3412370) immediately to the west of the object (Fig. \ref{fig:s3}). MPA1746-3412 was included in Tab. 1 of Miszalski et al. (2009a), but its peculiar spectrum is not a symbiotic star. It closely resembles other [WC10-11] Wolf-Rayet central stars of PNe (Leuenhagen et al. 1996), in particular Hen2-113 (Lagadec et al. 2006) and Hen3-1333 (Chesneau et al. 2006), but shows less stellar emission lines. The strongest emission lines of the Balmer series, [N~II] and [S~II] originate from the PN ionized by the cool [WC10-11] central star. The stellar spectrum is dominated by C~II $\lambda$7235 and C~III $\lambda$5696. Several other Galactic Bulge PNe are known to show similar late-[WC] spectra (G\'orny et al. 2004, 2009) as well as one LMC PN (Miszalski et al. 2011d). Late-[WC] types were originally thought to be over-represented in the Bulge (G\'orny et al. 2004), but this no longer seems to be the case following the discovery of several [WO] type central stars by DePew et al. (2011).

Most interesting about MPA1746-3412 is the OGLE lightcurve that shows a pseudo-periodic decline of 0.35 mag that repeats $\sim$1360 days and lasts for $\sim$950 days. This variability is most likely to be caused by dust obscuration events, as found in the [WC10] central star of Hen3-1333 (Pollacco et al. 1992; Jones et al. 1999; Cohen et al. 2002; Chesneau et al. 2006; Miszalski et al. 2011a). The recurrence timescale between events in MPA1746-3412 is comparable to those seen in Hen3-1333 (Miszalski et al. 2011a).

\subsection{WN6 Wolf-Rayet stars: Th3-28 and 359.88$-$03.58}
Acker \& Stenholm (1990) reclassified Th3-28 to be a WR star and it is part of the van der Hucht (2001) catalogue as WR93a with a spectral type WN2.5-3. We reclassify the subtype as WN6 under the Smith et al. (1996) scheme since CIV $\lambda$5808 is only about 30\% greater than HeI $\lambda$5876 in the peak/continuum ratio and NV is weak or absent.

359.88$-$03.58 was listed as an H$\alpha$ emitter KW083 by Kohoutek \& Wehmeyer (2003). The 2dF/AAOmega spectrum closely resembles Th3-28 and we also classify it as a WN6 star. The OGLE lightcurve is variable and shows smaller scale variations of $\sim$0.5 mag, probably pulsations, on top of a broader dip of $\sim$1 mag. A periodogram analysis found no periodicity in the variations. A dust obscuration event as in WC9 stars (e.g. Veen et al. 1998) is the most likely explanation for the observed variability and may indicate that a binary companion is present.

\subsection{H2-32}
\begin{figure*}
   \begin{center}
      \includegraphics[scale=0.6,angle=270]{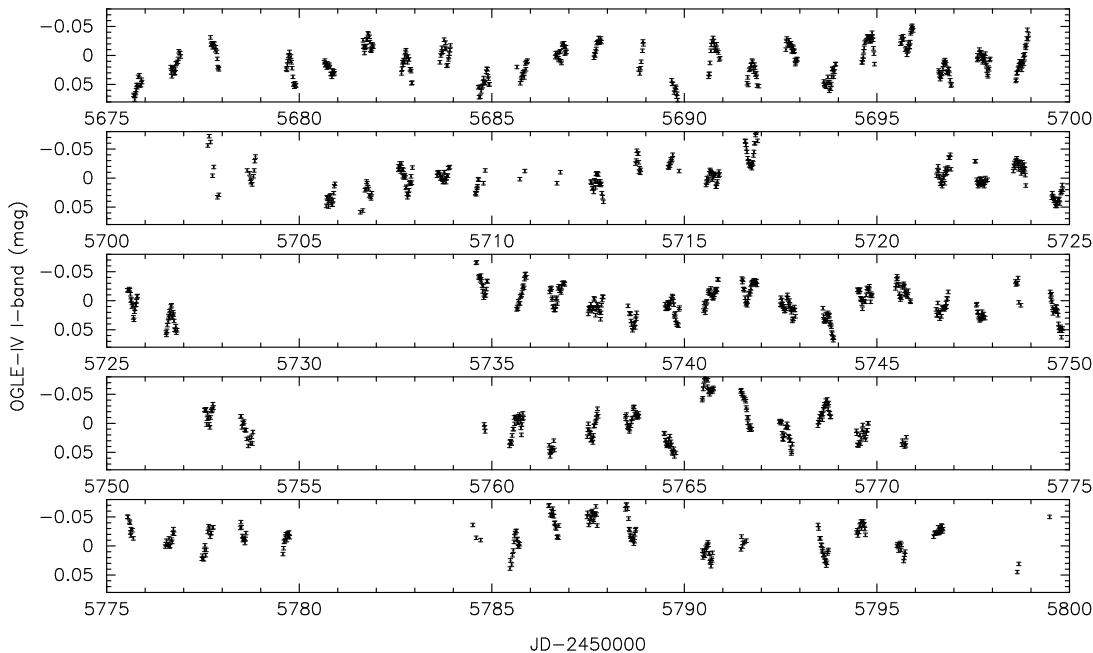}
   \end{center}
   \caption{Part of the OGLE-IV $I$-band lightcurve of H2-32 showing intra-day variations about the mean.}
   \label{fig:H2-32}
\end{figure*}

The OGLE lightcurve shows a gradual rise of 0.3 mag and also shows small timescale variability in very highly sampled parts of the OGLE-III and OGLE-IV observations. Figure \ref{fig:H2-32} shows one such part where a $\sim$5 day pseudo-periodic variation appears to be present, in addition to intra-day variability. The former may be an orbital period, while the latter is probably due to pulsations. These variations may be similar to those seen in Be stars (Porter \& Rivinius 2003), but we find no clear signs of a B-type star in the low-resolution AAOmega spectrum. It may still be possible that this object is a PN with a relatively cool central star, passing through an instability strip, as seen in several PNe (Handler 2003). The 2MASS colours are not a helpful discriminant between PNe and Be stars, since both occupy a similar part in diagnostic diagrams (Corradi et al. 2008). At this stage we favour a possible Be star interpretation, based on the very unusual lightcurve which is atypical for PNe. 

\subsection{PPA1758-2628}
The emission line spectrum is rather ordinary besides the exceptionally strong He~I emission lines. It is probably a highly reddened (compact) HII region. It is a source in both NVSS (Condon et al. 1998), $F(1.4 \mathrm{Ghz})=13.7\pm0.6$ mJy, and is very bright in GLIMPSE, $F(8 \mu\mathrm{m})=(1.148\pm0.015)\times10^3$ mJy. The 8$\mu$m/NVSS flux ratio, a diagnostic developed for distinguishing PNe from HII regions (Cohen et al. 2007), has a very high value of 84, in the ultracompact HII region regime of Cohen et al. (2011). This classification is consistent with the emission line spectrum that has unusually strong He~I emission lines for a PN relative to other emission lines. This object is also listed as a radio compact HII region in Giveon et al. (2005).

\subsection{Sa3-104}
The OGLE-II and OGLE-III lightcurves both show a strong spurious periodicity of exactly one year, most probably due to a non-stellar point-spread function. Similar systematic effects were seen in other objects studied by Miszalski et al. (2009a). The periodicity does not appear in the higher quality OGLE-IV data. On the other hand, it is an interesting possibility that the slow decline may represent a real secular change, but we cannot prove this one way or the other. 
The AAOmega spectrum presents a low-excitation nebula with log (H$\alpha$/[NII])=0.35 and log (H$\alpha$/[SII])=1.91 consistent with Galactic PNe (Frew \& Parker 2010). A weak stellar continuum shows absorption features of Ca II H \& K and relatively broad H$\gamma$ and H$\beta$ in absorption. These characteristics on their own are not remarkable, but we include it in our study because of its red appearance in the 2MASS colour-composite image and especially because of its exceptional brightness in GLIMPSE ([8.0]=5.05 mag). As such the source may be a reddened HII region or possibly a Be star (Porter \& Rivinius 2003). The non-detection of this source in the NVSS (Condon et al. 1998) with the bright GLIMPSE 8.0 $\mu$m flux seem to be consistent with this interpretation according to the MIR/Radio ratio diagnostic (Cohen et al. 2007, 2011).

\section{Discussion}
\label{sec:discussion}
\subsection{Completeness and Depth}
\label{sec:comp}
\begin{figure*}
   \begin{center}
      \includegraphics[scale=0.7,angle=270]{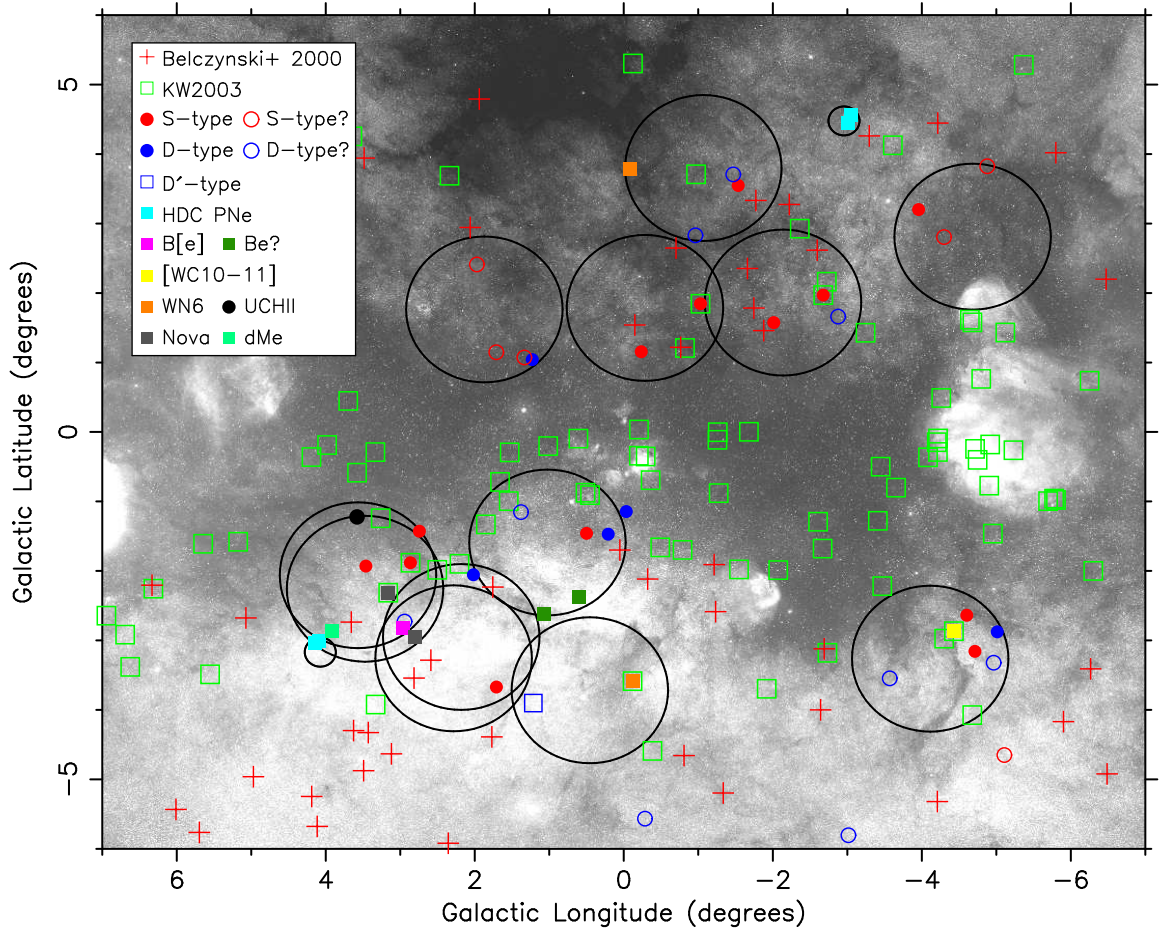}
   \end{center}
   \caption{Mosaic of blocked-down SHS H$\alpha$ fields overlaid with our sample, known symbiotics (Belczy\'nski et al. 2000) and H$\alpha$ emitters (KW2003). Fields observed with 2dF-AAOmega (Tab. \ref{tab:aaomega}) and FLAMES (Tab. \ref{tab:flames}) are shown by large and small circles, respectively. The D-type symbiotic star Hen2-375 and the HDC PN K2-17 are located outside this region at ($\ell$,$b$)=($-$22.7,$-$18.3) and ($-$23.2,$-$07.2), respectively.}
   \label{fig:bulge}
\end{figure*}

The completeness of the method used to select H$\alpha$ emission line candidates (Sect. \ref{sec:obs}) is best judged against lists of similar objects. We recovered 11/13 catalogued symbiotic stars coinciding with our AAOmega fields (Belczy\'nski et al. 2000). This is depicted graphically in Figure \ref{fig:bulge}. Only two of these were not observed, Bl3-14 and Th3-31, the former can be explained by an overlapping star in the optical that essentially removed the apparent excess (the excess is comparable to residuals around field stars), while the latter had no optical counterpart to a very red source visible in 2MASS ($J-K_s=3.6$, 2MASS J173426666$-$2928029). AAOmega spectra of these known symbiotic stars will be studied elsewhere.

A more comparable sample of H$\alpha$ emitters are those discovered by the objective prism survey of Kohoutek \& Wehmeyer (2003, hereafter KW2003). Table 1 of KW2003 lists 98 new H$\alpha$ emitters and 9 probable H$\alpha$ emitters that are more numerous at lower latitudes ($|b|\la1^\circ$). These sources are included in Fig. \ref{fig:bulge}. Table \ref{tab:kw} delineates the subset of KW2003 sources located in our AAOmega fields and their observation status. Of the objects we missed, inspection of the SHS and SSS images found that both KW081 and KW022 are small, i.e. easily missed when viewing large areas, while only KW023 shows a modest H$\alpha$ excess that might have been selected. The remainder show only weak H$\alpha$ excesses that explain their exclusion from our AAOmega observations. 

The completeness of our survey is therefore reasonably high, though it is difficult to quantify this further. For example, we identified H$\alpha$ emitters that were not found by KW2003 and only the best candidates were selected in the densest fields (e.g. S18) where the H$\alpha$ emitters were more numerous and it was more difficult to judge the level of H$\alpha$ excess. Nevertheless, we are the first to systematically combine deep spectra and $I$-band lightcurves to discover new symbiotics, greatly enhancing the completeness of our survey over previous works. In the $\sim$35 deg$^2$ considered in our survey, we have more than doubled the number of confirmed symbiotics. Combining our new discoveries with Belczy\'nski et al. (2000), we can also calculate an average symbiotic star surface density in Northern ($b>0^\circ)$ and Southern ($b<0^\circ)$ AAOmega fields based on five fields each of radius 1.05 degrees. We found $0.9\pm0.7$ deg$^{-2}$ and $1.0\pm0.4$ deg$^{-2}$ for confirmed symbiotics in Northern and Southern fields, respectively, where the errors represent the standard deviation of measurements from each of the five fields. If we include the possible symbiotic stars these values are $1.3\pm0.4$ deg$^{-2}$ and $1.3\pm0.6$ deg$^{-2}$, respectively. This constitutes a first estimate of the surface density of symbiotic stars towards the Galactic Bulge, which may in the future be compared with stellar population models, that also take into account the complex interstellar extinction towards that Bulge, to better constrain the total number of Galactic symbiotic stars (see e.g. Corradi et al. 2010a and Corradi 2012).

\begin{table}
   \centering
   \caption{The 17 H$\alpha$ emission line objects from Table 1. of KW2003 that are located within our AAOmega fields.}
   \label{tab:kw}
   \begin{tabular}{llll}
      \hline
      Field & KW & Name & Type \\
      \hline
      N5 & 022 & - & -  \\
      N2 & 023 & - & -  \\
      N2 & 025 & 357.32+01.97 & S \\
      N3 & 037 & NSV22840 & S  \\
      N3 & 042 &  - & -  \\
      S3 & 057 & MPA1746-3412 & [WC10-11] \\
      S3 & 059 & - & -\\
      S18 & 065 & - & -\\
      S18 & 066 &- & -  \\
      S18 & 073 &- & -  \\
      S18 & 075 &- & -  \\
      S21 & 079 & - & - \\ 
      S21 & 081 & - & - \\
      S21 & 082 & 002.86$-$01.88 & S\\
      S22 & 083 & 359.88$-$03.58 & WN6\\
      S3 & 102 & - & - \\
      S21 & 105 & 003.16$-$02.31 & Nova\\
      \hline
   \end{tabular}
\end{table}

\begin{figure}
   \begin{center}
      \includegraphics[scale=0.5]{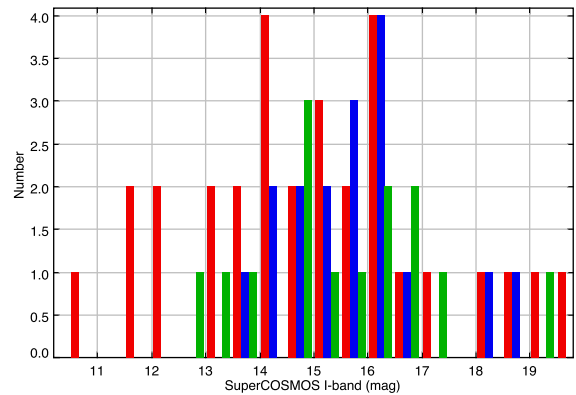}
   \end{center}
   \caption{SuperCOSMOS Science Archive $I$-band magnitude comparison between known symbiotics (red, Belczy\'nski et al. 2000), our new symbiotics (blue, this work), and symbiotics discovered so far by IPHAS (green, Corradi 2012 and ref. therein). The bin size is 0.5 mag with each data series shown in three adjacent columns per bin.}
   \label{fig:ssa}
\end{figure}

A general comparison can also be made with other symbiotic stars to assess the depth of our survey. Modern CCD-based photometry is not generally available for the Galactic Bulge at optical wavelengths, making a uniform comparison challenging. The depth of 2MASS is insufficient for the faintest objects in the NIR, while not all objects have OGLE-III $I$-band coverage. Uniform coverage is however available with digitised photographic all sky survey catalogue photometry available online in the SuperCOSMOS Science Archive\footnote{http://www-wfau.roe.ac.uk/ssa/} (SSA, Hambly et al. 2004). Here we make use only of the $I$-band magnitudes since bluer filters are strongly influenced by reddening and strong H$\alpha$ emission. A nearest neighbour cross-match was performed with the SSA for our new symbiotic star sample (excluding Hen2-375), the known symbiotic stars (Belczy\'nski et al. 2000) bounded by the Galactic coordinate range in Fig. \ref{fig:bulge}, and the IPHAS-selected symbiotic stars (Corradi 2012). Figure \ref{fig:ssa} displays the histogram of $I$-band magnitudes and no attempt was made to correct the data for photometric variability.  The data show our sample to fall towards the fainter end of the magnitude distribution of known symbiotics, suggesting our improved survey techniques have allowed us to identify relatively fainter symbiotic stars than previous surveys. Our sample is however comparable to the symbiotics discovered by IPHAS that are located outside the Galactic Bulge (Corradi 2012 and ref. therein).

\subsection{X-ray properties}
\label{sec:xray}
Symbiotic stars may also be detected at X-ray wavelengths due to a variety of emission mechanisms (e.g. M{\"u}rset et al. 1997; Luna et al. 2012). Supersoft sources are an indication of massive white dwarfs and are usually associated with low metallicity environments (e.g. Kato, Hachisu \& Miko{\l}ajewska 2013). We searched for an overlap between our sample and the Chandra Galactic Bulge Survey X-ray catalogue of Jonker et al. (2011). The survey consists of a series of relatively shallow 2 ks pointings in which 10 sources overlap (000.49$-$01.45, 357.98$+$01.57, 359.76$+$01.15, JaSt2-6, JaSt79, NSV22840, PHR1757-2718, 001.33$+$01.07, 001.37$-$01.15 and 001.71$+$01.14). None, however, had a corresponding detection within 30\arcsec\ under the selection criteria applied by Jonker et al. (2011). Several long period variables were identified by Udalski et al. (2012) to match with other sources catalogued by Jonker et al. (2011). 

There are two objects that may have X-ray detections in other surveys. The \emph{ROSAT} All-Sky Survey Faint Source Catalogue (Voges et al. 2000) detected a source, 1RXS J175001.8-293316, located 8.6\arcsec\ from JaSt2-6 with a countrate of $5\times10^{-2}$ counts s$^{-1}$ and positional error of 37.0\arcsec. Such a source would probably have been detected by Jonker et al. (2011), so this may be an indication of X-ray variability that needs further monitoring. Hong et al. (2009) obtained much deeper (100 ks) Chandra observations towards the Bulge and catalogued an X-ray source (J175153.6-293054) 2.0\arcsec\ away from JaSt79 with a median energy of 1.127 keV and a $2\sigma$ positional error of 1.67\arcsec. While this detection seems reliable, it would be unusual given the relatively large distance ($\sim$6.7 kpc) and high reddening ($E(B-V)\sim2.5$ mag) to JaSt79. As we detect the Mira both spectroscopically and its pulsations in the OGLE lightcurve, the bulk of the reddening must be interstellar and is likely to suppress such X-ray emission if it originated from JaSt79. Further dedicated observations are needed to confirm whether it is a true detection. 

\subsection{H1-45: The first probable Galactic Bulge carbon Mira}
\label{sec:cmira}
A longstanding problem in stellar astrophysics is the lack of luminous carbon stars belonging to the Galactic Bulge (e.g. Whitelock 1993; Whitelock et al. 1999; Ng 1999; Wahlin et al. 2007; Feast 2007). A small sample of intrinsically faint, relatively blue carbon stars have been documented (Azzopardi et al. 1985, 1991; Westerlund et al. 1991; Alksnis et al. 2001), but none rival the luminosities of bona-fide carbon stars found elsewhere and they may be the product of binary evolution (Whitelock et al. 1999). The luminosity of H1-45 is instead typical of carbon Miras ($M_K=-8.06\pm0.12$ mag) and its estimated period-luminosity relation distance of $6.2\pm1.4$ kpc places it within reach of the Galactic centre ($d\sim8.0\pm0.5$ kpc, Matsunaga et al. 2013). At the short end of 4.8 kpc its distance would be consistent with membership of the Galactic Disk. While we cannot rule out Disk membership, it seems more likely, however, that it belongs to the Galactic Bulge. At the Galactic coordinates of H1-45, studies of red clump stars have shown that the inner Galactic Bar extends nearer to us than at negative longitudes, in some cases reaching a distance of 6 kpc from the Sun (Stanek et al. 1997; Babusiaux \& Gilmore 2005; Rattenbury et al. 2007; Cabrera-Lavers et al. 2007; Gonzalez et al. 2011). It is therefore possible that H1-45 is situated at the near-end of the inner bar. One way to test this further would be to identify the source of the carbon enhancement in H1-45, whether it is intrinsic, produced by the visible carbon Mira, or extrinsic, produced by the AGB progenitor of the WD companion. If the Galactic Bulge cannot produce carbon stars, as seems to be the case, then a Galactic Disk membership would be more consistent with an intrinsic rather than extrinsic origin. 

  \subsection{Tendencies amongst each symbiotic star type}
  \label{sec:tendencies}
The discoveries made in this paper reinforce the observed properties of each symbiotic star type and highlight 
the challenges in proving a symbiotic star classification. A telling statistic is the ratio of the number of S-types to D-types (S/D)
in the new (S/D=2.2) and possible (S/D=0.7) samples. Clearly, the S-types are more straightforward to confirm, owing to the generally unobscured
cool stellar photospheres of the donor being readily observable at optical wavelengths. The lightcurves of S-types in both our new and possible sample
readily demonstrate semi-regular pulsations from the donor star, a trend also noticed in known S-type symbiotics (Gromadzki et al. in preparation).

At optical wavelengths D-types are considerably more difficult to confirm, owing to the presence of warm dust as seen in their NIR colours. In most cases this dust forms an obscuring shell or coccoon around the Mira component, rendering the spectral signatures and pulsations of the Mira invisible at far-red and shorter wavelengths. This is best demonstrated by RX Pup, where the Mira pulsations are only seen at NIR wavelengths and its absorption spectrum has never been observed at optical wavelengths (Miko{\l}ajewska et al. 1999). Without these signatures at optical wavelengths the D-types typically show their PN-like nebular spectra, originating near the relatively unobscured hot WD, and optical lightcurves that are either flat (e.g. 354.98-02.87 and K5-33) or only slowly variable due to dust obscuration (e.g. Al2-G, PPA1746-3454, PHR1751-3349 and PPA1807-3158; see also Gromadzki et al. 2009). Pulsations from the Mira components in these D-type systems should be readily visible in NIR lightcurves, but these are not readily available. Nevertheless, we were fortunate that some systems are less reddened, allowing for their Mira pulsations to be detected, confirming them as D-type symbiotics (H1-45, JaSt2-6 and JaSt79). 

\section{Conclusions}
\label{sec:conclusion}
We have conducted a survey of H$\alpha$ emission line stars towards the Galactic Bulge covering $\sim$35 deg$^2$. Deep optical spectroscopy and $I$-band long-term lightcurves were presented for most of the sample, revealing 20 new and 15 possible symbiotic stars, as well as several other unusual emission line stars. The novel incorporation of lightcurves has proven particularly effective in proving the symbiotic status of several objects whose classification may have remained ambiguous from optical spectroscopy alone. 

The main conclusions are as follows:
\begin{itemize}
   \item A total of 20 symbiotic stars were confirmed, consisting of 13 S-types, 6 D-types and 1 D$'$-type, of which 35\% show the Raman scattered OVI features. Orbital periods were determined for 5 S-types and Mira pulsation periods for 3 D-types. Further study of a list of 15 possible symbiotics (6 S-types and 9 D-types), particularly at NIR wavelengths to search for Mira pulsations in the case of D-types, should allow for more symbiotics to be confirmed. 
   \item H1-45 was found to be a carbon symbiotic Mira and is only the fourth now known in the Galaxy. Using the Whitelock et al. (2008) period-luminosity relation, the carbon Mira pulsation period of 408.6 days corresponds to a distance of $6.2\pm1.4$ kpc and an absolute magnitude of $M_K=-8.06\pm0.12$ mag. If it belongs to the near side of the Galactic Bulge, then it would be the first luminous carbon star identified in this environment, potentially offering a means to solve the longstanding missing Galactic Bulge carbon star puzzle (e.g. Feast 2007). Further observations of H1-45 are essential to further refine the distance to this unique object and understand its origin in the context of carbon star formation as a function of environment (e.g. Nowotny et al. 2013). The D$'$-type symbiotic ShWi5 may also be carbon-rich, but it requires further study.
   \item Dust obscuration events were discovered in the central stars of PHR1806-2652 and MPA1746-3412, increasing the sample of PNe showing such behaviour to 5 PNe. The central star of PHR1806-2652 exhibits [FeVII] emission and belongs to the high density core (HDC) PN family, i.e. it has symbiotic-like characteristics, but these are thought to be associated with main sequence companions and circumbinary dust disks. Three other HDC core PNe were also presented. MPA1746-3412 has a [WC10-11] Wolf-Rayet central star and therefore strongly resembles the very similar Hen3-1333. A dust obscuration event was also noticed in the massive WN6 star 359.88-03.58, suggesting it may be a binary system.
   \item Comparison with other catalogues of H$\alpha$ emitters (Kohoutek \& Wehmeyer 2003) and symbiotic stars (Belczy\'nski et al. 2000) demonstrated the high completeness of our survey. The apparent surface density of all symbiotic stars in the $\sim$35 deg$^2$ of our survey amounts to $\sim$1.0--1.3 deg$^{-2}$. Much further work remains to create a complete consensus of Galactic Bulge symbiotic stars in order to improve estimates of the total Galactic symbiotic star population (Corradi et al. 2010a; Corradi 2012).
\end{itemize}

\section{Acknowledgements}
This work benefited from the framework of the European Associated Laboratory ``Astrophysics Poland-France''. JM is supported by the Polish National Science Center grant number DEC-2011/01/B/ST9/06145.
The OGLE project has received funding from the European Research Council under the European Community's Seventh Framework Programme (FP7/2007-2013)/ERC grant agreement No. 246678. We would like to thank the referee, Romano Corradi, for insightful comments that have helped improve some aspects of this paper.

   We thank AAO staff for performing the AAOmega service observations and especially R.G. Sharp for taking additional observations. This work is partly based on observations made at the South African Astronomical Observatory (SAAO) and we thank R. Manick for his assistance in taking the observations of Hen2-375.

   This research has made use of the SIMBAD database and VizieR catalogue access tool, operated at CDS, Strasbourg, France. It also used observations made with the NASA/ESA Hubble Space Telescope, and obtained from the Hubble Legacy Archive, which is a collaboration between the Space Telescope Science Institute (STScI/NASA), the Space Telescope European Coordinating Facility (ST-ECF/ESA) and the Canadian Astronomy Data Centre (CADC/NRC/CSA).

   This paper utilizes public domain data obtained by the MACHO Project, jointly funded by the US Department of Energy through the University of California, Lawrence Livermore National Laboratory under contract No. W-7405-Eng-48, by the National Science Foundation through the Center for Particle Astrophysics of the University of California under cooperative agreement AST-8809616, and by the Mount Stromlo and Siding Spring Observatory, part of the Australian National University.

This research has made use of SAOImage DS9, developed by Smithsonian Astrophysical Observatory, and Montage, funded by the National Aeronautics and Space Administration's Earth Science Technology Office, Computation Technologies Project, under Cooperative Agreement Number NCC5-626 between NASA and the California Institute of Technology. Montage is maintained by the NASA/IPAC Infrared Science Archive.

\appendix
\clearpage
\newpage
\section[]{Near-infrared and mid-infrared photometry}
\label{sec:nirmir}
\begin{table*}
\centering
\caption{2MASS and GLIMPSE magnitudes of new and possible symbiotic stars. The full version is available online.} 
\label{tab:newmags}
\begin{tabular}{llrrrrrrrrrrr}
\hline
Name & Type & $J$ & $H$ & $K_s$ & $J-H$ & $H-K_s$ & $J-K_s$ & Qflag & [3.6] & [4.5] & [5.8] & [8.0] \\
\hline
000.49$-$01.45 & S & 9.86 & 8.48 & 7.73 & 1.38 & 0.75 & 2.13 & AAA & 7.34 & 7.57 & 7.28 & 7.25\\
001.70$-$03.67 & S & 9.00 & 8.12 & 7.78 & 0.88 & 0.34 & 1.22 & AAA & 7.64 & 7.76 & 7.62 & 7.47\\
002.86$-$01.88 & S & 9.98 & 8.68 & 8.10 & 1.30 & 0.58 & 1.88 & AAA & 7.77 & 7.90 & 7.69 & 7.64\\
\ldots & \ldots & \ldots & \ldots & \ldots & \ldots & \ldots & \ldots & \ldots & \ldots & \ldots & \ldots & \ldots \\
\hline
\end{tabular}
\end{table*}

\begin{table*}
\centering
\caption{2MASS and GLIMPSE magnitudes of other objects. The full version is available online.}
\label{tab:othermags}
\begin{tabular}{llrrrrrrrrrrr}
\hline
Name & Type & $J$ & $H$ & $K_s$ & $J-H$ & $H-K_s$ & $J-K_s$ & Qflag & [3.6] & [4.5] & [5.8] & [8.0] \\
\hline
003.16$-$02.31 & Nova & - & - & - & - & - & - & - & 13.74 & 12.75 & 11.99 & 10.43\\
359.88$-$03.58 & WN6 & 13.72 & 13.28 & 12.83 & 0.44 & 0.44 & 0.88 & AAA & 12.27 & 11.79 & 11.56 & 11.36\\
H2-32 & Be? & 13.80 & 13.43 & 12.57 & 0.38 & 0.86 & 1.24 & AAA & 9.80 & 9.06 & 8.49 & 7.57\\
\ldots & \ldots & \ldots & \ldots & \ldots & \ldots & \ldots & \ldots & \ldots & \ldots & \ldots & \ldots & \ldots \\
\hline
\end{tabular}
\end{table*}

\section[]{Additional spectroscopy}
\label{sec:spec}

\begin{figure}
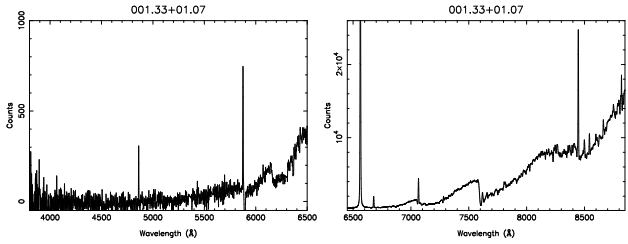

\begin{center}
\includegraphics[angle=270,scale=0.17]{syfigs/001.33+01.07b.ps}
\includegraphics[angle=270,scale=0.17]{syfigs/001.33+01.07r.ps} 
\end{center}
\caption{AAOmega spectra of possible S-type symbiotic stars. The full version is available online.}
\label{fig:ps1}
\end{figure}

\begin{figure}
\begin{center}
\end{center}
\caption{AAOmega spectra of possible S-type symbiotic stars (continued). The full version is available online.}
\label{fig:ps2}
\end{figure}

\begin{figure}
\begin{center}
\end{center}
\caption{AAOmega spectra of possible D-type symbiotic stars. The full version is available online.}
\label{fig:pd1}
\end{figure}

\begin{figure}
\begin{center}
\end{center}
\caption{AAOmega spectra of possible D-type symbiotic stars (continued). The full version is available online.}
\label{fig:pd2}
\end{figure}

\begin{figure}
\begin{center}
\end{center}
\caption{AAOmega spectra of possible D-type symbiotic stars (continued). The full version is available online.}
\label{fig:pd3}
\end{figure}

\begin{figure}
   \begin{center}
   \end{center}
   \caption{AAOmega spectra of other objects excluding PNe with high density cores (see Fig. \ref{fig:egb61}) and superpositions (see Appendix \ref{sec:superpos}). The full version is available online.}
   \label{fig:other1}
\end{figure}

\begin{figure}
   \begin{center}
   \end{center}
   \caption{AAOmega spectra of other objects (continued). The full version is available online.}
   \label{fig:other2}
\end{figure}

\begin{figure}
   \begin{center}
   \end{center}
   \caption{AAOmega spectra of other objects (continued). The full version is available online.}
   \label{fig:other3}
\end{figure}

\section[]{Lightcurves}
\label{sec:lc}

\begin{figure}
   \begin{center}
      \includegraphics[angle=270,scale=0.4]{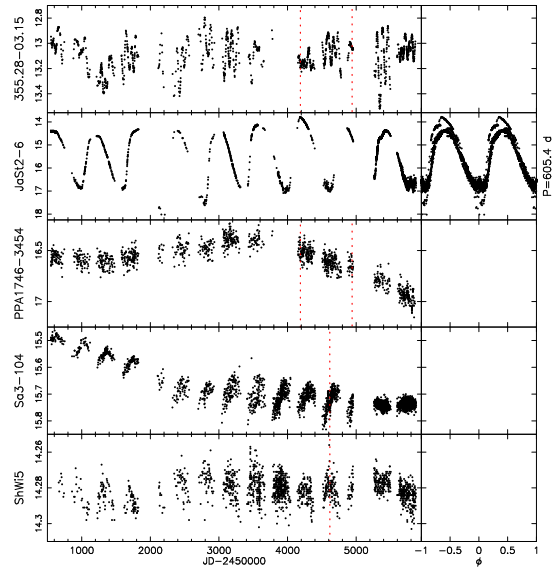}
   \end{center}
   \caption{Combined OGLE-II, OGLE-III and OGLE-IV lightcurves. The full version is available online.}
   \label{fig:ogletwothreefour}
\end{figure}

\begin{figure}
   \begin{center}
   \end{center}
   \caption{Combined OGLE-III and OGLE-IV lightcurves of objects not observed during previous OGLE phases. The full version is available online.}
   \label{fig:oglethreefour1}
\end{figure}

\begin{figure}
   \begin{center}
   \end{center}
   \caption{Combined OGLE-III and OGLE-IV lightcurves (continued). The full version is available online.}
   \label{fig:oglethreefour2}
\end{figure}

\begin{figure}
   \begin{center}
   \end{center}
   \caption{Combined OGLE-III and OGLE-IV lightcurves (continued). The full version is available online.}
   \label{fig:oglethreefour3}
\end{figure}

\begin{figure}
   \begin{center}
   \end{center}
   \caption{Combined OGLE-III and OGLE-IV lightcurves (continued). The full version is available online.}
   \label{fig:oglethreefour4}
\end{figure}

\begin{figure}
   \begin{center}
\includegraphics[angle=270,scale=0.3]{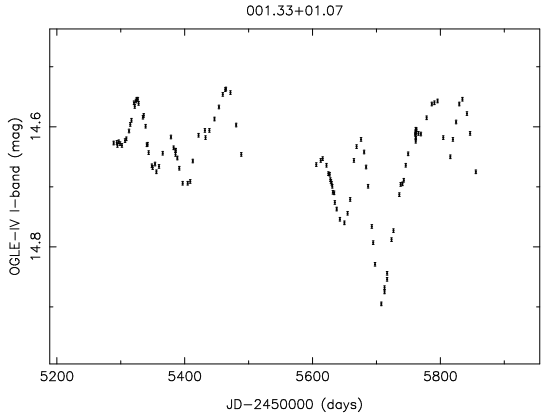}
   \end{center}
   \caption{OGLE-IV lightcurves of objects not observed during previous OGLE phases. The full version is available online.}
   \label{fig:oglefour1}
\end{figure}

\begin{figure}
   \begin{center}

   \end{center}
   \caption{OGLE-IV lightcurves (continued). The full version is available online.}
   \label{fig:oglefour2}
\end{figure}

\clearpage
\begin{figure}
   \begin{center}
\includegraphics[angle=270,scale=0.31]{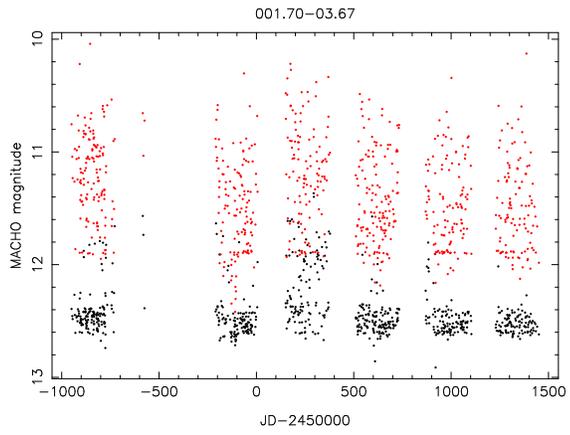}
   \end{center}
   \caption{MACHO $V$ (black) and $R$ (red) lightcurves. The full version is available online.}
   \label{fig:macho}
\end{figure}
\clearpage
\section[]{Images}
\label{sec:images}

\begin{figure}
   \begin{center}
\includegraphics[scale=0.25]{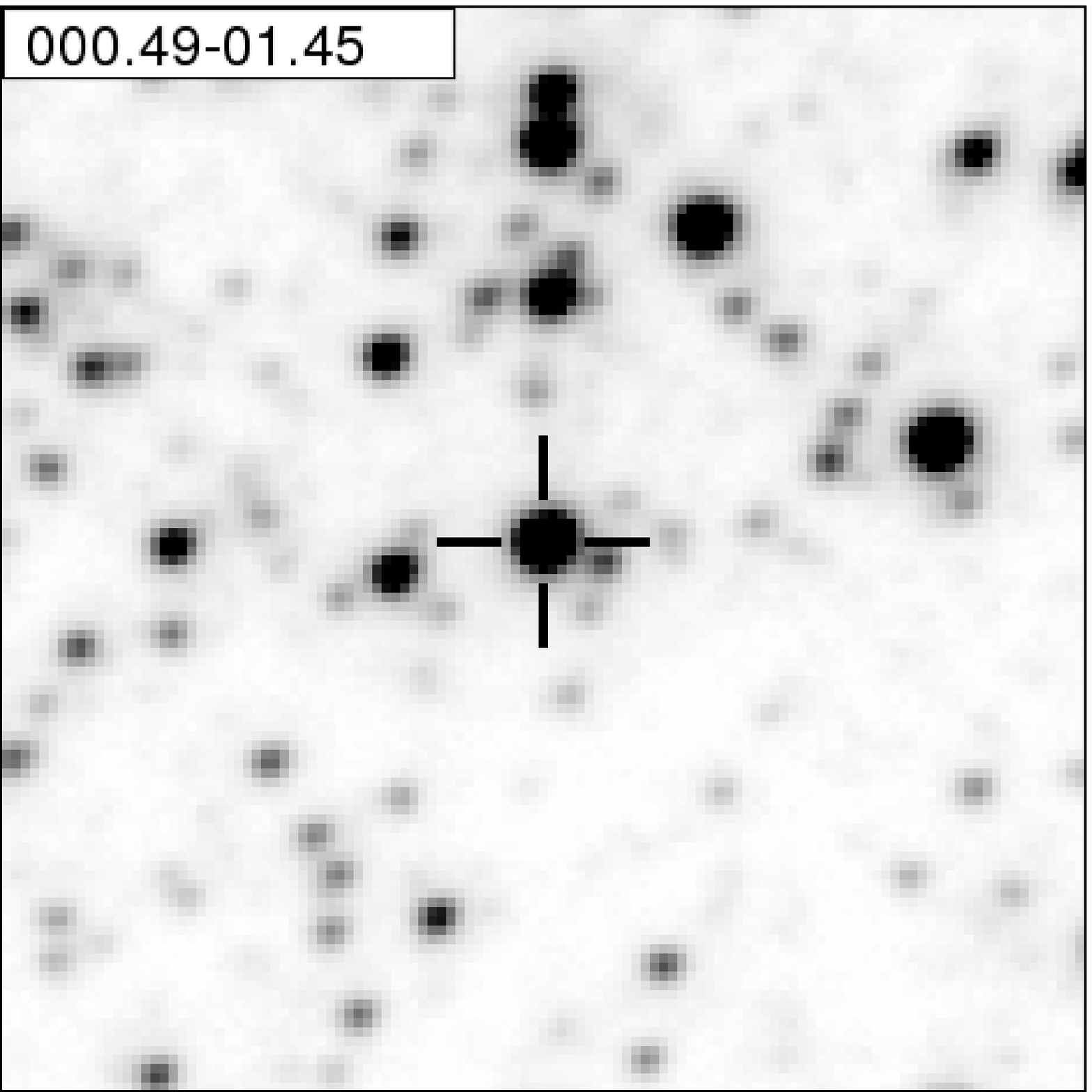}
   \end{center}
   \caption{OGLE-III and OGLE-IV $I$-band images. The full version is available online.}
   \label{fig:ogleim1}
\end{figure}

\begin{figure}
   \begin{center}
   \end{center}
   \caption{OGLE-III and OGLE-IV $I$-band images (continued). The full version is available online.}
   \label{fig:ogleim2}
\end{figure}

\clearpage

\begin{figure}
\begin{center}
\includegraphics[scale=0.285]{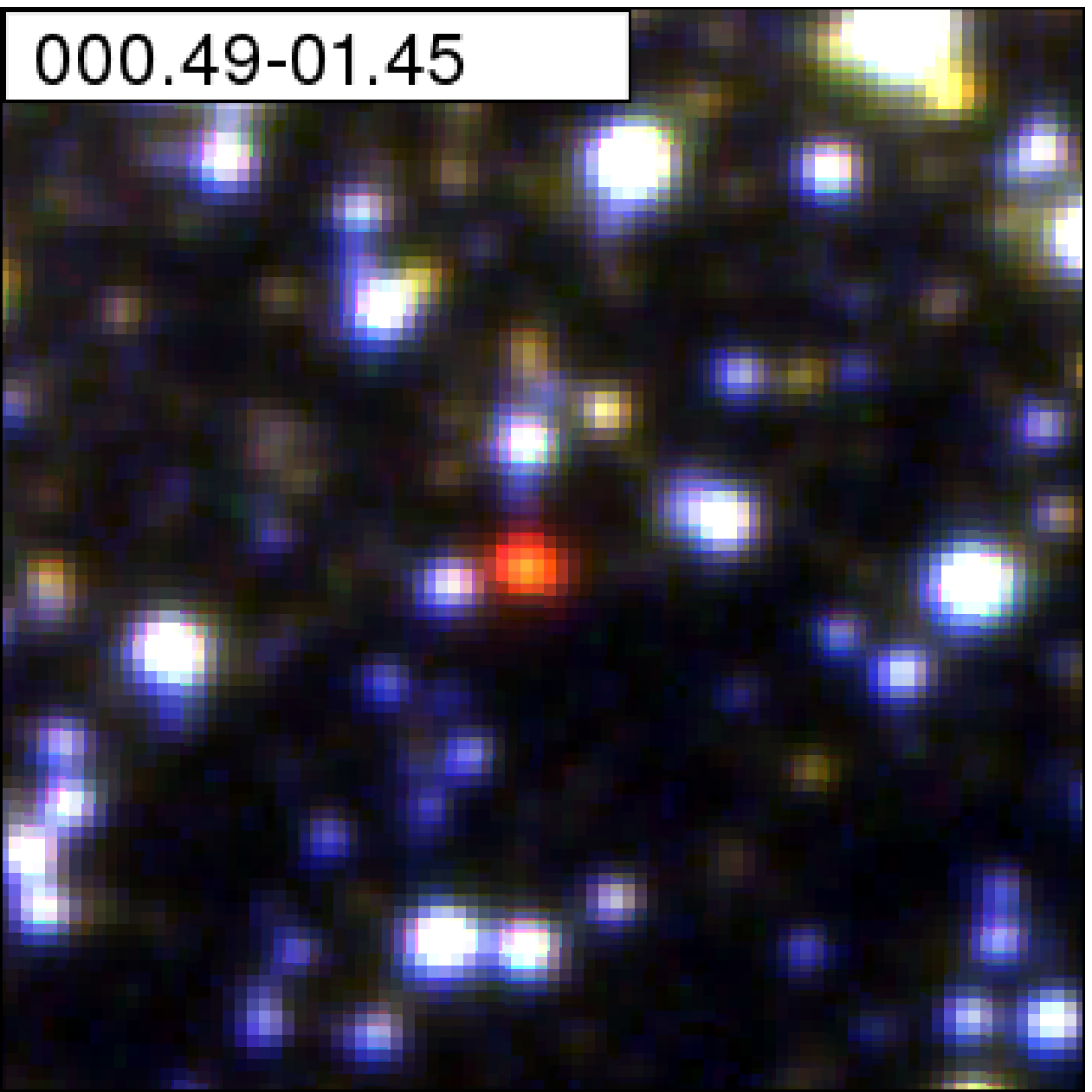}
\includegraphics[scale=0.2128]{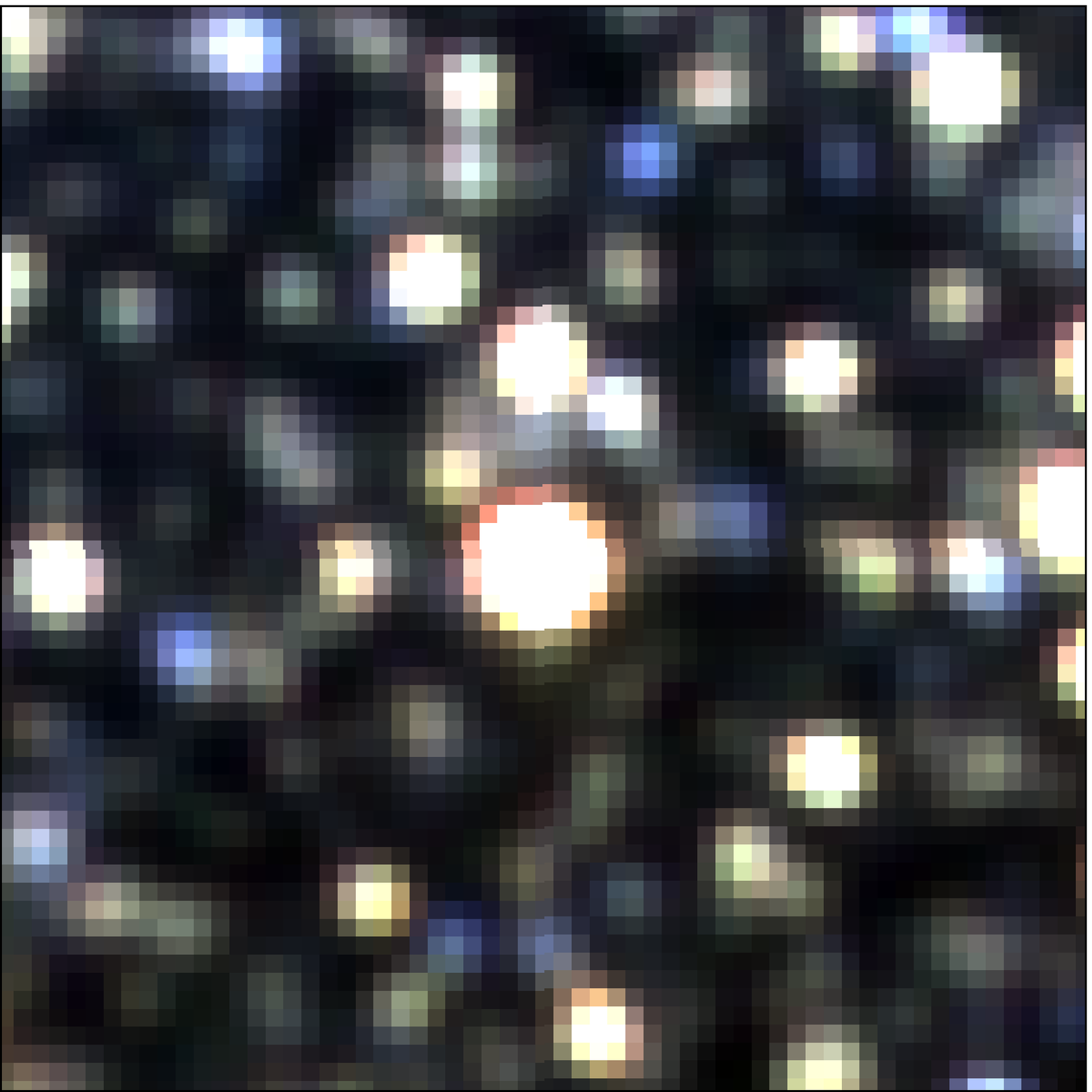}
\end{center}
\caption{Colour-composite images of our sample. The full version is available online.}
\label{fig:c1}
\end{figure}

\begin{figure}
\begin{center}
\end{center}
\caption{Colour-composite images (continued). The full version is available online.}
\label{fig:c2}
\end{figure}

\begin{figure}
\begin{center}
\end{center}
\caption{Colour-composite images (continued). The full version is available online.}
\label{fig:c3}
\end{figure}

\begin{figure}
\begin{center}
\end{center}
\caption{Colour-composite images (continued). The full version is available online.}
\label{fig:c4}
\end{figure}

\begin{figure}
\begin{center}
\end{center}
\caption{Colour-composite images (continued). The full version is available online.}
\label{fig:c5}
\end{figure}

\clearpage
\section[]{Superpositions}
\label{sec:superpos}

 The full text of this section is available online.

\begin{figure}
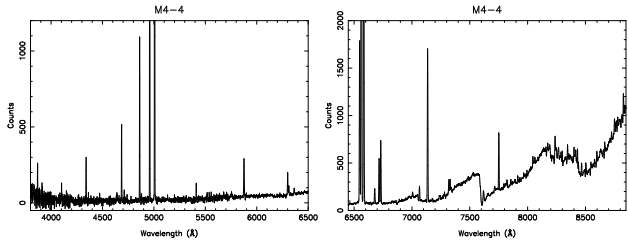

\begin{center}
\includegraphics[angle=270,scale=0.17]{syfigs/M4-4b.ps}
\includegraphics[angle=270,scale=0.17]{syfigs/M4-4r.ps}\\
\end{center}
\caption{Spectra of two superpositions. The full version is available online.}
\label{fig:superspec}
\end{figure}

\begin{figure}
   \begin{center}
      \includegraphics[scale=0.4]{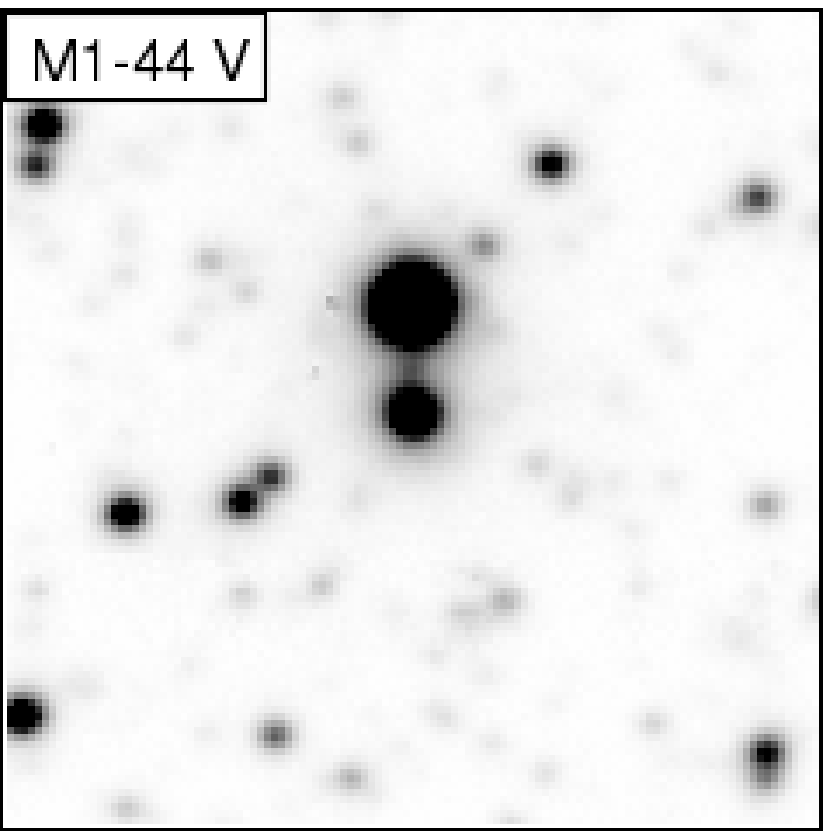}
      \includegraphics[scale=0.4]{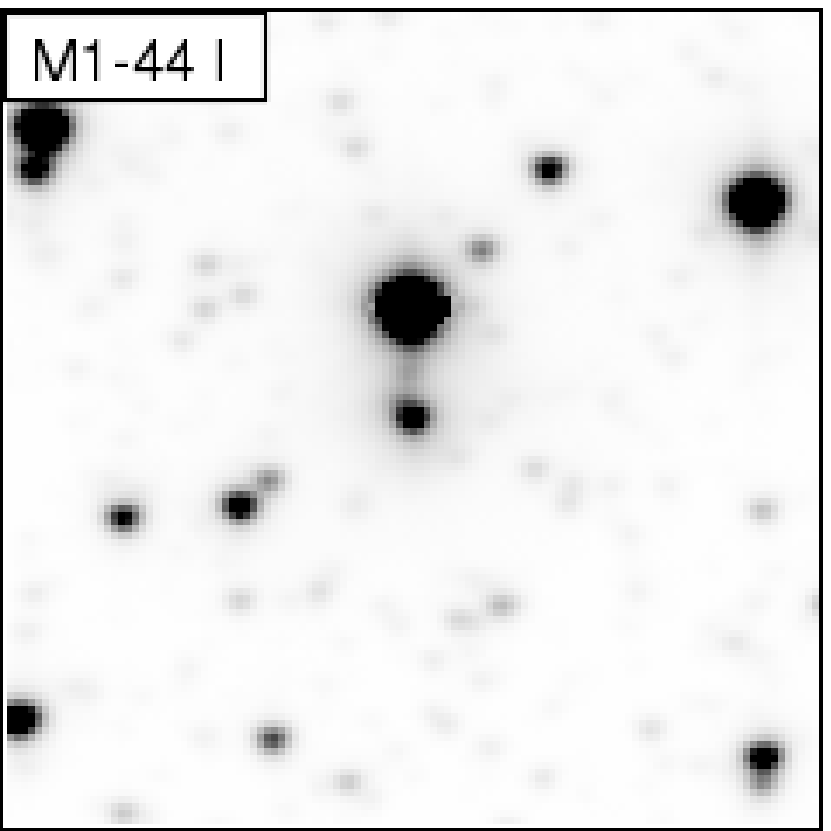}
   \end{center}
   \caption{Images of superpositions in Bulge PNe. The full version is available online.}
   \label{fig:superpos}
\end{figure}

\end{document}